\documentclass[pdflatex,sn-mathphys]{sn-jnl}

\jyear{2021}%

\usepackage{amsmath,amsfonts,bm}

\def\1{\bm{1}}

\def\rvg{{\mathbf{g}}}

\def\rvv{{\mathbf{v}}}

\DeclareMathAlphabet{\mathsfit}{\encodingdefault}{\sfdefault}{m}{sl}
\SetMathAlphabet{\mathsfit}{bold}{\encodingdefault}{\sfdefault}{bx}{n}

\theoremstyle{thmstyleone}%
\theoremstyle{thmstyletwo}%

\theoremstyle{thmstylethree}%

\usepackage[nomathsf,nomathsfit,noupgreek,notheorems,nocompactproof]{changdefs}

\usepackage{xcolor}
\newcommand{\simple}{\textnormal{simple}}
\newcommand{\base}{\textnormal{base}}

\newcommand{\data}{\textnormal{data}}

\newcommand{\SO}{\mathrm{SO}}
\newcommand{\so}{\mathfrak{so}}
\newcommand{\IG}{\mathcal{IG}}
\let\oldAA\AA
\renewcommand{\AA}{\textnormal{\oldAA}}

\usepackage{algorithm}
\usepackage{algpseudocode}
\usepackage{caption}
\usepackage{subcaption}
\usepackage{makecell}
\usepackage{ulem}
\usepackage{float}

\DeclareCaptionLabelFormat{adja-page}{\hrulefill\\#1 #2}

\begin{document}

\title[Distributional Graphormer]{Towards Predicting Equilibrium Distributions for Molecular Systems with Deep Learning}


\author*[1]{\fnm{Shuxin} \sur{Zheng}}\email{\{shuxin.zheng, chang.liu, haiguang.liu, tie-yan.liu\}@microsoft.com}
\equalcont{These authors contributed equally to this work.}

\author[1]{\fnm{Jiyan} \sur{He}}
\equalcont{These authors contributed equally to this work.}

\author*[1]{\fnm{Chang} \sur{Liu}}
\equalcont{These authors contributed equally to this work.}

\author[1]{\fnm{Yu} \sur{Shi}}
\equalcont{These authors contributed equally to this work.}

\author[1]{\fnm{Ziheng} \sur{Lu}}
\equalcont{These authors contributed equally to this work.}

\author[1]{\fnm{Weitao} \sur{Feng}}

\author[1]{\fnm{Fusong} \sur{Ju}}

\author[1]{\fnm{Jiaxi} \sur{Wang}}

\author[1]{\fnm{Jianwei} \sur{Zhu}}

\author[1]{\fnm{Yaosen} \sur{Min}}

\author[1]{\fnm{He} \sur{Zhang}}

\author[1]{\fnm{Shidi} \sur{Tang}}

\author[1]{\fnm{Hongxia} \sur{Hao}}

\author[1]{\fnm{Peiran} \sur{Jin}}

\author[2]{\fnm{Chi} \sur{Chen}} 

\author[1]{\fnm{Frank} \sur{No{\'e}}}

\author*[1]{\fnm{Haiguang} \sur{Liu}}
\equalcont{These authors contributed equally to this work.}

\author*[1]{\fnm{Tie-Yan} \sur{Liu}}

\affil[1]{\large Microsoft Research AI4Science}
\affil[2]{\large Microsoft Quantum}
\affil{\url{https://DistributionalGraphormer.github.io}}

\abstract{
Advances in deep learning have greatly improved structure prediction of molecules. However, many macroscopic observations that are important for real-world applications are not functions of a single molecular structure, but rather determined from the equilibrium distribution of structures. Traditional methods for obtaining these distributions, such as molecular dynamics simulation, are computationally expensive and often intractable. In this paper, we introduce a novel deep learning framework, called Distributional Graphormer (DiG), in an attempt to predict the equilibrium distribution of molecular systems. Inspired by the annealing process in thermodynamics, DiG employs deep neural networks to transform a simple distribution towards the equilibrium distribution, conditioned on a descriptor of a molecular system, such as a chemical graph or a protein sequence. This framework enables efficient generation of diverse conformations and provides estimations of state densities. We demonstrate the performance of DiG on several molecular tasks, including protein conformation sampling, ligand structure sampling, catalyst-adsorbate sampling, and property-guided structure generation. DiG presents a significant advancement in methodology for statistically understanding molecular systems, opening up new research opportunities in molecular science.
}


\keywords{Equilibrium Distribution, Statistical Mechanics, Deep Learning, Molecular States}

\maketitle


\newpage

\section{Main}\label{sec:main}

\begin{figure}[!ht]
     \centering
     \includegraphics[width=\textwidth]{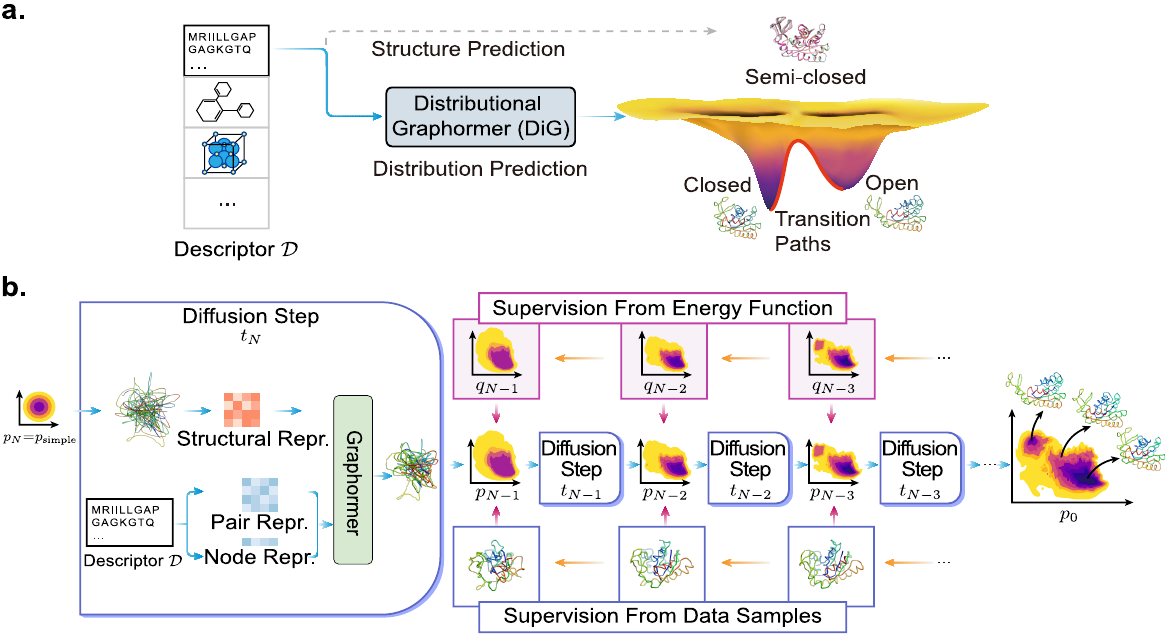}
    \caption{\textbf{Predicting conformational distributions with the Distributional Graphormer (DiG) framework.} 
    (a) DiG takes the basic descriptor $\clD$ of a target molecular system as input, e.g., amino acid sequence, to generate a probability distribution of structures which aims at approximating the equilibrium distribution and sampling different metastable states or intermediate states. In contrast, static structure prediction methods, such as AlphaFold~\citep{jumper2021highly}, aim at predicting one single high-probability structure of a molecule. 
    (b) The DiG framework for predicting distributions of molecular structures.
    A deep-learning model (Graphormer~\cite{ying2021transformers}) is used as modules to predict a diffusion process ({\color{cyan}$\rightarrow$}) that gradually transforms a simple distribution towards the target distribution.
    The model is learned so that the derived distribution $p_i$ in each intermediate diffusion time step $i$ matches the corresponding distribution $q_i$ in a predefined diffusion process ({\color{orange}$\leftarrow$}) that is set to transform the equilibrium distribution to the simple distribution.
    Supervision can be obtained from both samples (lower row), and a molecular energy function (upper row).
    }
    \label{fig:1}
\end{figure}

Deep learning methods are now state of the art to predict structures of molecular systems with high efficiency. For example, AlphaFold achieves atomic-level accuracy in protein structure predictions~\cite{jumper2021highly}, and has enabled new applications in structural biology~\cite{cramer2021alphafold2,akdel2022structural,pereira2021high}; fast docking methods based on deep neural networks have been developed and applied to predict ligand binding structures~\cite{stark2022equibind,corso2023diffdock}, supporting virtual screening in drug discovery~\cite{diaz2022deep,scardino2023good}; deep learning models predict the relaxed structures of adsorbates on catalyst surfaces~\cite{chanussot2021open, ying2021transformers, chen2022universal, schaarschmidt2022learned}. All these developments demonstrate the potential of deep learning approaches in modeling molecular structures and states.

However, accurate prediction of the most probable structure only reveals a small portion of the information needed to understand a molecular system in equilibrium. In reality, molecules can be highly flexible and the equilibrium distribution is crucial for studying statistical mechanical properties.
For example, functions of some biomolecules can be inferred from the probabilities associated with structures to identify metastable states; also based on probabilistic densities in the structure space, thermodynamic properties, such as entropy and free energies, can be computed by applying statistical mechanics methods.

Fig.~\ref{fig:1}a illustrates the difference between conventional structure prediction and the prediction of distributions of molecular structures. Although adenylate kinase has two distinct experimentally known conformations (open and closed states), a predicted structure usually corresponds to
a highly probable metastable state or a low-probability intermediate state (as shown in this figure).
A method is desired to allow us to sample the equilibrium distribution of adenylate kinase structures containing both functional states and their relative probabilities.

In contrast to the prediction of single structures, the prediction of equilibrium distributions still relies on classical and computationally expensive simulation methods while the development of deep learning methods for this task is still in its infancy. Most commonly, equilibrium distributions are sampled with molecule dynamics simulations which are computationally costly or even intractable~\cite{lindorff2011fast}. Enhanced sampling simulations~\cite{barducci2011metadynamics,kastner2011umbrella} and Markov state modeling~\cite{chodera2014markov} can speed up rare event sampling, but rely on system-specific choices such as collective variables along which the sampling is enhanced, and is thus not an easily generalizable approach. A popular approach is coarse-grained molecular dynamics~\cite{monticelli2008martini,clementi2008coarse} for which deep learning approaches have recently been developed~\cite{wang2019machine,arts2023two} that have shown promising results for individual molecular systems but not yet demonstrated generalization. Boltzmann Generators~\cite{noe2019boltzmann} are a deep learning approach to generate equilibrium distributions by constructing a probability flow from an easy-to-sample reference state, but due to the flow architecture~\cite{kingma2018glow} this approach is also difficult to generalize to different molecules. Generalization has been demonstrated for flows generating long timesteps for small peptides, but these methods have not yet scaled to large proteins~\cite{klein2023timewarp}.

In this work, we develop the \uline{Di}stributional \uline{G}raphormer (DiG), a new deep learning approach aiming to approximately predict the equilibrium distribution and efficiently sample diverse and chemically plausible structures of molecular systems. We show that DiG can generalize across molecular systems and propose diverse structures for molecules not used during training that resemble experimentally known structures.
DiG draws inspiration from simulated annealing~\cite{kirkpatrick1983optimization,neal2001annealed, del2006sequential,doucet2022annealed}, which produces a complex distribution by gradually reﬁning a simple uniform distribution through the simulation of an annealing process. Following this idea, DiG reduces the difficulty in the equilibrium distribution prediction problem by simulating a diffusion process that gradually transforms a simple distribution to the target distribution that aims at approximating the equilibrium distribution of the given molecular system~\citep{sohl2015deep,ho2020denoising} (\figref{1}b, {\color{cyan}$\rightarrow$}). 
The diffusion process is realized by a deep-learning model that is based upon the Graphormer architecture (\figref{1}b,~\citep{ying2021transformers}), and that is conditioned on a descriptor of the target molecule, such as a chemical graph or an amino acid sequence.
DiG can be trained using structure data from MD simulations and experiments. For cases where such data are not sufficient, we develop a novel Physics-Informed Diffusion Pre-training (PIDP) method to train DiG directly under the supervision from energy functions (force fields) of the systems. 
In both modes, the model receives a training signal in each diffusion step independently (\figref{1}b, {\color{orange}$\leftarrow$}), enabling efficient training that avoids backpropagating through the entire diffusion process.

The performance of DiG is evaluated on three prediction tasks: protein conformation distribution, ligand conformation distribution, and molecular adsorption distribution on catalyst surfaces. We demonstrate that DiG is capable of generating realistic and diverse molecular structures in these tasks. For the proteins shown in this paper, DiG efficiently generated structures to resemble major functional states, but with orders of magnitude less time than required for MD simulation.
We also demonstrate that DiG can facilitate inverse design of molecular structures by applying biased distributions that favor structures with desired properties. This capability has the potential to broaden the scope of molecular design for properties that lack adequate data to guide the design process. 
These results indicate that DiG significantly advances deep learning methodology for molecules from predicting a single structure towards predicting probability distributions of molecular structures, paving the way for efficient prediction of thermodynamic properties of molecules.

\section{The Framework of Distributional Graphormer} \label{sec:framework}

Deep neural networks have been demonstrated to predict accurate molecular structures from descriptors $\clD$ for many molecular systems~\citep{jumper2021highly,stark2022equibind,corso2023diffdock,chanussot2021open, ying2021transformers, chen2022universal, schaarschmidt2022learned}. Here, DiG aims to take one step further to predict not only the most probable structure, but also diverse structures with probabilities under the equilibrium distribution. 
To tackle this challenge, inspired by the heating-annealing paradigm, we break down the difficulty of this problem into a series of simpler problems. The heating-annealing paradigm can be viewed as a pair of reciprocal stochastic processes on the structure space that simulate the transformation between the equilibrium distribution and a system-independent simple distribution $p_\simple$.
Following this idea, we employ an explicit diffusion process (forward process; \figref{1}b orange arrows) that gradually transforms the target distribution of the molecule $q_{\clD,0}$, as the initial distribution, towards $p_\simple$ through a time period $\tau$.
The corresponding reverse diffusion process then transforms $p_\simple$ back to the target distribution $q_{\clD,0}$. This is the generation process of DiG (\figref{1}b, blue arrows).
The reverse process is performed by updates predicted by deep neural networks from the given $\clD$, which are trained to match the forward process.
Compared to directly predicting the equilibrium distribution from $\clD$, the heating-annealing paradigm significantly reduces the difficulty of this problem. As $p_\simple$ is chosen to enable independent sampling and have a closed-form density function, DiG enables independent sampling of the equilibrium distribution by simulating the reverse process started from $p_\simple$, and also provides a density function for the distribution by tracking the process.

Specifically, we choose $p_\simple := \clN(\bfzro,\bfI)$ as the standard Gaussian distribution in the state space, and the forward diffusion process as the Langevin diffusion process targeting this $p_\simple$ (Ornstein–Uhlenbeck process)~\citep{langevin1908theorie,uhlenbeck1930theory,roberts1996exponential}.
A time dilation scheme $\beta_t$~\citep{wibisono2016variational} is introduced for approximate convergence to $p_\simple$ after a finite time $\tau$. The result is written as the following stochastic differential equation (SDE):
\begin{align}
    \ud \bfR_t = -\frac{\beta_t}2 \bfR_t \dd t + \sqrt{\beta_t} \dd \bfB_t,
    \label{eqn:fwd-cont}
\end{align}
where $\bfB_t$ is the standard Brownian motion (a.k.a Wiener process). Choosing this forward process leads to a $p_\simple$ that is more concentrated than a heated distribution hence it is easier to draw high-density samples, and the form of the process enables efficient training and sampling.

Following stochastic process theory (e.g.,~\citep{anderson1982reverse}), the reverse process is also a stochastic process, written as the following SDE:
\begin{align}
    \ud \bfR_{\ttb} = \frac{\beta_{\ttb}}2 \bfR_{\ttb} \dd \ttb + \beta_{\ttb} \nabla \log q_{\clD,\ttb}(\bfR_{\ttb}) \dd \ttb + \sqrt{\beta_{\ttb}} \dd \bfB_{\ttb},
    \label{eqn:rev-cont}
\end{align}
where $\ttb := \tau-t$ is the reversed time, $q_{\clD,\ttb} := q_{\clD, t = \tau-\ttb}$ is the forward-process distribution at the corresponding time, and $\bfB_{\ttb}$ is the Brownian motion in reversed time.
To recover $q_{\clD,0}$ from $p_\simple$ by simulating this reverse process, deep neural networks are employed to construct a score model $\bfss^\theta_{\clD,t}(\bfR)$, which is trained to predict the true score function $\nabla \log q_{\clD,t}(\bfR)$ of each instantaneous distribution $q_{\clD,t}$ from the forward process.
This formulation is called diffusion-based generative model and has been demonstrated to be able to generate high-quality samples of images and other content~\citep{sohl2015deep,ho2020denoising,song2021score,dhariwal2021diffusion,ramesh2022hierarchical}.
As our score model is defined in molecular conformational space, we employ our previously developed Graphormer model~\citep{ying2021transformers} as the neural network architecture backbone of DiG, to leverage its capabilities in modeling molecular structures and to generalize to a range of molecular systems.

With the $\bfss^\theta_{\clD,t}(\bfR)$ model, drawing a sample $\bfR_0$ from the equilibrium distribution of a system $\clD$ can be done by simulating the reverse process \eqnref{rev-cont} on $N+1$ steps that uniformly discretizes $[0,\tau]$ with step size $h = \tau/N$ (\figref{1}b, blue arrows):
\begin{align}
    & \bfR_N \sim p_\simple, \\*
    & \bfR_{i-1} = \frac{1}{\sqrt{1-\beta_i}} \Big( \bfR_i + \beta_i \bfss^\theta_{\clD,i}(\bfR_i) \Big) + \clN(\bfzro, \beta_i \bfI), \, i = N, \cdots, 1,
    \label{eqn:rev-disc}
\end{align}
where the discrete step index $i$ corresponds to time $t = i h$, and $\beta_i := h \beta_{t = i h}$.
Note that the reverse process does not need to be ergodic. The way that DiG models the equilibrium distribution is using the instantaneous distribution at the instant $t=0$ (or $\ttb = \tau$) on the reverse process, but not using a time average.
As $\bfR_N$ samples can be drawn independently, DiG can generate statistically independent $\bfR_0$ samples for the equilibrium distribution. In contrast to Molecular Dynamics (MD) or Markov Chain Monte Carlo (MCMC) simulations, generation of DiG samples does not suffer from rare events, and can thus be far more computationally efficient.

\subsubsection*{\textit{Physics-Informed Diffusion Pre-training}} \label{sec:pidp-method}

DiG can be trained by conformation data sampled over a range of molecular systems. However, collecting sufficient experimental or simulation data to characterize the equilibrium distribution for various systems is extremely costly. To address this data scarcity problem, we propose a novel pre-training algorithm, called Physics-Informed Diffusion Pre-training (PIDP), which effectively optimizes DiG on an initial set of candidate structures that need not to be sampled from the equilibrium distribution. 
The supervision comes from the energy function $E_\clD$ of each system $\clD$, which defines the equilibrium distribution $q_{\clD,0}(\bfR) \propto \exp(-\frac{E_\clD(\bfR)}{k_\rmB T})$ at the target temperature $T$.

The key idea is that the true score function $\nabla \log q_{\clD,t}$ from the forward process \eqnref{fwd-cont} obeys a partial differential equation, known as the Fokker-Planck equation (e.g.,~\citep{risken1996fokker}). We then pre-train the score model $\bfss^\theta_{\clD,t}$ by minimizing the following loss function that enforces the equation to hold:
{\small
\begin{align}
    & \sum_{i=1}^N \frac1M \sum_{m=1}^M \Big\Vert
        \frac{\beta_i}{2} \Big(
            \nabla \big( \bfR_{\clD,i}^{(m)} \cdot \bfss^\theta_{\clD,i}(\bfR_{\clD,i}^{(m)}) \big)
            + \nabla \lrVert*[\big]{\bfss^\theta_{\clD,i}(\bfR_{\clD,i}^{(m)})}^2
         + \nabla \big( \nabla \cdot \bfss^\theta_{\clD,i}(\bfR_{\clD,i}^{(m)}) \big)
        \Big)
        \\*
        & \quad\quad {} - \fracpartial{}{t} \bfss^\theta_{\clD,i}(\bfR_{\clD,i}^{(m)})
    \Big\Vert^2 
    {}+ \frac{\lambda_1}M \sum_{m=1}^M \Big\Vert
        \frac1{k_\rmB T} \nabla E_\clD(\bfR_{\clD,1}^{(m)}) + \bfss^\theta_{\clD,1}(\bfR_{\clD,1}^{(m)})
    \Big\Vert^2.
    \label{eqn:pinnloss}
\end{align} 
}%
Here, the second term, weighted by $\lambda_1$, matches the score model at the final generation step to the score from the energy function, and the first term implicitly propagates the energy-function supervision to intermediate time steps (\figref{1}b, upper row).
The structures $\{\bfR_{\clD,i}^{(m)}\}_{m=1}^M$ to evaluate the loss are points on a grid spanning the structure space. What is favorable is that, these structures do not have to obey the equilibrium distribution (as is required by data structures), since they are only used to discretize functions in the structure space, therefore the cost of preparing these structures can be much lower. As structure spaces of molecular systems are often very high-dimensional (e.g., thousands for proteins), a regular grid would have intractably many points. Fortunately, the space of actual interest is only a low-dimensional manifold of physically reasonable structures (structures with low energy) relevant to the problem.
This allows us to effectively train the model only on these relevant structures as $\bfR_0$ samples, and pass them through the forward process for $\bfR_i$ samples.
See Supplementary Sec.~\ref{sec:training-details-protein} for an example on acquiring relevant structures for protein systems.

We also leverage stochastic estimators including Hutchinson's estimator~\citep{hutchinson1989stochastic,grathwohl2019ffjord} to reduce the complexity in calculating derivatives of high-order and for high-dimensional vector-valued functions. Note that for each step $i$, the corresponding model $\bfss^\theta_{\clD,i}$ receives a training loss independent of other steps and can be directly back-propagated. This step-by-step supervision pattern helps to achieve efficient pre-training.

\subsubsection*{\textit{Training DiG with Data}} \label{sec:training-with-data-method} 

In addition to using the energy function for information on the probability distribution of the molecular system, DiG can also be trained with molecular structure samples which can be obtained from experimental structure determination methods, molecular dynamics, or other simulation methods. See Supplementary Sec.~\ref{sec:training-details} for data collection details.
Even when the simulation data is limited, they still provide information about the regions the distribution needs to cover and the local shape of the distribution, hence are helpful to improve a pre-trained DiG. 
To train DiG on data, the score model $\bfss^\theta_{\clD,i}(\bfR_i)$ is matched to the corresponding score function $\nabla \log q_{\clD,i}$ demonstrated by data samples. This can be done by minimizing $\bbE_{q_{\clD,i}(\bfR_i)} \lrVert{\bfss^\theta_{\clD,i}(\bfR_i) - \nabla \log q_{\clD,i}(\bfR_i)}^2$ for each diffusion time step $i$. Although the precise calculation of $\nabla \log q_{\clD,i}$ is impractical, the loss function can be equivalently reformulated into denoising score-matching form~\cite{vincent2011connection,alain2014regularized}:
\begin{align}
    \frac1N \sum_{i=1}^N \bbE_{q_{\clD,0}(\bfR_0)} \bbE_{p(\bfeps_i)} \lrVert{\sigma_i \bfss^\theta_{\clD,i}(\alpha_i \bfR_0 + \sigma_i \bfeps_i) + \bfeps_i}^2, \label{eqn:scoreloss}
\end{align}
where $\alpha_i := \prod_{j=1}^i \sqrt{1-\beta_j}$, $\sigma_i := \sqrt{1 - \alpha_i^2}$, and $p(\bfeps_i)$ is the standard Gaussian distribution. The expectation under $q_{\clD,0}$ can be estimated using the simulation dataset.
Note that this function allows direct loss estimation and backpropagation for each $i$ in constant (w.r.t $i$) cost, recovering the efficient step-by-step supervision again (\figref{1}b, lower row).

\subsubsection*{\textit{Density Estimation by DiG}}\label{sec:density}

Many thermodynamic properties of a molecular system (e.g., free energy, entropy) also require calculating the density function of the equilibrium distribution, which is another aspect of the distribution besides a sampling method.
DiG allows for this by tracking the distribution change along the diffusion process~\citep{song2021score}:
\begin{align}
    \log p^\theta_{\clD,0}(\bfR_0)
    ={} & \log p_\simple \lrparen*[\big]{ \bfR^\theta_{\clD,\tau}(\bfR_0) } \\
        & {}- \int_0^\tau \frac{\beta_t}2 \nabla \cdot \bfss^\theta_{\clD,t} \lrparen*[\big]{ \bfR^\theta_{\clD,t}(\bfR_0) } \dd t
    - \frac{D}{2} \int_0^\tau \beta_t \dd t,
    \label{eqn:logdensity}
\end{align}
where $D$ is the dimension of the state space, and $\bfR^\theta_{\clD,t}(\bfR_0)$ is the solution to the ordinary differential equation (ODE):
\begin{align}
    \ud \bfR_t
    = -\frac{\beta_t}{2} \lrparen*[\Big]{ \bfR_t + \bfss^\theta_{\clD,t}(\bfR_t) } \dd t,
    \label{eqn:rev-ode}
\end{align}
with initial condition $\bfR_0$, which can be solved using standard black box ODE solvers or more efficient specific solvers (Supplementary Sec.~\ref{appx:accel-infer}).

\subsubsection*{\textit{Property-Guided Structure Generation with DiG}} 
\label{sec:method-inverse}

There is a growing demand for inverse design of materials and molecules. The goal is to find structures with desired properties, such as intrinsic electronic band gaps, elastic modulus, and ionic conductivity, without going through a forward searching process.
DiG provides a feature to enable such property-guided structure generation, by directly predicting the conditional structural distribution given a value $c$ of a microscopic property.

To achieve this, regarding the data-generating process in \eqnref{rev-cont}, we only need to adapt the score function, from $\nabla \log q_{\clD,t}(\bfR)$ to $\nabla_\bfR \log q_{\clD,t}(\bfR \mid c)$. Using Bayes' rule, the latter can be reformulated as $\nabla_\bfR \log q_{\clD,t}(\bfR \mid c) = \nabla \log q_{\clD,t}(\bfR) + \nabla_\bfR \log q_\clD(c \mid \bfR)$, where the first term can be approximated by the learned (unconditioned) score model, i.e. the new score model is:
\begin{align}
    \bfss^\theta_{\clD,i}(\bfR_i \mid c) = \bfss^\theta_{\clD,i}(\bfR_i) + \nabla_{\bfR_i} \log q_\clD(c \mid \bfR_i).
    \label{eqn:conditional-score}
\end{align}
Hence, only a $q_\clD(c \mid \bfR)$ model is additionally needed~\cite{song2021score,dhariwal2021diffusion}, which is a property predictor or classifier that is much easier to train than a generative model. 

It is noted that in a normal workflow for machine-learning (ML) inverse design, a dataset must be generated to meet the conditional distribution, then an ML model will be trained on this dataset for structure predictions. The ability to generate structures for conditional distribution without requiring a conditional dataset places DiG in an advantageous position when compared to the normal workflow in terms of efficiency and computational cost. 

\subsubsection*{\textit{Interpolation between States}} 
Given two states, DiG can approximate a reaction path that corresponds to reaction coordinates or collective variables, and find intermediate states along the transition pathway.
This is achieved through the fact that the distribution transformation process described in \eqnref{fwd-cont} is equivalent to the process in \eqnref{rev-ode} if $\bfss^\theta_{\clD,i}$ is well learned, which is deterministic and invertible hence establishes a correspondence between the structure and latent space. 
We can then uniquely map the two given states in the structure space to the latent space, approximate the path in the latent space by linear interpolation, and then map the path back to the structure space. Since the distribution in the latent space is Gaussian which has a convex contour, the linearly interpolated path goes through high-probability or low-energy regions, so it gives an intuitive guess of the real reaction path.

\section{Results}\label{sec:result}

Here, we demonstrate that DiG can be applied to study protein conformations, protein-ligand interactions, and molecule adsorption on catalysis surfaces. In addition, we investigate the inverse design capability of DiG, through its application to carbon polymorph generation for desired electronic band gaps.

\subsection{Protein Conformation Sampling}
\label{sec:result-prot-sampling}

\begin{figure} [b!]
    \centering
    \includegraphics[width=\textwidth]{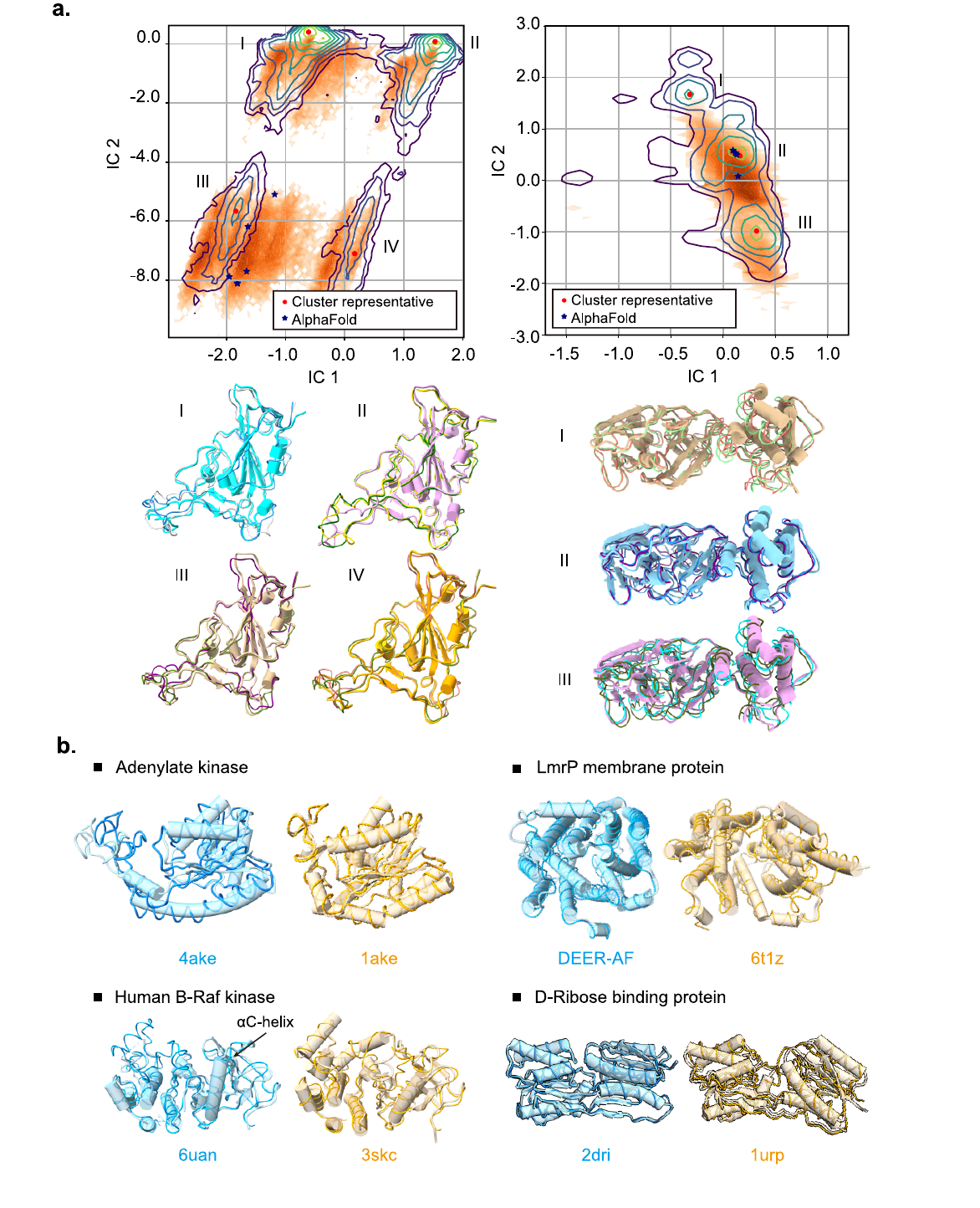}
    \caption{\textbf{Distribution and sampling results for protein conformations.}}
\end{figure}
\begin{figure} [t!]
    \captionsetup{labelformat=adja-page}
    \ContinuedFloat
    \caption{(a) Structures generated by DiG resemble the diverse conformations of millisecond MD simulations. MD simulated structures are projected onto the reduced 2D space spanned by TICA coordinates, and the probability densities are depicted using contour lines. For RBD protein, MD simulation reveals four highly populated regions in the 2D space spanned by TICA coordinates (left panel). Structures generated by DiG are mapped to this 2D space shown as orange dots, whose distributions are reflected by the color intensity. Below the distribution map, structures generated by DiG (thin ribbons) are superposed to representative structures of four clusters. AlphaFold predicted structures ({\color{blue} $\star$}) are also shown in the plot. Right panel shows the results of the main protease of SARS-CoV-2, compared with MD simulations and AlphaFold prediction results. The contour map reveals three clusters, DiG generates highly similar structures in cluster II \& III, while structures in cluster-I are accurately generated.  (b) The performance of DiG on generating multiple conformations of proteins (each structure is labeled by its PDB ID, except the DEER-AF, which is AlphaFold predicted model that is consistent with experimental observations). Structures generated by DiG (thin ribbons) are compared with the experimentally determined structures (cylindrical cartoons) in each case. For the four proteins (adenylate kinase, Lmrb membrane protein, human B-Raf kinase, and D-ribose binding protein), structures in two functional states (distinguished by {\color{cyan}cyan} and {\color{brown}brown}) are well reproduced by DiG (ribbons).}
    \label{fig:2}
\end{figure}

At physiological conditions, most protein molecules exhibit dynamical behaviors, rather than existing as rigid objects in their most energetically favorable states. The sampling of these conformations is crucial for the comprehensive understanding of protein properties and their interactions with other molecules in cells. Recently, it has been reported that AlphaFold~\cite{jumper2021highly} can generate alternative conformations for certain proteins, by manipulating input information such as multiple sequence alignments (MSA)~\citep{del2022sampling}. However, this approach is developed on the basis of varying the depth of MSA, it is hard to generalize to all proteins (especially for those with a small number of similar sequences). Therefore, it is highly desirable to have advanced AI models that can sample diverse structures consistent with the energy landscape in the conformational space~\cite{del2022sampling}. Here, we show that DiG is capable of generating diverse and functionally relevant protein structures, which is a key capability for being able to efficiently sample equilibrium distributions.

It is noted that the equilibrium distribution of protein conformations is difficult to obtain experimentally or computationally, so in contrast to protein structure prediction, there is a lack of high-quality data for training or benchmarking. To train this model, we collect experimental and simulated structures from public databases. In order to mitigate the data scarcity issue, besides the structures from the protein databank, we also generated an in-house simulation dataset and developed the PIDP training method (See Supplementary Sec.~\ref{appx:method-pidp} and~\ref{appx:md-prot-complex} for training procedure and the dataset).
The performance of DiG was assessed at two levels: (1) comparing the conformational distributions against those obtained from extensive (millisecond timescale) atomistic MD simulations; (2) validating on proteins with multiple known conformations.
As shown in Fig.~\ref{fig:2}a, the conformational distributions are obtained from MD simulations for two proteins from the SARS-CoV-2 virus~\cite{zimmerman2021sars} (the receptor-binding-domain (RBD) of spike protein and the main protease, also known as 3CL protease, see Supplementary Sec.~\ref{appx:eval_method} for details on MD simulation data). These two proteins are the crucial components of the SARS-CoV-2 virus and key targets for drug development in the treatment of COVID-19~\cite{zhang2020crystal,tai2020characterization}. The millisecond timescale MD simulations extensively sample conformation space, and we therefore regard the resulting distribution as a proxy to the equilibrium distribution.
Taking protein sequences as the descriptor inputs for DiG, structures were generated for these two proteins. Although MD simulation data of these proteins were not used for DiG training, the generated structures resemble the conformational distributions explored by MD in the reduced dimension space spanned by collective variables (Fig.~\ref{fig:2}a). In the 2D projection shown here, the MD simulations of RBD populate four regions, which are also sampled by DiG (see Fig.~\ref{fig:2}a, left panel). The four representative structures corresponding to the cluster centers are well generated by DiG. Similarly, three representative structures for main protease were obtained by clustering analysis on MD simulation trajectories, and then the generated structures were aligned to these three representatives (Fig.~\ref{fig:2}a). We noticed that conformations in cluster-I region are not well recovered by DiG, indicating room for improvement. In terms of conformational space coverage, we compared the DiG sampled regions with those explored by MD simulations in the conformation manifold spanned by the TICA variables (Fig.~\ref{fig:2}a). For example, on the 2D manifold, about 70\% of the RBD conformations sampled by millisecond-scale MD simulations can be covered with just 10,000 DiG-generated structures (see Supplementary Fig.~\ref{fig:figs1} for details).

Atomistic MD simulations are computationally very expensive, therefore millisecond time scale simulations of proteins are rarely reported in literature, except for simulations on special-purpose hardware such as the Anton supercomputer~\cite{lindorff2011fast} or extensive distributed simulations combined in Markov state models~\cite{chodera2014markov}. In order to get an additional assessment on the diversity of protein structures generated by DiG, we turn to proteins for which multiple structures have been experimentally determined. Although it is a less stringent test, the capability of sampling alternative conformations can facilitate the research of protein dynamics and functional mechanisms.
We analyzed four proteins, each with two distinguishable conformations corresponding to different functional states (Fig.~\ref{fig:2}b). Remarkably, the conformations sampled by DiG have good coverage in the conformational space near the two states for each protein. The experimentally determined conformations are shown in cylinder cartoons, each aligned with two structures generated by DiG (shown in ribbon representations). For example, the adenylate kinase protein has two conformations (PDB IDs 1ake and 4ake), each with high-quality structures in their vicinity (backbone RMSD $<$ 1.0 \r{A} for the structure superposed to the closed state, 1ake; backbone RMSD $<$ 3.0 \r{A} for the structures superposed to the open state, 4ake). Similarly, for the drug transport protein LmrP, DiG generated structures resembling both states. We note that one structure is experimentally determined, and the other (denoted as DEER-AF) is the AlphaFold predicted structure~\cite{del2022sampling} supported by double electron electron resonance (DEER) experimental data~\cite{masureel2014protonation}. For the case of human B-Raf kinase, the overall RMSD difference between the two experimentally determined states is not as pronounced as in the other three proteins. The major structural difference is in the A-loop region and a nearby helix ($\alpha$ C-helix, indicated in the figure)~\cite{nussinov2022alphafold}. Structures generated by DiG accurately recover such regional structural differences in this kinase protein. Another interesting case is the D-Ribose binding protein with two separated domains, which can be packed in two distinct conformations. DiG correctly generates structures corresponding to both the {\color{cyan}straight-up conformation} (cylinder cartoon) and the {\color{brown}twisted/tilted conformation}. It is noted that if we align one domain of D-ribose binding protein, the other domain only partially matches the twisted conformation as an `intermediate' state. Furthermore, for a pair of structures of the same protein, DiG can be applied to generate transition pathways by latent space interpolations (see demonstration cases in the DiG webpage: \url{https://DistributionalGraphormer.github.io}). 
The dynamics revealed by such pathways can inspire hypotheses on molecular mechanisms for experimental validation. In summary, DiG is capable of generating diverse protein structures corresponding to different functional states, thus going beyond the capabilities of current static structure prediction methods.

\begin{figure} [h!]
    \centering
    \includegraphics[width=\textwidth]{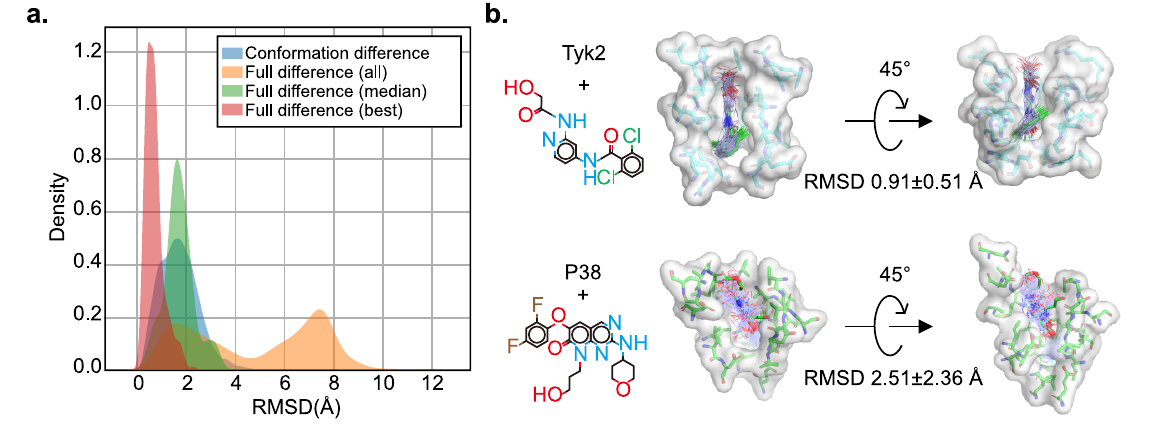}
    \caption{\textbf{Results of DiG for ligand structure sampling around protein pockets.} (a) The results of DiG on poses of ligands bound to protein pockets. DiG generates ligand structures and binding poses, with good accuracy compared to the crystal structures (reflected by the RMSD statistics shown in red histogram for the best matching cases, and the green histogram for the median RMSD statistics). When considering all 50 predicted binding poses for each system, diversity is observed, as reflected in the RMSD histogram (yellow color, normalized) compared to the references. (b) Representative systems show that the diversity in ligand binding poses is related to the binding pocket properties. For deep and narrow binding pocket such as for the Tyk2 protein (shown in the surface representation, top panel), DiG predicts highly similar binding poses for the ligand (in atom-bond representations, top panel). For the P38 protein the binding pocket is relatively flat and shallow and predicted ligand poses are highly diverse and have large conformational flexibility (bottom panel, in the same representations as in the Tyk2 case).}
    \label{fig:ligand}
\end{figure}

\subsection{Ligand Structure Sampling around Binding Sites}

An immediate extension of protein conformational sampling is to predict protein-ligand interactions, such as ligand binding positions in druggable pockets.  
To model the interactions between protein and ligand, we mainly use a simulation dataset of about 1500 complexes for training (See Supplementary Sec.~\ref{appx:md-prot-complex} for the dataset). 
We evaluated the performance of DiG in ligand binding to protein pockets for 409 protein-ligand systems~\cite{schindler2020large,wang2015accurate} (not in the training dataset). By providing atomic positions surrounding a pocket and a ligand descriptor (here, a SMILES string), DiG generates ligand structures to fit the pocket. During the ligand structure sampling, DiG models the atomic coordinate distribution of both binding pocket and the ligand. The flexible binding pockets were observed in the testing, with changes in atomic positions up to 1.0 \r{A} in terms of RMSD compared to the input atomic positions. For the ligand structures, the deviation comes from two sources: (1) the conformational difference between generated structures and experimental structures; and (2) the difference in the binding pose due to ligand translation and rotation. Among all tested cases, the conformational differences are small, with an RMSD value of 1.74 \r{A} on average, indicating that generated ligand structures are highly similar to the bound ligands resolved in crystal structures (Fig.~\ref{fig:ligand}a). When including the binding pose deviations originated from ligand positions and orientations, larger alignment discrepancies are observed for ligand structures. Yet, the DiG is still capable of predicting at least one correct structure for each ligand out of 50 generated structures. In a retrospective measurement, the best-matched structure among 50 generated structures for each ligand is within 2.0 \r{A} RMSD compared to the experimental data for nearly all 409 testing systems (Fig.~\ref{fig:ligand}a for the RMSD distribution, with more cases shown in Supplementary Fig.~\ref{fig:figs3}). The accuracy of generated structures for ligand is related to the characteristic of binding pockets. For example, the ligand binding to the target protein Tyk2 showed an average deviation of 0.91 \r{A} (RMSD) from the crystal structure (see Fig.~\ref{fig:ligand}b, top). In another example for target P38, the ligand exhibited more diverse binding poses, likely due to the shallow pocket of this target. Under such circumstances, the most stable binding pose may be less dominant compared to other favorable poses (Fig.~\ref{fig:ligand}b, bottom). MD simulations reveal similar trends as DiG-generated structures, with ligand binding to Tyk2 more tightly than the case of P38 (Supplementary Fig.~\ref{fig:figs2}). Overall, we observed that the generated structures indeed resemble experimentally observed poses. 

\subsection{Catalyst-Adsorbate Sampling}

\begin{figure}[!ht]
         \centering
         \includegraphics[width=\textwidth]{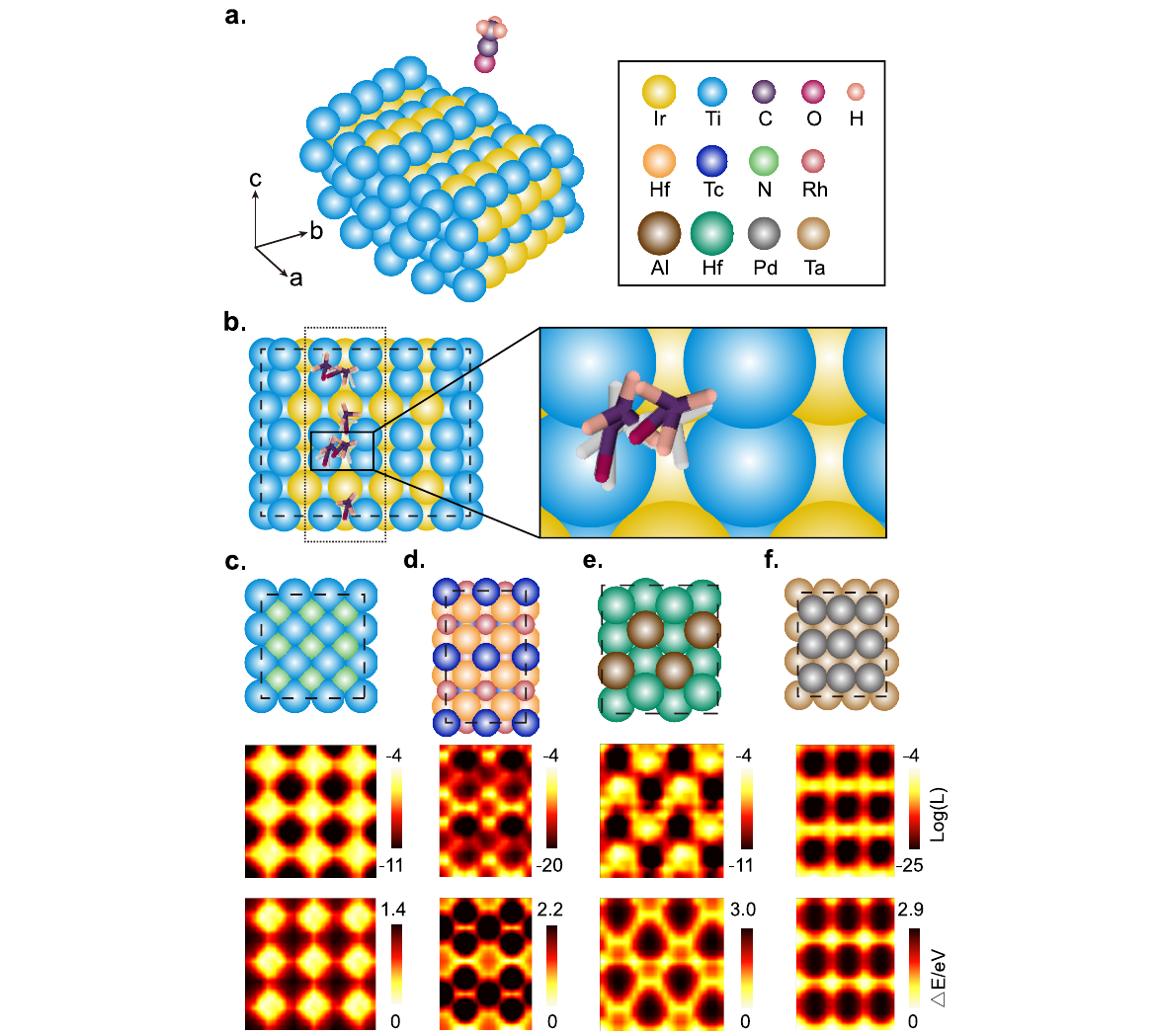}
         \caption{\textbf{Results of DiG for catalyst-adsorbate sampling problems.}
         (a) The problem setting: prediction of the adsorption configuration distribution of an adsorbate on a catalyst surface.
         (b) The adsorption sites and corresponding configurations of the adsorbate found by DiG (in color), compared with DFT results (in white). DiG finds all the adsorption sites, with adsorbate structures close to the DFT baseline. For all adsoprtion sites and configurations, refer to Appendix E.
         (c-f) Adsorption prediction results of single N and O atoms on catalyst surfaces, compared to DFT calculations. Top panels show the catalyst surface; the probability distribution of adsorbate molecules on the corresponding catalyst surfaces are shown in the middle panels in log-scale; the bottom panels show the calculated interactions between the adsorbate molecule and the catalyst using DFT methods. The adsorption sites and predicted probabilities are highly consistent with the energy landscape obtained by DFT computations.
         }
        \label{fig:catalyst_adsorbate}
\end{figure}

Identifying active adsorption sites is a central task in heterogeneous catalysis. Due to complex surface-molecular interactions, such tasks rely heavily on a combination of quantum chemistry methods such as density functional theory (DFT) and sampling techniques such as MD and grid-search. These lead to large and sometimes intractable computational costs, especially when it comes to surfaces with complex chemical environments. We evaluate DiG's capability for this task by training it on the MD trajectories of catalyst-adsorbate systems from the Open Catalyst Project and carrying out further evaluations on random combinations of adsorbates and surfaces that are not included in the training set~\cite{chanussot2021open}. By feeding the model with a substrate and a molecular adsorbate, DiG can predict adsorption sites and stable adsorbate configurations, along with the probability for each configuration (see Supplementary Sec.~\ref{sec:catalyst-adsorbate-sampling-details} for training details and Supplementary Sec.~\ref{appx:eval_method} for evaluation details). \figref{catalyst_adsorbate}a-b shows the adsorption configurations of an acyl group on a stepped TiIr alloy surface. Multiple adsorption sites are predicted by DiG. To test the plausibility of these predicted configurations and evaluate the coverage of the predictions, we carry out a grid-search using DFT methods. The results confirm that DiG predicts all stable sites found by the grid-search and the adsorption configurations are in close agreement with an RMSD of $0.5\sim0.8\,\AA$ (Fig.~\ref{fig:catalyst_adsorbate}b). It should be noted that the combination of substrate and adsorbate shown in Fig.~\ref{fig:catalyst_adsorbate}b is not included in the training data set. Therefore, the result demonstrates the cross-system generalization capability of DiG in catalyst adsorption predictions. Here we show only the top view, and Fig.~\ref{fig:figs4} in addition shows the front view of the adsorption configurations. 

DiG not only predicts the adsorption sites with correct configurations, but also provides a probability estimate for each adsorption configuration. This capability is illustrated in the systems with single-atom adsorbates (including H, N, and O) on 10 randomly chosen metallic surfaces. For each combination of adsorbate and catalyst substrate, the DiG is applied to predict the adsorption sites and the probability distributions. Then for the same systems, grid-search DFT calculations were carried out to find all adsorption sites and the corresponding energies.
Taking the adsorption sites identified by grid-search as references, DiG achieved 81\% site coverage for single-atom adsorbates on the 10 metallic catalyst surfaces.
Fig.~\ref{fig:catalyst_adsorbate}(c-f) show closer examinations on adsorption predictions for four systems, namely C, H, N, and O on TiN, RhTcHf, AlHf, and TaPd metallic surfaces (top panels). The predicted adsorption probabilities projected on the surface in parallel with the catalyst surface are shown in the middle panels. The log-scaled heatmaps of the probabilities show excellent accordance with the adsorption energies calculated using DFT methods (bottom panels). 
It is worth noting that the speed of DiG is much faster compared to DFT, i.e., it only takes about 1 minute to sample all adsorption sites for a catalyst-adsorbate system for DiG on a single modern GPU, but at least $2$ hours for a single DFT relaxation with VASP, which number will be further multiplied by a factor of $>100$ depending on the resolution of the searching grid~\cite{hafner2008ab}.
Such fast and accurate prediction of adsorption sites and the corresponding distributional features can be useful in identifying the catalytic mechanisms and guiding the search of new catalysts.

\subsection{Property-Guided Structure Generation}
While DiG by default generates structures following the learned training data distribution, the output distribution can be biased to steer the structure generation to meet particular requirements. Here we leverage this capability by employing DiG for inverse design (described in Sec.~\ref{sec:method-inverse}).
As a proof-of-concept, we search for carbon polymorphs with desired electronic band gaps. Similar tasks are critical to the discovery of novel photovoltaic and semi-conductive materials~\cite{lu2021computational}. 
To train this model, we prepared a structure dataset composed of carbon 
atoms by carrying out random structure search based on energy profiles obtained from DFT calculations~\cite{lucarbondataset}. The structures corresponding to energy minima form the dataset used to train 
DiG, which in turn are applied to generate carbon structures.  
We use a neural network model based on the M3GNet architecture~\cite{chen2022universal} as the property predictor for band gap, which is fed to the property-guided structure generation of carbon structures.

\begin{figure}[!ht]
    \centering
    \includegraphics[width=\textwidth]{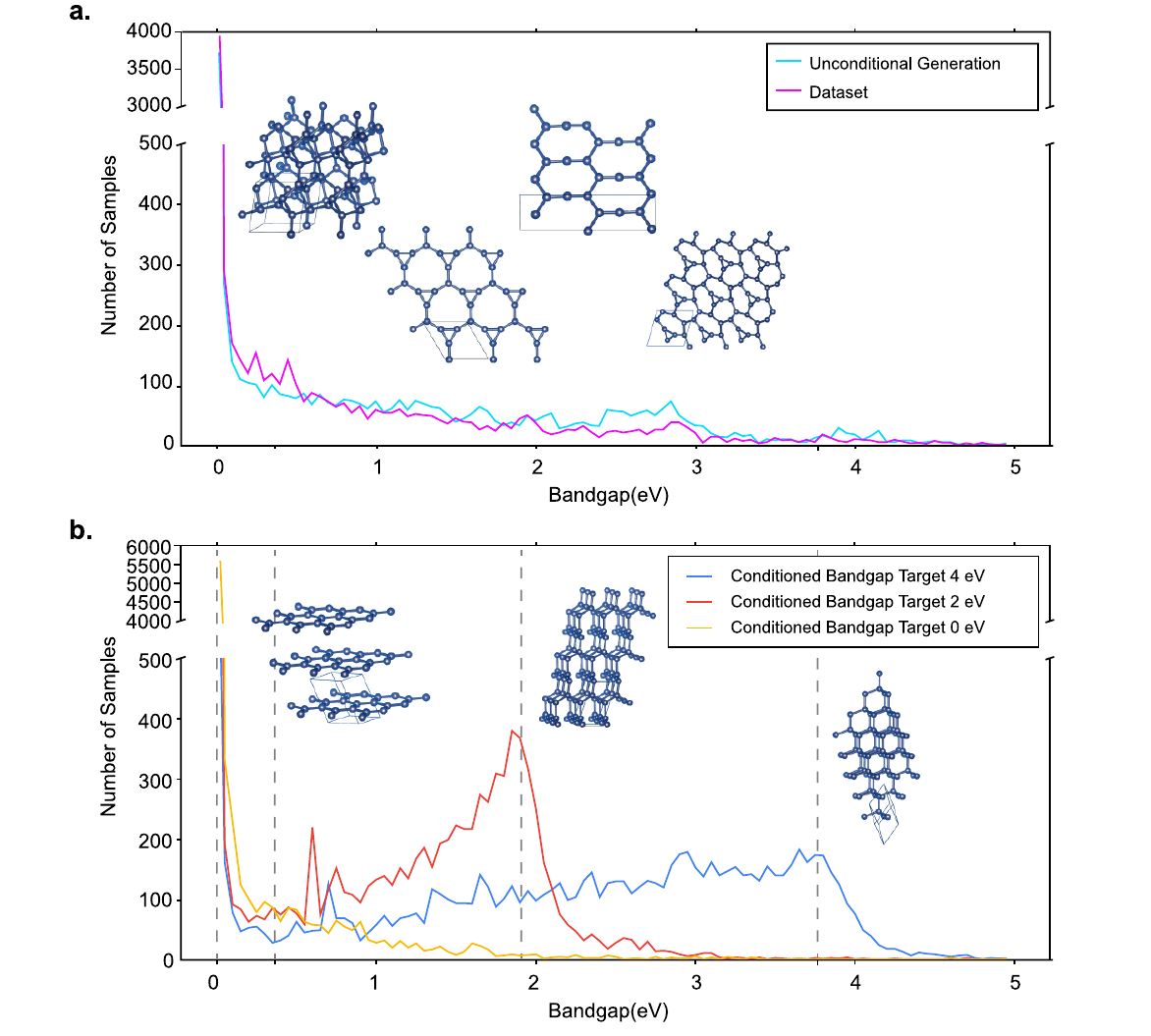}
    \caption{\textbf{Property-guided structure generation of carbon structures with particular band gaps.} (a)  Electronic band gaps of generated structures from trained DiG with no specification on the desired band gap. The generated structures do not show any obvious preference on band gaps, closely resembling the distribution of the training dataset. (b) Structure generated for three band gaps (0, 2, and 4 eV). The distributions of band gaps for generated structures peak at the desired values. In particular, DiG generates graphite-like structures when desired band gap is 0 eV; while at 4 eV band gap, the generated structures are most similar to diamonds. Representative structures are shown in the inset of the plot.}
    \label{fig:conditioned_sampling}
\end{figure}

Fig.~\ref{fig:conditioned_sampling} shows the distributions of band gaps calculated from generated carbon structures. In the original training dataset, most structures have a band gap around 0 eV (see Fig.~\ref{fig:conditioned_sampling}a). When the target band gaps are supplied to DiG, the  structures are generated with the desired band gaps. With the guidance of a band gap model in conditional generation, the distribution is biased towards the targets, showing pronounced peaks around the target band gaps. Representative structures are shown in Fig.~\ref{fig:conditioned_sampling}. For conditional generation with a target band gap of 4 eV, DiG generates stable carbon structures similar to diamond, which has large band gaps. In the case of 0 eV band gap, we obtain graphite-like structures with low band gaps. In Fig.~\ref{fig:conditioned_sampling}a, we show some structures by unconditional generation. To evaluate the quality of carbon crystal structures generated by DiG, 
we calculate the ratio of structures that match one of the relaxed structures in the dataset by using the \texttt{StructureMatcher} in the PyMatgen package~\cite{ong2013python}. For unconditional generation, the match rate is 99.87\%, and the average matched normalized RMSD computed from fractional coordinates over all sampled structures is 0.16. For conditional generation, the match rate is 99.99\%, but with a higher average normalized RMSD of 0.22. While increasing the possibility of generating structures with target band gap, conditional generation can influence the quality of the structures (see Supplementary Sec.~\ref{appx:limit-data} for more discussions). This proof-of-concept study shows that DiG not only captures the probability distributions with complex features in large configurational space, but also can be applied for inverse materials design, when combined with a property quantifier, such as an ML predictor. Since the training of the property prediction model (e.g., the M3GNet band gap model) and the diffusion model of DiG are fully decoupled, our approach can be readily extended to inverse design for other properties.

\section{Discussion}

Predicting the equilibrium distribution of molecular states is a formidable challenge in the molecular sciences, with far-reaching implications for deciphering structure-function relationships, computing macroscopic properties, and designing novel molecules and materials. With existing methods, a vast number of measurements or simulated samples of single molecules are required to gather sufficient data for characterizing the equilibrium distribution. 
We introduce Distributional Graphormer (DiG), a deep generative framework capable of predicting probability distributions which enables efficiently sampling diverse conformations and estimating their state densities across molecular systems.
Drawing inspiration from the annealing process, DiG employs a sequence of deep neural networks to progressively transform state distributions from a simplistic mathematical form to the target distributions which can be trained to approximate the equilibrium distribution with suitable training data.

We have applied DiG to several molecular prediction tasks, including protein conformation sampling, protein-ligand binding structure generation, molecular adsorption on catalyst surfaces, and property-guided structure generation. The results show that DiG is capable of generating chemically realistic and diverse structures, and distributions resembling those of extensive MD simulations in low dimensional projections in some cases. By harnessing the power of advanced deep learning architectures, DiG can learn the representation of molecular conformations that are transformed from molecular descriptors, such as amino acid sequences for proteins or chemical formulas for compound molecules. Furthermore, its capacity to model complex, multimodal distributions using diffusion models enables it to capture equilibrium distributions in high-dimensional space.

DiG has been demonstrated to be capable to generalize across molecules within the same class, such as in the case of proteins, small molecules, and catalyst structures. 
Consequently, the framework opens the door to a multitude of research opportunities and applications in molecular science. Thus, when fed with suitably distributed training data, DiG can provide insights into the statistical understanding of molecules, enabling the computing of macroscopic properties, such as free energies and thermodynamic stability. These insights are critical for investigating the physical and chemical phenomena of molecular systems.

Finally, with its capability in generating independent and identically distributed (i.i.d.) conformations from equilibrium distributions, DiG offers a significant computational advantage over traditional sampling or simulation approaches that suffer from rare events, such as MCMC or MD simulations. 
DiG achieves similar conformation space coverage as millisecond-timescale MD simulations do in the two tested protein cases. Based on the OpenMM benchmark performance of modern GPU devices, it would require about 7-10 GPU years on Nvidia A100s to complete a simulation of 1.8 ms for RBD of the spike protein; while generating 50k structures using DiG only takes about 10 days on a single A100 GPU without any inference acceleration (see more discussion in Supplementary Sec.~\ref{appx:accel-infer}). Similar levels of speedup can be achieved in the case of predicting adsorbate distribution on the catalyst surface, as elaborated in the result section. If such order-of-magnitude speed-up can be combined with generating high-accuracy probability distributions, this will be transformative for molecular simulation and design.

While the quantitative prediction of equilibrium distributions at given thermodynamic states will hinge upon the availability of training data,
the capacity of DiG to explore vast and diverse conformational spaces contributes to the discovery of novel and functional molecular structures, including protein structures, ligand conformers, and adsorbate configurations. DiG can therefore help to bridge the gap between microscopic descriptors and macroscopic observations of molecular systems, with potential impact on various areas of molecular sciences including life sciences, drug design, catalysis research, and materials sciences.


\bibliography{sn-bibliography}

\section{Acknowledgements}

We thank Nathan A. Baker, Lixin Sun, Bas Veeling, Victor García Satorras, Andrew Foong and Cheng Lu for insightful discussions; Shengjie Luo for helping with dataset preparations; Jingjie Su for  managing the project; Jingyun Bai for helping with figure design; colleagues at Microsoft for their encouragement and support.

%

\section{Author information}
\subsection*{\textit{Contributions}}
S.Zheng and TY.Liu led the research. S.Zheng, J.He, C.Liu, Z.Lu and H.Liu conceived the project. J.He, C.Liu, Y.Shi, W.Feng and F.Ju and J.Wang developed the diffusion model and training pipeline. J.He, Y.Shi, Z.Lu, J.Zhu, F.Ju, H.Zhang and H.Liu developed data and analytics systems. H.Liu, Y.Shi, Z.Lu, Y.Min and S.Tang conducted simulations. H.Hao, P.Jin, C.Chen, and F.No{\'e} contributed technical advice and ideas. S.Zheng, J.He, C.Liu, Y.Shi, Z.Lu, F.No{\'e}, H.Zhang and H.Liu wrote the paper with the inputs from all authors.

\subsection*{\textit{Corresponding authors}}
Correspondence to \href{shuxin.zheng@microsoft.com}{Shuxin Zheng}, \href{chang.liu@microsoft.com}{Chang Liu}, \href{haiguang.liu@microsoft.com}{Haiguang Liu} and \href{tie-yan.liu@microsoft.com}{Tie-Yan Liu}.


\newpage
\begin{appendices}
\renewcommand{\thefigure}{S\arabic{figure}}




\section{Technical Details} \label{appx:tech-details}
\newcommand{\bfRb}{\bar{\bfR}}

\begin{table}[]    \centering
    \caption{Notations.}
    \label{tab:notation-1}
    \begin{tabular}{ll}
        \toprule
        \multicolumn{2}{c}{General formulation}\\
        \cmidrule{1-2}
        $\clD$ & System descriptor \\
        $\bfR$ & Molecular structure \\
        $\{\bfR_{\clD,0}^{(m)}\}_{m=1}^M$ & \makecell[l]{Molecular structures of system $\clD$ in the dataset for \\ PIDP training} \\
        $\{\bfR_{\clD,0}^{(n)}\}_{n=1}^{N_{\data}}$ & \makecell[l]{Molecular structures of system $\clD$ in the dataset for\\ data-based (denoising score matching) training} \\
        $D$ & Dimension of $\bfR$ \\
        $E_\clD(\bfR)$ & (Potential) Energy function of system $\clD$ \\
        $k_\rmB$ & Boltzmann constant \\
        $T$ & Temperature \\
        $c$ & A property of molecular structure \\
        $I$ & \makecell[l]{Length of descriptor / number of individual elements \\ in a system} \\
        $\imath,\jmath \in \{1, 2, \cdots, I\}$ & Index for individual elements in a system \\
        \midrule
        \multicolumn{2}{c}{Diffusion process} \\
        \cmidrule{1-2}
        $\tau$ & Total time length/duration of the forward diffusion process \\
        $t \in [0,\tau]$ & Time variable (continuous) \\
        $N$ & \makecell[l]{Number of time discretization steps for the diffusion process} \\
        $i \in \{1, 2, \cdots, N\}$ & Time step (discrete) \\
        $h$ ($=\tau/N$) & Time discretization step size \\
        $\ttb$ or $\iib$ & Reverse time or step \\
        $\bfB_t$ and $\bar{\bfB}_t$ & \makecell[l]{Standard Brownian motion in dimension $D$ and \\its reverse process} \\
        $\bfff_t(\bfR_t)$ & Drift function in a general diffusion process \\
        $g_t$ & Diffusion rate scheme in a general diffusion process \\
        $q_{\clD,0}$ & \makecell[l]{Equilibrium distribution of system $\clD$ \\(under a certain temperature)} \\
        $\bfR_0$ & \makecell[l]{Molecular structure variable following\\ equilibrium distribution $q_{\clD,0}$} \\
        $q_{\clD,t}$ or $q_{\clD,i}$ & \makecell[l]{Distribution of molecular structure in intermediate \\time or step in the forward diffusion process} \\
        $\bfR_t$ or $\bfR_i$ & \makecell[l]{Molecular structure variable in intermediate time or step, \\following $q_{\clD,t}$ or $q_{\clD,i}$} \\
        $q(\bfR_t \mid \bfR_0)$ or $q(\bfR_i \mid \bfR_0)$ & Marginal transition kernel of the forward diffusion process \\
        $\beta_t$ or $\beta_i$ & \makecell[l]{Time dilation scheme.\\ Note $\beta_i := h \beta_{t_i}$ which is different from others} \\
        $\sigma_t$ or $\sigma_i$ & \makecell[l]{Noise variance scheme\\ (standard deviation of $q(\bfR_t \mid \bfR_0)$ or $q(\bfR_i \mid \bfR_0)$)} \\
        $\bfeps_t$ or $\bfeps_i$ & Standard Gaussian noise variable \\
        $\bfss^\theta_{\clD,t}$ or $\bfss^\theta_{\clD,i}$ & Score model for $q_{\clD,t}$ or $q_{\clD,i}$ \\
        $\bfeps^\theta_{\clD,t}$ or $\bfeps^\theta_{\clD,i}$ & Noise-predicting model for $q_{\clD,t}$ or $q_{\clD,i}$ \\
        $p_\simple$ & \makecell[l]{The simple distribution to which the forward diffusion\\ process converges} \\
        $p^\theta_{\clD,t}$ or $p^\theta_{\clD,i}$ & \makecell[l]{Distribution of molecular structure in intermediate\\ time or step in the reverse diffusion process simulated \\by $\bfss^\theta_{\clD,t}$ or $\bfss^\theta_{\clD,i}$ or $\bfeps^\theta_{\clD,t}$ or $\bfeps^\theta_{\clD,i}$} \\
        \bottomrule
    \end{tabular}
\end{table}

\subsection{Formulation of DiG}

The forward process \eqnref{fwd-cont} is constructed using the Langevin dynamics that takes the simple distribution $p_\simple := \clN(\bfzro,\bfI)$ as its stationary distribution:
$\ud \bfR_t = \frac{1}{2} \nabla \log p_\simple(\bfR_t) \dd t + \ud \bfB_t = -\frac{1}{2} \bfR_t \dd t + \ud \bfB_t.$
From any conformational distribution of any system as the initial distribution $q_{\clD,0}(\bfR_0)$, the distribution evolves under this process and converges to $p_\simple$ exponentially~\cite{roberts1996exponential,durmus2016high,cheng2017convergence,dalalyan2017theoretical}.
For a faster simulation convergence, it is preferred to introduce a time dilation scheme $\beta_t$ that increases in $t$~\cite{wibisono2016variational}. This gives \eqnref{fwd-cont}.

To draw structure samples using DiG, we simulate the reverse process \eqnref{rev-cont} from samples from $p_\simple$ the standard Gaussian distribution, which can be easily drawn independently. Note that this ``reverse'' is not a sample-level point-to-point inverse, but a distribution-level inverse: denoting its induced distribution as $p_{\ttb}$, if $p_{\ttb = 0} = q_{\clD,\tau}$, then $p_{\ttb = \tau} = q_{\clD,0}$.
When employing a trained score model $\bfss^\theta_{\clD,t}$ to approximate the score function $\nabla \log q_{\clD,t}$, the reverse process can be simulated by the Euler-Maruyama discretization using the step size $h$ up to $o(h)$ local error:
$\bfR_{\iib + 1} = \Big( 1 + \frac{1}{2} \beta_{\iib} \Big) \bfR_{\iib} + \beta_{\iib} \bfss^\theta_{\clD,\iib}(\bfR_{\iib}) + \clN(\bfzro, \beta_{\iib} \bfI)$,
where the indexed quantities are evaluated at $\ttb := \iib h$, and $\beta_{\iib} := h \beta_{\ttb = \iib h}$, and ``$+ \clN(\cdots)$'' denotes adding a randomly drawn sample from the denoted Gaussian distribution.
In the original step index $i$, this becomes:
$\bfR_{i-1} = \Big( 1 + \frac{1}{2} \beta_i \Big) \bfR_i + \beta_i \bfss^\theta_{\clD,i}(\bfR_i) + \clN(\bfzro, \beta_i \bfI)$.
By design $h$ is chosen small for accurate simulation, so is each $\beta_i$. Hence we can leverage the approximation $\frac{1}{\sqrt{1-\beta_i}} = 1 + \frac{1}{2} \beta_i + o(h)$ which does not increase the dicretization local error. This gives \eqnref{rev-disc} for generating samples from the equilibrium distribution.

\subsubsection{Physics-Informed Diffusion Pre-training}
\label{appx:method-pidp}
The goal of the score model $\bfss^\theta_{\clD,t}(\bfR_t)$ is to match the corresponding true score function $\nabla \log q_{\clD,t}(\bfR_t)$ from the forward process \eqnref{fwd-cont} for each $t \in [0,\tau]$. To better leverage the diffusion-process construction of DiG, we use a partial differential equation governing the true score function and construct a loss function that enforces the equation to hold for training the score model.

Under a general diffusion process $\ud \bfR_t = \bfff_t(\bfR_t) \dd t + g_t \dd \bfB_t$, the instantaneous distribution transformation is given by the Fokker-Planck equation (FPE; in logarithm form):
\begin{align}
    \fracpartial{}{t} \log q_t(\bfR_t)
    ={} & -\nabla \cdot \bfff_t(\bfR_t) - \nabla \log q_t(\bfR_t) \cdot \bfff_t(\bfR_t) \\
    & {}+ \frac{g_t^2}2 \Big(\nabla^2 \log q_t(\bfR_t) + \lrVert{\nabla \log q_t(\bfR_t)}^2 \Big),
    \label{eqn:fpe}
\end{align}
For the specific diffusion process \eqnref{fwd-cont}, the evolving distribution $q_t$ from the forward process satisfies:
$\fracpartial{}{t} \log q_{\clD,t}(\bfR_t)
    = \frac{\beta_t}2 \Big( D + \bfR_t \cdot \nabla \log q_{\clD,t}(\bfR_t)
    + \nabla^2 \log q_{\clD,t}(\bfR_t) + \lrVert{\nabla \log q_{\clD,t}(\bfR_t)}^2 \Big),$
where $D$ is the dimension of $\bfR$.
Taking the gradient of the above equation gives:
$\fracpartial{}{t} \nabla \log q_{\clD,t}(\bfR_t)
    = \frac{\beta_t}2 \Big(
        \nabla \cdot \big( \bfR_t \cdot \nabla \log q_{\clD,t}(\bfR_t) \big)
        + \nabla \big( \nabla \cdot \nabla \log q_{\clD,t}(\bfR_t) \big)
         {}+ \nabla \lrVert{\nabla \log q_{\clD,t}(\bfR_t)}^2
    \Big),$
which becomes an equation of the score function $\nabla \log q_{\clD,t}(\bfR_t)$. To well approximate $\nabla \log q_{\clD,t}(\bfR_t)$, the score model $\bfss^\theta_{\clD,t}(\bfR_t)$ also needs to satisfy this equation. To enforce it, we follow the idea of physics-informed neural networks~\cite{raissi2019physics} that converts a differential equation into a loss function of the to-be-solved function. The loss function is typically taken as the squared norm of the equality residual, which in our case is:
\begin{align}
    \lrVert*[\Big]{
        \frac{\beta_t}2 \Big(
            \nabla \cdot \big( \bfR_t \cdot \bfss^\theta_{\clD,t}(\bfR_t) \big)
            + \nabla \big( \nabla \cdot \bfss^\theta_{\clD,t}(\bfR_t) \big)
            + \nabla \lrVert{\bfss^\theta_{\clD,t}(\bfR_t)}^2
        \Big)
        - \fracpartial{}{t} \bfss^\theta_{\clD,t}(\bfR_t)
    }^2,
\end{align}
for each $t$.
By time discretization and evaluating the loss on a set of samples $\{\bfR_{\clD,i}^{(m)}\}_{m=1}^M$, this gives the first term in \eqnref{pinnloss}.

The FPE does not have (nor need) a boundary condition as long as each $q_{\clD,t}$ is normalized. 
For the initial condition, we know that the score of the target equilibrium distribution $\nabla \log q_{\clD,0}(\bfR_0) = -\nabla E_\clD(\bfR_0) / (k_\rmB T)$ is exactly given by the gradient of the energy function of the system. This is where the energy function comes to supervise the model, and this supervision propagates to other time steps via the first term of the loss.
To implement this initial condition, we minimize
$\lrVert{\bfss^\theta_0(\bfR_0) - \nabla \log q_{\clD,0}(\bfR_0)}^2 = \lrVert{\bfss^\theta_0(\bfR_0) + \nabla E_\clD(\bfR_0) / (k_\rmB T)}^2$, which leads to the second term in \eqnref{pinnloss}.
Note that in \eqnref{pinnloss} the loss term is not imposed on $t_0 = 0$ (i.e., $i=0$). This is because in the actual implementation, the score model is expressed using a noise-predicting model $\bfeps^\theta_{\clD,t}$ as $\bfss^\theta_{\clD,t} = -\sigma_t \bfeps^\theta_{\clD,t}$ (explained in Supplementary \secref{details-data-loss}), which, at $t = 0$, the vanishing $\sigma_0 = 0$ causes an ill-defined score model. This is commonly solved by starting the diffusion simulation from an infinitesimal initial time step~\cite{song2021score}, which corresponds to $t=h$ or $i=1$ here. On the other hand, from the data-generation process \eqnref{rev-disc}, the last required time step for the model is $i=1$, where the sample needs to be updated to follow the equilibrium distribution. So it is reasonable to supervise $\bfss^\theta_{\clD,i=1}$ or $\bfeps^\theta_{\clD,i=1}$ with the energy function.

In comparison, we note that there are other 
common approaches to train a generative model using a given energy function, but they cannot leverage the advantage of the diffusion-process construction of DiG and thus do not enjoy the step-by-step supervision pattern and are not as effective to train large models.
The most popular way is to minimize the reverse Kullback-Leibler (KL) divergence $\KL(p^\theta_{\clD,0} \Vert q_{\clD,0})$ between the model-defined equilibrium distribution and the true equilibrium distribution, which is equivalent to minimizing the (Helmholtz) free energy:
\begin{align}
    \mathrm{FreeEng}_\clD = \bbE_{p^\theta_{\clD,0}(\bfR_0)} [E_\clD(\bfR_0) + k_\rmB T \log p^\theta_{\clD,0}(\bfR_0)].
    \label{eqn:free-eng}
\end{align}
In the expression, no sample from $q_{\clD,0}$ is required, so the access to the energy function $E_\clD$ suffices for training. This approach is known as variational inference~\citep{jordan1999introduction,wainwright2008graphical,kingma2014auto,rezende2015variational,kingma2016improved} in machine learning, and the negative (Helmholtz) free energy is also called evidence lower bound (ELBO). This method is recently used to train a generative model for the equilibrium distribution of molecular systems~\citep{noe2019boltzmann}. A more modern approach minimizes the alpha divergence between the model and the equilibrium distributions~\citep{li2016renyi,hernandez2016black}, which generalizes the reverse KL divergence and ameliorates the mode-collapse tendency to some extent. It is also applied to molecular systems recently~\citep{midgley2022flow}.
These methods can be directly applied to DiG given the density evaluation method \eqnref{logdensity}, but it loses step-by-step supervision as it only supervises the end distribution $p^\theta_{\clD,0}$, which makes training large models hard. Moreover, evaluating the density function requires an ODE solver, so the optimization requires backpropagation through the ODE solver, which is very costly.

\subsubsection{Training DiG with Data} \label{sec:details-data-loss}

To develop a method to train the model step-by-step using data from $q_{\clD,0}$, we start by score matching for each step $i$, that is to minimize 
\begin{align}
\bbE_{q_{\clD,i}(\bfR_i)} \lrVert{\bfss^\theta_{\clD,i}(\bfR_i) - \nabla \log q_{\clD,i}(\bfR_i)}^2.
\end{align}
Although this loss can be made tractable (i.e., to get rid of the unknown true score function $\nabla \log q_{\clD,i}$) using the standard score matching technique~\citep{hyvarinen2005estimation}, the resulting loss function involves the divergence of the score model $\nabla \cdot \bfss^\theta_{\clD,i}$ which is expensive to evaluate and optimize.
Another way to make it tractable is via the denoising score matching technique~\citep{vincent2011connection,alain2014regularized}. The method first reforms the intermediate marginal distribution in terms of the marginal transition kernel $q(\bfR_i \mid \bfR_0)$ from the forward process (which does not depend on a specific system hence no $\clD$ subscript),
$q_{\clD,i}(\bfR_i) = \int q_{\clD,0}(\bfR_0) q(\bfR_i \mid \bfR_0) \dd \bfR_0$,
and then decompose the score function as:
\begin{align}
    \nabla \log q_{\clD,i}(\bfR_i)
    ={} & \frac1{q_{\clD,i}(\bfR_i)} \int q_{\clD,0}(\bfR_0) \nabla_{\bfR_i} q(\bfR_i \mid \bfR_0) \dd \bfR_0 \\
    ={} & \int q_{\clD,0}(\bfR_0) \frac{q(\bfR_i \mid \bfR_0)}{q_{\clD,i}(\bfR_i)} \nabla_{\bfR_i} \log q(\bfR_i \mid \bfR_0) \dd \bfR_0 \\
    ={} & \bbE_{q_\clD(\bfR_0 \mid \bfR_i)} [\nabla_{\bfR_i} \log q(\bfR_i \mid \bfR_0)],
    \label{eqn:fisher-id}
\end{align}
a.k.a Fisher's identity~\cite{cappe2005inference}. The score-matching loss then becomes:
\begin{align}
    & \bbE_{q_{\clD,i}(\bfR_i)} \lrVert{\bfss^\theta_{\clD,i}(\bfR_i) - \nabla \log q_{\clD,i}(\bfR_i)}^2 \\
    ={} & \bbE_{q_{\clD,i}(\bfR_i)} \lrVert{\bfss^\theta_{\clD,i}(\bfR_i)}^2 
        - 2 \bbE_{q_{\clD,i}(\bfR_i)} \lrbrack{\bfss^\theta_{\clD,i}(\bfR_i) \cdot \nabla \log q_{\clD,i}(\bfR_i)} \\
        & + \bbE_{q_{\clD,i}(\bfR_i)} \lrVert{\nabla \log q_{\clD,i}(\bfR_i)}^2 \\
    \stackrel{\text{\eqnref{fisher-id}}}{=}{} &
        \bbE_{q_{\clD,i}(\bfR_i)} \bbE_{q(\bfR_0 \mid \bfR_i)} \lrVert{\bfss^\theta_{\clD,i}(\bfR_i)}^2 \\
        & {}- 2 \bbE_{q_{\clD,i}(\bfR_i)} \bbE_{q(\bfR_0 \mid \bfR_i)} \lrbrack{\bfss^\theta_{\clD,i}(\bfR_i) \cdot \nabla_{\bfR_i} \log q(\bfR_i \mid \bfR_0)} \\
        & {}+ \bbE_{q_{\clD,i}(\bfR_i)} \lrVert{\nabla \log q_{\clD,i}(\bfR_i)}^2 \\
    ={} & \bbE_{q_{\clD,i}(\bfR_i)} \bbE_{q(\bfR_0 \mid \bfR_i)} \lrVert{\bfss^\theta_{\clD,i}(\bfR_i) - \nabla_{\bfR_i} \log q(\bfR_i \mid \bfR_0)}^2 \\
        & {}- \bbE_{q_{\clD,i}(\bfR_i)} \bbE_{q(\bfR_0 \mid \bfR_i)} \lrVert{\nabla_{\bfR_i} \log q(\bfR_i \mid \bfR_0)}^2 \\
        & {}+ \bbE_{q_{\clD,i}(\bfR_i)} \lrVert{\nabla \log q_{\clD,i}(\bfR_i)}^2 \\
    ={} & \bbE_{q_{\clD,0}(\bfR_0)} \bbE_{q(\bfR_i \mid \bfR_0)} \lrVert{ \bfss^\theta_{\clD,i}(\bfR_i) - \nabla_{\bfR_i} \log q(\bfR_i \mid \bfR_0) }^2 \\
        & {}+ \bbE_{q_{\clD,0}(\bfR_0)} \bbE_{q(\bfR_i \mid \bfR_0)} \lrbrack*[\big]{ \lrVert{\nabla \log q_{\clD,i}(\bfR_i)}^2 - \lrVert{\nabla_{\bfR_i} \log q(\bfR_i \mid \bfR_0)}^2 }.
\end{align}
Noting that the second term in the last expression is a constant of $\theta$, optimizing the score-matching loss for step $i$ is equivalent to minimizing the first term:
\begin{align}
    \bbE_{q_{\clD,0}(\bfR_0)} \bbE_{q(\bfR_i \mid \bfR_0)} \lrVert{ \bfss^\theta_{\clD,i}(\bfR_i) - \nabla_{\bfR_i} \log q(\bfR_i \mid \bfR_0) }^2.
    \label{eqn:dsmloss}
\end{align}
This is the denoising score matching loss.
To explain the name, in the original context, $q(\bfR_i \mid \bfR_0) = \clN(\bfR_i \mid \bfR_0, \sigma_i^2 \bfI)$ which adds noise to the data sample $\bfR_0$ to get a noisy version $\bfR_i$, and the resulting loss
\begin{align}
& \bbE_{q_{\clD,0}(\bfR_0)} \bbE_{q(\bfR_i \mid \bfR_0)} \lrVert{\bfss^\theta_{\clD,i}(\bfR_i) + \frac{\bfR_i - \bfR_0}{\sigma_i^2}}^2 \\
& {}= \frac{1}{\sigma_i^2} \bbE_{q_{\clD,0}(\bfR_0)} \bbE_{q(\bfR_i \mid \bfR_0)} \lrVert{\bfR_i + \sigma_i^2 \bfss^\theta_{\clD,i}(\bfR_i) - \bfR_0}^2
\end{align}
drives the ``decoder'' $\bfR_i + \sigma_i^2 \bfss^\theta_{\clD,i}(\bfR_i)$ to recover the original clean data point $\bfR_0$ by ``denoising'' $\bfR_i$.

Optimizing the denoising score matching loss \eqnref{dsmloss} is tractable once we know the conditional distribution $q(\bfR_i \mid \bfR_0)$, which is fortunately available in closed form for the forward process \eqnref{fwd-cont}. Under continuous-time, the result is $q(\bfR_t \mid \bfR_0) = \clN(\bfR_t \mid \alpha_t \bfR_0, \sigma_t^2 \bfI)$~\citep{song2021score,karras2022elucidating}, where $\alpha_t := \exp(-\frac12 \int_0^t \beta_{t'} \dd t')$ and $\sigma_t := \sqrt{1 - \alpha_t^2}$. For a discretized expression, recall that the time interval $[0,\tau]$ is uniformly divided into $N+1$ points with step size $h = \tau/N$, step $i$ corresponds to time $t = ih$, and $\beta_i := h \beta_{t=ih}$. This leads to 
\begin{align}
\alpha_{t=ih}
& {}= \sqrt{\exp(-\sum_{j=1}^i \beta_j + o(h))}
= \sqrt{\exp(o(h)) \prod_{j=1}^i \exp(-\beta_j)}\\
& {}= \sqrt{(1 + o(h)) \prod_{j=1}^i (1 - \beta_j + o(h))}
= \sqrt{\prod_{j=1}^i (1 - \beta_j) + o(h)} \\
& {}= \prod_{j=1}^i \sqrt{1 - \beta_j} + o(h),
\end{align}
so we can take $\alpha_i := \prod_{j=1}^i \sqrt{1 - \beta_j}$. Correspondingly, $\sigma_i = \sqrt{1-\alpha_i^2}$. The required conditional distribution is then:
\begin{align}
    q(\bfR_i \mid \bfR_0) = \clN(\bfR_i \mid \alpha_i \bfR_0, \sigma_i^2 \bfI).
    \label{eqn:q_il0}
\end{align}
The loss \eqnref{dsmloss} for time step $i$ then becomes
$\bbE_{q_{\clD,0}(\bfR_0)} \bbE_{q(\bfR_i \mid \bfR_0)} \lrVert*[\big]{ \bfss^\theta_{\clD,i}(\bfR_i) + \frac1{\sigma_i^2} (\bfR_i - \alpha_i \bfR_0) }^2$.
Using the reparameterization of the Gaussian distribution $q(\bfR_i \mid \bfR_0)$ as $\bfR_i = \alpha_i \bfR_0 + \sigma_i \bfeps_i$ where $\bfeps_i \sim p(\bfeps_i) := \clN(\bfzro,\bfI)$, the loss is further reformed as:
$\bbE_{q_{\clD,0}(\bfR_0)} \bbE_{p(\bfeps_i)} \lrVert*[\big]{\bfss^\theta_{\clD,i}(\alpha_i \bfR_0 + \sigma_i \bfeps_i) + \frac1{\sigma_i} \bfeps_i}^2
= \frac1{\sigma_i^2} \bbE_{q_{\clD,0}(\bfR_0)} \bbE_{p(\bfeps_i)} \lrVert{\sigma_i \bfss^\theta_{\clD,i}(\alpha_i \bfR_0 + \sigma_i \bfeps_i) + \bfeps_i}^2$.
To balance the scale of the loss for different $i \in \{ 1, \cdots, N \}$, the loss \eqnref{dsmloss} for step $i$ is normalized by the scale of
$\bbE_{q(\bfR_i \mid \bfR_0)} \lrVert{\nabla_{\bfR_i} \log q(\bfR_i \mid \bfR_0)}^2
= \bbE_{p(\bfeps_i)} \lrVert*[\big]{\frac{\bfeps_i}{\sigma_i}}^2 = \frac1{\sigma_i^2}$~\cite{song2021score}, which finally leads to \eqnref{scoreloss}.

From the expression of this loss \eqnref{scoreloss}, we find that the ``model'' $-\sigma_i \bfss^\theta_{\clD,i}(\alpha_i \bfR_0 + \sigma_i \bfeps_i)$ can be seen as to ``predict the noise label'' $\bfeps_i$, whose distribution is well centered and scaled. This is the range that a 
deep learning model works the best. So to make a comfortable and friendly learning task, we implement the model to directly output the vector value for $-\sigma_i \bfss^\theta_{\clD,i}$, which we denote as $\bfeps^\theta_{\clD,i}$ and call it the noise-predicting model. The score model can still be recovered by:
\begin{align}
    \bfss^\theta_{\clD,i}(\bfR_i) = -\bfeps^\theta_{\clD,i}(\bfR_i) / \sigma_i, \label{eqn:noise-and-score}
\end{align}
as an approximation to the true score function $\nabla \log q_{\clD,i}(\bfR_i)$.
The training loss \eqnref{scoreloss} then becomes:
\begin{align}
    \frac1N \sum_{i=1}^N \bbE_{q_{\clD,0}(\bfR_0)} \bbE_{p(\bfeps_i)} \lrVert{\bfeps^\theta_{\clD,i}(\alpha_i \bfR_0 + \sigma_i \bfeps_i) - \bfeps_i}^2. \label{eqn:noiseloss}
\end{align}
This recovers the formulation in~\cite{ho2020denoising,song2021score}.
To understand the loss, note the marginal transition kernel \eqnref{q_il0} of the forward process means $\bfR_i = \alpha_i \bfR_0 + \sigma_i \bfeps_i$ where $\bfeps_i \sim \clN(\bfzro,\bfI)$. So the $\bfeps^\theta_{\clD,i}$ model tries to recover the noise variable $\bfeps_i$ from $\bfR_i$ that were used to generate $\bfR_i$.

\subsubsection{Density Evaluation using DiG} \label{sec:details-density}

Viewed in the continuous-time limit, DiG defines a distribution via transforming $p_\simple$ through the reverse process \eqnref{rev-cont}, where the score function is approximated by the model. Written in forward time $t$, this process follows the following SDE:
\begin{align}
    \ud \bfR_t = -\frac{\beta_t}2 \bfR_t \dd t - \beta_t \bfss^\theta_{\clD,t}(\bfR_t) \dd t + \sqrt{\beta_t} \dd \bar{\bfB}_t,
    \label{eqn:rev-sde}
\end{align}
where $\bar{\bfB}_t$ is the reverse of the Brownian motion. The distribution transformation under this process is given by its FPE in \eqnref{fpe}:
\begin{align}
    & \fracpartial{}{t} \log p^\theta_{\clD,t}(\bfR_t) \\
    ={} & -\nabla \cdot \Big( -\frac{\beta_t}2 \bfR_t - \beta_t \bfss^\theta_{\clD,t}(\bfR_t) \Big)
    - \nabla \log p^\theta_{\clD,t}(\bfR_t) \cdot \Big( -\frac{\beta_t}2 \bfR_t - \beta_t \bfss^\theta_{\clD,t}(\bfR_t) \Big) \\
    & {}- \frac{\beta_t}2 \Big(\nabla^2 \log p^\theta_{\clD,t}(\bfR_t) + \lrVert{\nabla \log p^\theta_{\clD,t}(\bfR_t)}^2 \Big),
    \label{eqn:fpe-rev-sde}
\end{align}
where the last term has a negative sign in correspondence to the reverse Brownian motion.
When the model is well-learned, $\bfss^\theta_{\clD,t}$ well approximates $\nabla \log q_{\clD,t}$ and $p^\theta_{\clD,t}$ well approximates $q_{\clD,t}$, hence we can approximate $\nabla \log p^\theta_{\clD,t}$ also using $\bfss^\theta_{\clD,t}$. This turns \eqnref{fpe-rev-sde} into:
\footnote{When $\bfss^\theta_t \ne \nabla \log q_{\clD,t}$ or $q_{\clD,\tau} \ne p_\tau := p_\simple$, \eqnref{fpe-rev-sde} and \eqnref{fpe-rev-ode} (or \eqnref{rev-sde} and \eqnref{rev-ode}) give different evolving densities. See~\citep{song2021maximum,lu2022maximum} for more discussions.}
\begin{align}
    \fracpartial{}{t} \log p^\theta_{\clD,t}(\bfR_t)
    ={} & -\nabla \cdot \Big( -\frac{\beta_t}2 \bfR_t - \frac{\beta_t}2 \bfss^\theta_{\clD,t}(\bfR_t) \Big) \\
    & -\nabla \log p^\theta_{\clD,t}(\bfR_t) \cdot \Big( -\frac{\beta_t}2 \bfR_t - \frac{\beta_t}2 \bfss^\theta_{\clD,t}(\bfR_t) \Big).
    \label{eqn:fpe-rev-ode}
\end{align}
Comparing this equation with the general-form FPE in \eqnref{fpe}, we can find that this equation is exactly the FPE of the ``deterministic diffusion process'' defined by the ODE in \eqnref{rev-ode}. In other words, the ODE in \eqnref{rev-ode}, and the SDE in \eqnref{rev-sde}, render the same $\fracpartial{}{t} \log p^\theta_{\clD,t}(\bfR_t)$ hence the same marginal distribution $p^\theta_{\clD,t}(\bfR_t)$ in each time step $t$ (since they have the same terminal distribution $p_\tau = p_\simple$; note the mentioned requirement $\bfss^\theta_{\clD,t} = \nabla \log p^\theta_{\clD,t}$ for this claim to hold).
Since the SDE in \eqnref{rev-sde} is the same as \eqnref{rev-cont} and in turn leads to the sampling/generation process in \eqnref{rev-disc}, this finding indicates that we can also generate equilibrium-distribution samples by simulating the ODE in \eqnref{rev-ode}. This kind of deterministic process or ODE sampling process is used in protein conformation sampling (see the end of Supplementary Sec.~\ref{sec:acarbon-so3-diffusion-sampling-details}) and property-guided structure generation (Supplementary Sec.~\ref{sec:inverse-design-diffusion-details}).

Back to density evaluation using DiG, we can estimate the density function of the model-defined equilibrium distribution $p^\theta_{\clD,0}$ by integrating w.r.t the diffusion time step $t$ following the above ODE in \eqnref{fpe-rev-ode}, which does not contain any unknown objects (recall that we made the assumption that $\bfss^\theta_{\clD,t} = \nabla \log p^\theta_{\clD,t}$ to use \eqnref{fpe-rev-ode}).
Let $\bfR_t$ be a solution to \eqnref{rev-ode}, which is a deterministic curve in the state space. Then we find the total derivative w.r.t time $t$ (a.k.a material/particle derivative) is:
\begin{align}
    \fracdiff{}{t} \log p^\theta_{\clD,t}(\bfR_t)
    ={} & \fracpartial{}{t} \log p^\theta_{\clD,t}(\bfR_t) + \nabla \log p^\theta_{\clD,t}(\bfR_t) \cdot \fracdiff{\bfR_t}{t} \\
    ={} & \fracpartial{}{t} \log p^\theta_{\clD,t}(\bfR_t) + \nabla \log p^\theta_{\clD,t}(\bfR_t) \cdot \Big( -\frac{\beta_t}2 \bfR_t - \frac{\beta_t}2 \bfss^\theta_{\clD,t}(\bfR_t) \Big).
\end{align}
Compared with \eqnref{fpe-rev-ode}, we find:
\begin{align}
    \fracdiff{}{t} \log p^\theta_{\clD,t}(\bfR_t)
    = -\nabla \cdot \Big( -\frac{\beta_t}2 \bfR_t - \frac{\beta_t}2 \bfss^\theta_{\clD,t}(\bfR_t) \Big)
    = \frac{D}2 \beta_t + \frac{\beta_t}2 \nabla \cdot \bfss^\theta_{\clD,t}(\bfR_t).
\end{align}
By integration w.r.t $t$, this gives:
\begin{align}
    \log p^\theta_{\clD,0}(\bfR_0) = \log p^\theta_\tau(\bfR_\tau) - \int_0^\tau \frac{\beta_t}2 \nabla \cdot \bfss^\theta_{\clD,t}(\bfR_t) \dd t - \frac{D}2 \int_0^\tau \beta_t \dd t,
\end{align}
which gives \eqnref{logdensity}.

The equivalent deterministic process described by \eqnref{rev-ode} is called ``probabilistic flow ODE'' in machine learning literature~\citep{song2021score}. Since this deterministic process produces the same marginal distribution $p^\theta_{\clD,t}$ (particularly the equilibrium distribution $p^\theta_{\clD,t}$), it can also be used to generate samples. Due to the deterministic nature, this approach enables more techniques that could accelerate the sampling process (Supplementary Sec.~\ref{appx:accel-infer}).

\subsection{Protein Conformation Sampling}

\subsubsection{Diffusion Process on Coarse-Grained Representation of Protein}
\label{sec:acarbon-so3-diffusion-sampling-details}

Following the practice of successful protein structure prediction methods, e.g., AlphaFold~\citep{jumper2021highly}, we use the coarse-grained representation for protein as the $\bfR$ variable. With residues treated as rigid bodies, proteins are represented by the coordinates $\bfC$ in $\bbR^3$ of alpha-carbon atoms and the orientations $\bfQ$ in the 3-dimensional rotation group (a.k.a special orthogonal group) $\SO(3)$ of all the residues. Following AlphaFold~\cite{jumper2021highly} (Supplementary~1.8.1), the coordinates and orientations are constructed using backbone atom positions from the experimental structure, 
followed by a Gram–Schmidt process.

For the coordinates $\bfC$, the standard diffusion modeling can be applied. However, it is not straightforward for the orientation, as $\SO(3)$ is a non-Euclidean manifold. Therefore, the forward and reverse diffusion processes need to be generalized. For this treatment, we adopted the technique from~\citep{leach2022denoising,corso2023diffdock}.

\paragraph{Diffusion Process on the Orientations in the Special Orthogonal Group}
Noting that $\SO(3)$ is a Lie group (i.e., a manifold that is also an algebraic group), we can represent its elements in its Lie algebra (i.e., the tangent space at the identity element) $\so(3)$, which is a 3-dimensional linear space where vector addition and scaling are valid and random sampling is conventional.
Specifically, a 3-dimensional vector $\bfqq = (x,y,z) \in \so(3)$ can be interpreted as defining the rotation axis and the rotation angle (the amount of rotation) of the corresponding rotation transformation on $\bbR^3$ by the direction and the norm of $\bfqq$ as a usual vector in $\bbR^3$. The rotation matrix, as a form to represent an element in $\SO(3)$, can be constructed by:
\begin{align}
    \bfQ = \Exp\begin{psmallmatrix} 0 & z & -y \\ -z & 0 & x \\ y & -x & 0 \end{psmallmatrix} \in \SO(3), \;
    \bfqq = (x,y,z) \in \so(3),
    \label{eqn:Q-and-q}
\end{align}
in the conventional sense of a matrix exponential map.

To ease the calculation on $\SO(3)$, the forward diffusion process on it is taken as the corresponding Brownian motion (i.e., no drift term),
\begin{align}
    \ud \bfQ_t = \sqrt{\fracdiff{\sigma_t^2}{t}} \dd \tilde{\bfB}_t,
    \label{eqn:fwd-cont-so3}
\end{align}
where $\tilde{\bfB}_t$ denotes the Brownian motion on $\SO(3)$, and $\sqrt{\fracdiff{\sigma_t^2}{t}}$ (with $\sigma_t$ strictly increasing) is a time-dilation factor. This process converges to the uniform distribution (maximal entropy distribution on a compact space; $\SO(3)$ is compact) as the corresponding $p_\simple$.
Simulation of the Brownian motion, say, from time step $t_{i-1}$ to $t_i$, can be analogously done (up to $o(t_i - t_{i-1})$ discretization error) by adding a noise variable from the isotropic Gaussian distribution on $\SO(3)$ with variance $\sigma_{t_i}^2 - \sigma_{t_{i-1}}^2$, denoted as $\IG_{\SO(3)}(\bfzro, \sigma_{t_i}^2 - \sigma_{t_{i-1}}^2)$.
Drawing samples from $\IG_{\SO(3)}(\bfzro, \sigma^2)$ can be done in $\so(3)$ by uniformly sampling a direction in $\bbR^3$ for $\bfqq$, and sampling the length of $\bfqq$ (the length is within $[0,\pi]$) from the 1-dimensional distribution with the following density function:
\begin{align}
    & p_{\IG,\sigma^2}(\|\bfqq\|) = \frac{1 - \cos \|\bfqq\|}{\pi} \ppt_{\IG,\sigma^2}(\|\bfqq\|), \\
    \text{where }
    & \ppt_{\IG,\sigma^2}(\|\bfqq\|) := \sum_{l=0}^\infty (2l+1) \ue^{-l(l+1)(\sigma_{t_i}^2 - \sigma_{t_{i-1}}^2)} \frac{\sin((l+\frac12) \|\bfqq\|)}{\sin(\frac12 \|\bfqq\|)}.
    \label{eqn:igso3-density}
\end{align}
The density function written in $\bfqq$ under the Lebesgue measure in $\so(3)$ is then $\IG_{\SO(3)}(\bfqq \mid \bfzro, \sigma^2) \propto p_{\IG,\sigma^2}(\|\bfqq\|)$.

We learn the score model (instead of the noise-predicting model) for this setting.
As a gradient, the output of the score at time $t$ becomes an element in the tangent space at $\bfQ_t$. Again thanks to the group structure, they can be mapped to the tangent space at the identity element, i.e. $\so(3)$. The norm in $\so(3)$ consistent with the metric on $\SO(3)$ (i.e., the amount of rotation) is just the Euclidean 2-norm on the vector form $\bfqq$. Hence in the Lie algebra $\so(3)$, metric-related objects are the common Euclidean ones, including norm, gradient and divergence, which are to be used in the FPE \eqnref{fpe} hence the corresponding PIDP loss \eqnref{pinnloss} and data-based loss \eqnref{scoreloss}.
Nevertheless, there is a subtlety regarding the measure. To make the score function consistent with the diffusion process via the FPE, the density function should be taken w.r.t the uniform distribution on $\SO(3)$, which does not project to the Lebesgue measure (``uniform distribution'') in $\so(3)$. Instead, the $\SO(3)$ uniform distribution has the density:
\begin{align}
    p_\Unif(\bfqq) := \frac{1 - \cos \|\bfqq\|}{\pi}, \; (\|\bfqq\| \le \pi)
    \label{eqn:so3-unif}
\end{align}
under the Lebesgue measure in $\so(3)$.
Note this is also what $p_\simple$ takes.
So the required score function of a distribution should be $\nabla \log \frac{p(\bfqq)}{p_\Unif(\bfqq)}$, where $p(\bfqq)$ is the density function of the distribution under the Lebesgue measure in $\so(3)$.
In particular, the score function of the isotropic Gaussian on $\SO(3)$ is $\nabla_\bfqq \log \ppt_{\IG,\sigma^2}(\|\bfqq\|)$.

With these facts, we are ready to develop PIDP and data-based training for a diffusion model that involves the $\SO(3)$ space.
Recall that the coarse-grained representation $\bfR$ for proteins comprises alpha-carbon coordinates $\bfC$ and the orientations $\bfQ$ of residues. The orientations can be equivalently represented in $\so(3)$ as $\bfqq$, so we have $\bfR = (\bfC, \bfqq)$. Hence, the score model and the energy gradient (appearing in PIDP) take both $\bfC$ and $\bfqq$ as input, and output vectors for both $\bfC$ and $\bfqq$. Since the output vectors are in different spaces and have different losses, we split the output: $\bfss^\theta_{\clD,t} = (\bfss^{(\bfC),\theta}_{\clD,t}, \bfss^{(\bfqq),\theta}_{\clD,t})$, and $\nabla E_\clD = (\nabla_\bfC E_\clD, \nabla_\bfqq E_\clD)$.

Now consider the PIDP loss for $\bfss^{(\bfqq),\theta}_{\clD,t}$.
Following the forward process in \eqnref{fwd-cont-so3} and adopting the above definition of score function, the FPE \eqnref{fpe} leads to the loss:
\begin{align}
    & \lrVert{\frac12 \fracdiff{\sigma_t^2}{t} \Big( \nabla \big( \nabla \cdot \bfss^{(\bfqq),\theta}_{\clD,t}(\bfC_t,\bfqq_t) \big) + \nabla \lrVert{\bfss^{(\bfqq),\theta}_{\clD,t}(\bfC_t,\bfqq_t)}^2 \Big) - \fracpartial{}{t} \bfss^{(\bfqq),\theta}_{\clD,t}(\bfC_t,\bfqq_t)}^2 \\
    + & \lambda \lrVert{ \bfss^{(\bfqq),\theta}_{\clD,0}(\bfC_0,\bfqq_0) + \nabla_{\bfqq_0} E_\clD(\bfC_0,\bfqq_0) / (k_\rmB T) }^2.
    \label{eqn:pinnloss-protein-so3}
\end{align}
Following the pattern to run a PIDP loss as introduced in \secref{framework}, the sample of $(\bfC_0,\bfqq_0)$ is drawn from relevantly low-energy structures $\{(\bfC_{\clD,0}^{(m)},\bfqq_{\clD,0}^{(m)})\}_{m=1}^M$ for protein $\clD$ (not necessarily following the equilibrium distribution), and the corresponding $(\bfC_t,\bfqq_t)$ is sampled by letting $(\bfC_0,\bfqq_0)$ undergo the forward process.
We construct the forward process for $\bfC_t$ and $\bfqq_t$ independently, so the marginal transition kernel can be decomposed as:
\begin{align}
    q(\bfC_t, \bfqq_t \mid \bfC_0, \bfqq_0) = q(\bfC_t \mid \bfC_0) \, q(\bfqq_t \mid \bfqq_0).
    \label{eqn:so3-R3-marginal-transition-decomp}
\end{align}
Note that in the intermediate marginal distribution $q_t(\bfC_t,\bfqq_t)$, the two variables are not independent since they are not in the equilibrium distribution. Hence, both the $\bfss^{(\bfC),\theta}_{\clD,t}$ model and the $\bfss^{(\bfqq),\theta}_{\clD,t}$ model take both $\bfC_t$ and $\bfqq_t$ into their input.
For $\bfqq_t$ in \eqnref{so3-R3-marginal-transition-decomp}, recall that it is led by the Brownian motion on $\SO(3)$, whose marginal transition kernel is available in closed form:
\begin{align}
    q(\bfqq_t \mid \bfqq_0) = \IG_{\SO(3)}(\bfqq_t \mid \bfqq_0, \sigma_t^2),
    \label{eqn:so3-brownian-marginal-transition}
\end{align}
which turns sampling $\bfqq_t$ straightforward following the above description to sample an $\IG_{\SO(3)}$.
For $\bfC_t$ in \eqnref{so3-R3-marginal-transition-decomp}, it is sampled using \eqnref{R3-brownian-marginal-transition} detailed in the next part.

Data-based loss for $\bfss^{(\bfqq),\theta}_{\clD,t}$ is still based on the denoising score-matching loss \eqnref{dsmloss}. The score function to be matched can be simplified as $\nabla_{\bfqq_t} \log q(\bfC_t, \bfqq_t \mid \bfC_0, \bfqq_0) = \nabla_{\bfqq_t} \log q(\bfqq_t \mid \bfqq_0)$ from \eqnref{so3-R3-marginal-transition-decomp}. This $q(\bfqq_t \mid \bfqq_0)$ is an $\IG_{\SO(3)}$ from \eqnref{so3-brownian-marginal-transition}. Recalling the score function of $\IG_{\SO(3)}$, the data-based loss can then be written as:
\begin{align}
    \lrVert{ \bfss^{(\bfqq),\theta}_{\clD,t}(\bfC_t,\bfqq_t) - \nabla_{\bfqq_t} \log \ppt_{\IG,\sigma_t^2} (\|\bfqq_t\|) }^2.
    \label{eqn:databasedloss-protein-so3}
\end{align}
The function $\ppt_{\IG,\sigma^2} (\|\bfqq\|)$ is introduced in \eqnref{igso3-density}. The sample $(\bfC_t,\bfqq_t)$ for evaluating this loss is drawn following \eqnref{so3-R3-marginal-transition-decomp}, which again amounts to drawing $\bfqq_t$ following \eqnref{so3-brownian-marginal-transition} and $\bfC_t$ following \eqnref{R3-brownian-marginal-transition} below. The required $(\bfC_0,\bfqq_0)$ sample is drawn from the dataset $\{(\bfC_{\clD,0}^{(n)},\bfqq_{\clD,0}^{(n)})\}_{n=1}^{N_\data}$ that follows the equilibrium distribution of system $\clD$.

\paragraph{Diffusion Process on the Alpha-Carbon Coordinates}

To match the diffusion choice for $\SO(3)$ in \eqnref{fwd-cont-so3}, we also adopt the Brownian motion as the forward process in the Euclidean space for the alpha-carbon coordinates:
\begin{align}
    \ud \bfC_t = \sqrt{\fracdiff{\sigma_t^2}{t}} \dd \bfB_t.
    \label{eqn:fwd-cont-acarbon}
\end{align}
This coincides with the choice in noise-conditioned score network~\citep{song2019generative,song2021score}.

The PIDP loss for $\bfss^{(\bfC),\theta}_{\clD,t}$ from the FPE \eqnref{fpe} then becomes:
\begin{align}
    & \lrVert{\frac12 \fracdiff{\sigma_t^2}{t} \Big( \nabla \big( \nabla \cdot \bfss^{(\bfC),\theta}_{\clD,t}(\bfC_t,\bfqq_t) \big) + \nabla \lrVert{\bfss^{(\bfC),\theta}_{\clD,t}(\bfC_t,\bfqq_t)}^2 \Big) - \fracpartial{}{t} \bfss^{(\bfC),\theta}_{\clD,t}(\bfC_t,\bfqq_t)}^2 \\
    + & \lambda \lrVert{ \bfss^{(\bfC),\theta}_{\clD,0}(\bfC_0,\bfqq_0) + \nabla_{\bfC_0} E_\clD(\bfC_0,\bfqq_0) / (k_\rmB T) }^2.
    \label{eqn:pinnloss-protein-acarbon}
\end{align}
The sample of $(\bfC_0,\bfqq_0)$ to evaluate the loss is again drawn from relevant structures $\{(\bfC_{\clD,0}^{(m)},\bfqq_{\clD,0}^{(m)})\}_{m=1}^M$ for protein $\clD$, and the sample of $\bfqq_t$ following \eqnref{so3-brownian-marginal-transition}.
For the sample of $\bfC_t$, it is drawn from the marginal transition kernel of the diffusion process in \eqnref{fwd-cont-acarbon}, which is a Gaussian distribution thus easy to draw:
\begin{align}
    q(\bfC_t \mid \bfC_0) = \clN(\bfC_t \mid \bfC_0, \sigma_t^2 \bfI).
    \label{eqn:R3-brownian-marginal-transition}
\end{align}

Data-based loss for $\bfss^{(\bfC),\theta}_{\clD,t}$ also follows \eqnref{dsmloss}, where the required score function is $\nabla_{\bfC_t} \log q(\bfC_t,\bfqq_t \mid \bfC_0,\bfqq_0) = \nabla_{\bfC_t} \log q(\bfC_t \mid \bfC_0)$ due to \eqnref{so3-R3-marginal-transition-decomp}, which leads to:
\begin{align}
    \lrVert{ \bfss^{(\bfC),\theta}_{\clD,t}(\bfC_t,\bfqq_t) - \nabla_{\bfC_t} \log q(\bfC_t \mid \bfC_0) }^2.
    \label{eqn:databasedloss-protein-acarbon}
\end{align}
The sample of $(\bfC_0,\bfqq_0)$ here is drawn from the dataset $\{(\bfC_{\clD,0}^{(n)},\bfqq_{\clD,0}^{(n)})\}_{n=1}^{N_\data}$ that follows the equilibrium distribution, and sample of $(\bfC_t,\bfqq_t)$ is drawn following \eqnsref{so3-brownian-marginal-transition,R3-brownian-marginal-transition}.
If substituting \eqnref{R3-brownian-marginal-transition} and expressing the loss in terms of the standard Gaussian sample $\bfeps_t \sim \clN(\bfzro,\bfI)$ following the style of \eqnref{scoreloss}, then \eqnref{databasedloss-protein-acarbon} becomes:
$\lrVert{ \bfss^{(\bfC),\theta}_{\clD,t}(\bfC_0 + \sigma_t \bfeps_t, \bfqq_t) + \frac{\bfeps_t}{\sigma_t} }^2$.
Nevertheless, such a reformulation does not easily apply to substitute $\bfqq_t$ due to the complexity of $\IG_{\SO(3)}$ in \eqnref{so3-brownian-marginal-transition}.

\paragraph{Structure Sampling Using DiG}

To generate structure samples using DiG, we find rather than directly simulating the reverse SDE analogous to \eqnref{rev-disc}, it is better to simulate the equivalent deterministic process defined by an ODE analogous to \eqnref{rev-ode}. The rationale of the equivalent ODE is explained in Supplementary Sec.~\ref{sec:details-density}. Following the deduction there, the equivalent ODE for the diffusion processes in \eqnsref{fwd-cont-so3,fwd-cont-acarbon} can be derived as:
\begin{align}
    \ud \begin{psmallmatrix} \bfC_t \\ \bfqq_t \end{psmallmatrix}
    = -\frac12 \fracdiff{\sigma_t^2}{t} \begin{psmallmatrix}
    \bfss^{(\bfC),\theta}_{\clD,t} (\bfC_t,\bfqq_t) \\ \bfss^{(\bfqq),\theta}_{\clD,t} (\bfC_t,\bfqq_t)
    \end{psmallmatrix} \ud t
    = -\frac12 \begin{psmallmatrix}
    \bfss^{(\bfC),\theta}_{\clD,t} (\bfC_t,\bfqq_t) \\ \bfss^{(\bfqq),\theta}_{\clD,t} (\bfC_t,\bfqq_t)
    \end{psmallmatrix} \ud \sigma_t^2.
    \label{eqn:acarbon-so3-ode}
\end{align}
The simulation on $\bfC_t$ is thus:
\begin{align}
    \bfC_{t_{i-1}} = \bfC_{t_i} + \frac12 (\sigma_{t_i}^2 - \sigma_{t_{i-1}}^2) \bfss^{(\bfC),\theta}_{\clD,t_i}(\bfC_{t_i},\bfqq_{t_i}).
    \label{eqn:rev-ode-disc-acarbon}
\end{align}
The simulation on $\bfqq_t$ can be done similarly by $\bfqq_{t_{i-1}} = \bfqq_{t_i} + \frac12 (\sigma_{t_i}^2 - \sigma_{t_{i-1}}^2) \bfss^{(\bfqq),\theta}_{\clD,t_i}(\bfC_{t_i},\bfqq_{t_i})$.
This discretization on $\so(3)$ is equivalent to discretization on $\SO(3)$ in the sense that their one-step difference is $o(t_i - t_{i-1})$, but the simulation in $\so(3)$ directly faces the risk that the discretization error may lead the $\bfqq$ variable going out of the domain (i.e., $\|\bfqq\| > \pi$), as there is no mechanism to guarantee the constraint. We therefore carry out the simulation in $\SO(3)$ instead:
\begin{align}
    \bfQ_{t_{i-1}} = \Exp \Big( \frac12 (\sigma_{t_i}^2 - \sigma_{t_{i-1}}^2) \bfss^{(\bfQ),\theta}_{\clD,t_i}(\bfC_{t_i},\bfqq_{t_i}) \Big) \bfQ_{t_i},
    \label{eqn:rev-ode-disc-SO3}
\end{align}
where $\Exp$ is the conventional matrix exponent, and $\bfss^{(\bfQ),\theta}_{\clD,t_i}$ denotes the skew-symmetric matrix by organizing the outputs of $\bfss^{(\bfqq),\theta}_{\clD,t_i}$ in the same way as converting $\bfQ$ and $\bfqq$ \eqnref{Q-and-q}.
Note that the matrix exponent of a skew-symmetric matrix is a rotation matrix, \eqnref{rev-ode-disc-SO3} then guarantees $\bfQ_{t_{i-1}} \in \SO(3)$ whenever $\bfQ_{t_i} \in \SO(3)$.
Alg.~\ref{alg:protein_sampling} summarizes the sampling procedure.

\paragraph{Interpolation between Protein States Using DiG}
The deterministic nature of the ODE \eqnref{acarbon-so3-ode} establishes a deterministic map between a real state $\bfR_0$ and the corresponding latent state $\bfR_\tau$.
This enables the complicated interpolation between two given states, $\bfR_0^{(A)} = (\bfC_0^{(A)}, \bfqq_0^{(A)})$ and $\bfR_0^{(B)} = (\bfC_0^{(B)}, \bfqq_0^{(B)})$, by a simpler interpolation in the latent space of $\bfR_\tau$ where the distribution is simple.
For the alpha-carbon coordinates $\bfC_0^{(A)}$ and $\bfC_0^{(B)}$, we apply linear interpolation to their corresponding latent states $\bfC_{\tau}^{(A)}$ and $\bfC_{\tau}^{(B)}$ through the ODE, and then transform the line to the real-state space of $\bfC_0$ by the ODE reversely. Since the coordinate distribution in the latent space is standard Gaussian, which has a convex contour, linear interpolation there would pass through high-probability regions.
For the residue orientations $\bfqq_0^{(A)}$ and $\bfqq_0^{(B)}$, as the corresponding latent states $\bfqq_\tau^{(A)}$ and $\bfqq_\tau^{(B)}$ lie in the product space of $\so(3)$ which is non-Euclidean, we leverage spherical linear interpolation which gives the geodesic in $\so(3)$ between two given end states, in place of the linear interpolation which is the geodesic in the Euclidean space. Explicitly, the interpolation curves in the latent space are:
\begin{align}
    \bfC_{\tau}^{(\eta)} &= (1-\eta) \bfC_{\tau}^{(A)} + \eta \bfC_{\tau}^{(B)},
    \label{eqn:interp-acarboon} \\
    \bfqq_{\tau}^{(\eta)} &= (\bfqq_{\tau}^{(B)}(\bfqq_{\tau}^{(A)})^{-1})^{\eta}\bfqq_{\tau}^{(A)}
    \label{eqn:interp-so3},
\end{align}
where $\eta \in [0, 1]$ parameterizes the interpolation curve. Subsequently, $(\bfC_{\tau}^{(\eta)}, \bfqq_{\tau}^{(\eta)})$ in \eqnsref{interp-acarboon,interp-so3} are taken as the starting state to simulate the ODEs \eqnsref{rev-ode-disc-acarbon,rev-ode-disc-SO3} reversely to generate the interpolated structures at the parameter $\eta$ along the transition pathway between state $A$ and $B$.

\subsubsection{Model Specification}\label{sec:protein-model-spec}

Following the practice of AlphaFold~\citep{jumper2021highly}, we use amino acid sequences as the molecular descriptor $\clD$ for proteins. To process the sequence to generate informative abstract representations for the feed into DiG, we follow the data processing method in the training stage of AlphaFold and leverage the pre-trained Evoformer to produce node and pair representations. Conditioned on representations of proteins, DiG aims to gradually transform random noise to reasonable protein structures following the equilibrium distribution. Considering that protein simulation trajectories that are long enough and reach equilibrium distribution are very rare, for protein conformation sampling, as in Sec.~\ref{sec:framework}, we pre-train DiG with PIDP first, and then further improved the performance by training the model with a small amount of simulation data.

\begin{algorithm}[th]
\caption{Protein score model PIDP training (single step)}\label{alg:protein_pidp_training}
\begin{algorithmic}[1]
\Require Score model $\bfss^{\theta}_{\clD,t}(\bfR)$ to be trained, boundary loss weight $\lambda$, randomly sampled system $\clD$, full-atom energy function $E_\clD$ for system $\clD$, a randomly sampled relevant full-atom protein structure $\bar{\bfR}_{\clD,0}^{(m)}$ for system $\clD$, 
randomly sampled time step $t \in [0,\tau]$.
\State Construct $\bfR_{\clD,0}^{(m)} := ( \bfC_{\clD,0}^{(m)}, \bfqq_{\clD,0}^{(m)} ) $ from the sampled $\bar{\bfR}_{\clD,0}^{(m)}$;
\State Compute the energy gradient $\nabla_\bfC E_\clD(\bfC_{\clD,0}^{(m)}, \bfqq_{\clD,0}^{(m)})$ and $\nabla_\bfqq E_\clD(\bfC_{\clD,0}^{(m)}, \bfqq_{\clD,0}^{(m)})$ for the alpha-carbon coordinates and residue orientations, respectively, from the full-atom energy function $E_\clD$ using Alg.~\ref{alg:protein_pidp_full_to_cg_ff}; \label{line:force}
\State Sample $(\bfC_{\clD,t}^{(m)}, \bfqq_{\clD,t}^{(m)})$ from $(\bfC_{\clD,0}^{(m)}, \bfqq_{\clD,0}^{(m)})$ using \eqnref{R3-brownian-marginal-transition} and \eqnref{so3-brownian-marginal-transition};
\State Evaluate the PIDP loss $\clL_{\text{PIDP}}$ as \eqnref{pinnloss-protein-so3} + \eqnref{pinnloss-protein-acarbon}; \label{line:pidp} 
\State Update the model parameter $\theta$ by performing an optimization step on $\clL_{\text{PIDP}}$ with respect to $\theta$. \label{line:update}
\end{algorithmic}
\end{algorithm}

\begin{algorithm}[th]
\caption{Compute the energy gradient on orientations and coordinates}\label{alg:protein_pidp_full_to_cg_ff}
\begin{algorithmic}[1]
\Require Energy function $E$, full atom protein structure $\bar{\bfR}$ with $I$ amino acid residues.
\State Construct $\bfR := (\bfC, \bfQ)$ from $\bar{\bfR}$;
\State Compute the energy $E = E(\bar{\bfR})$;
\For {each residue $\imath$ in $1,\cdots,I$}
\State Set $\bfX_\imath$ as the coordinates of all $N_\imath$ atoms in the residue $\imath$;
\State $\bfX_{\imath,\mathrm{rel}} := (\bfX_\imath-\bfC_\imath)\bfQ_\imath\trs$ (c.f. \eqnref{rigid-body});
\State $\bfgg_{\bfC_\imath} := \sum\nolimits_{a=1}^{N_\imath} \nabla_{\bfX_{\imath,a}} E$ (c.f. \eqnref{grad-acarbon});
\State $\bfgg_{\bfqq_\imath} := \sum_{a=1}^{N_\imath} \bfX_{\imath,\mathrm{rel},a}\trs \nabla_{\bfqq_\imath} \bfQ_\imath \nabla_{\bfX_{\imath,a}} E$ (c.f. \eqnref{grad-so3});
\EndFor
\State Return $\nabla_\bfC E := \{\bfgg_{\bfC_\imath}\}_{\imath=1}^I$ and $\nabla_\bfqq E := \{ \bfgg_{\bfqq_\imath} \}_{\imath=1}^I$.
\end{algorithmic}
\end{algorithm}

\begin{algorithm}[th]
\caption{Estimate $\nabla (\nabla \cdot \bfss^\theta_{\clD,t}(\bfR_t))$ by Hutchinson's trace estimator}\label{alg:protein_pidp_evaluate_loss}
\begin{algorithmic}[1]
\Require Score model $\bfss^{\theta}_{\clD,t}(\bfR)$, protein structure $\bfR_t$, number of random vectors $N_\mathrm{est}$.
\State Sample $N_\mathrm{est}$ random vectors of the same dimension as $\bfR$: $\left\{ \rvv^{(n)} \right\}_{n=1}^{N_\mathrm{est}} \overset{\mathrm{i.i.d.}}{\sim} \mathrm{Rademacher}(0.5)$;
\State $\rvg := \frac{1}{N_\mathrm{est}} \sum\nolimits_{n=1}^{N_\mathrm{est}} \nabla_{\bfR_t} ( {\rvv^{(n)}}\trs \nabla_{\bfR_t} (\bfss^{\theta}_{\clD,t}(\bfR_t)\trs \rvv^{(n)}) )$;
\State Return $\rvg$ as an approximation to $\nabla (\nabla \cdot \bfss^\theta_{\clD,t}(\bfR_t))$.
\end{algorithmic}
\end{algorithm}

We first train DiG by minimizing a PIDP loss that aligns the score model with the gradients of the energy function and enforces the boundary conditions. 
Alg.~\ref{alg:protein_pidp_training} outlines the training process.
The PIDP training requires the energy gradient label. This is facilitated by an energy function from OpenMM~\cite{eastman2017openmm} at the full-atom level. But as we are adopting a coarse-grained representation for proteins, so the energy gradients w.r.t alpha carbon coordinates and residue orientations are expected. For this, we leverage the rigid-body assumption and the chain rule to derive the conversion, as shown in Alg.~\ref{alg:protein_pidp_full_to_cg_ff}.

Here we briefly explain the derivation in Alg.~\ref{alg:protein_pidp_full_to_cg_ff}. The rigid-body assumption states that for each residue $\imath$ with $N_\imath$ atoms, the relative coordinates $\bfX_{\imath,\mathrm{rel}} \in \bbR^{N_\imath \times 3}$ of all its atoms w.r.t its alpha-carbon at $\bfC_\imath \in \bbR^{1 \times 3}$ in a standard coordinate system is fixed. Under this assumption, if the coarse-grained representation of this residue is $(\bfC_\imath,\bfQ_\imath)$ where $\bfQ_\imath$ is the orientation of the residue relative to the standard coordinate system, then the absolute coordinates of these atoms are:
\begin{align}
    \bfX_\imath = \bfC_\imath + \bfX_{\imath,\mathrm{rel}} \bfQ_\imath,
    \label{eqn:rigid-body}
\end{align}
and if considering a protein with $I$ residues, the full-atom coordinates are $\bfRb := \{\bfC_\imath + \bfX_{\imath,\mathrm{rel}} \bfQ_\imath\}_{\imath=1}^I$.
So in this way, we can convert the full-atom energy function $E(\bfRb) = E(\{\bfX_\imath\}_{\imath=1}^I)$ as a function of the coarse-grained coordinates $\bfR = (\bfC,\bfQ)$ using \eqnref{rigid-body}.
The gradient w.r.t $\bfC_\imath$ is then:
\begin{align}
    \bfgg_{\bfC_\imath} &:= \nabla_{\bfC_\imath} E
    = (\nabla_{\bfC_\imath} \bfX_\imath)\trs \nabla_{\bfX_\imath} E
    = \sum_{a=1}^{N_\imath} \nabla_{\bfX_{\imath,a}} E,
    \label{eqn:grad-acarbon}
\end{align}
where $(\nabla_{\bfC_\imath} \bfX_\imath)_{a\mu, \nu} := \fracpartial{\bfX_{\imath,a,\mu}}{\bfC_{\imath,\nu}}$ is the Jacobian matrix (here $\mu,\nu \in \{1,2,3\}$ indices the spacial dimension), and the last equality holds since the Jacobian is $(\nabla_{\bfC_\imath} \bfX_\imath)_{a\mu, \nu} = \delta_{\mu\nu}$ meaning that this matrix element is one if $\mu = \nu$ or it is zero.

As for the gradient w.r.t the orientation, the $\so(3)$ representation denoted as $\bfqq$ is finally required. The conversion from $\bfqq$ and $\bfQ$ is given by \eqnref{Q-and-q}. Together with the rigid-body assumption \eqnref{rigid-body}, the gradient is:
\begin{align}
    \bfgg_{\bfqq_\imath} &:= \nabla_{\bfqq_\imath} E
    = (\nabla_{\bfqq_\imath} \bfX_\imath)\trs \nabla_{\bfX_\imath} E
    = (\nabla_{\bfQ_\imath} \bfX_\imath \nabla_{\bfqq_\imath} \bfQ_\imath)\trs \nabla_{\bfX_\imath} E \\
    &= \sum_{a=1}^{N_\imath} \bfX_{\imath,\mathrm{rel},a}\trs \nabla_{\bfqq_\imath} \bfQ_\imath \nabla_{\bfX_{\imath,a}} E,
    \label{eqn:grad-so3}
\end{align}
where the last term means $(\bfgg_{\bfqq_\imath})_\gamma = \sum_{a=1}^{N_\imath} \bfX_{\imath,\mathrm{rel},a}\trs \fracpartial{\bfQ_\imath}{\bfqq_{\imath,\gamma}} \nabla_{\bfX_{\imath,a}} E$, where $\gamma \in \{1,2,3\}$ indices one of the three dimensions of $\bfqq_\imath$, and $\fracpartial{\bfQ_\imath}{\bfqq_{\imath,\gamma}}$ is the $3 \times 3$ matrix composed of the partial derivatives from \eqnref{Q-and-q}.
In the equation, again $\nabla_{\bfqq_\imath} \bfX_\imath$, $\nabla_{\bfQ_\imath} \bfX_\imath$ and $\nabla_{\bfqq_\imath} \bfQ_\imath$ are Jacobian matrices, and the second last equality holds due to the chain rule of differentiation. From the rigid-body assumption \eqnref{rigid-body}, $(\nabla_{\bfQ_\imath} \bfX_\imath)_{a\nu', \mu\nu} = \fracpartial{\bfX_{\imath,a,\nu'}}{\bfQ_{\imath,\mu,\nu}} = \bfX_{\imath,\mathrm{rel},a,\mu} \delta_{\nu\nu'}$, so:
\begin{align}
    & [(\nabla_{\bfQ_\imath} \bfX_\imath \nabla_{\bfqq_\imath} \bfQ_\imath)\trs \nabla_{\bfX_\imath} E]_\gamma
    = \sum_{a,\nu',\mu,\nu} (\nabla_{\bfQ_\imath} \bfX_\imath)_{a\nu',\mu\nu} (\nabla_{\bfqq_\imath} \bfQ_\imath)_{\mu\nu,\gamma} (\nabla_{\bfX_\imath} E)_{a\nu'} \\
    ={}& \sum_{a,\mu,\nu} \bfX_{\imath,\mathrm{rel},a,\mu} (\nabla_{\bfqq_\imath} \bfQ_\imath)_{\mu\nu,\gamma} (\nabla_{\bfX_\imath} E)_{a\nu}
    = \sum_{a} \bfX_{\imath,\mathrm{rel},a}\trs \fracpartial{\bfQ_\imath}{\bfqq_{\imath,\gamma}} \nabla_{\bfX_\imath,a} E,
\end{align}
which gives the last equality.

Moreover, to avoid costly divergence evaluation, we use Hutchinson's trace estimator in Alg.~\ref{alg:protein_pidp_evaluate_loss} to handle $\nabla (\nabla \cdot \bfss^\theta_{\clD,t}(\bfR_t))$ (recall that $\bfss^\theta_{\clD,t}(\bfR_t) = (\bfss^{(\bfC),\theta}_{\clD,t}(\bfC_t,\bfqq_t), \bfss^{(\bfqq),\theta}_{\clD,t}(\bfC_t,\bfqq_t))$) in \eqnsref{pinnloss-protein-so3,pinnloss-protein-acarbon}.

Optimizing PIDP from random initialization of the deep learning model is extremely hard due to the complex landscapes of the training objectives in \eqnsref{pinnloss-protein-so3,pinnloss-protein-acarbon}. Therefore, a good initialization of the model and some training techniques are necessary to stabilize the optimization of PIDP. To this end, before performing PIDP, we train DiG on the experimental structures and use it as a more stable initialization for PIDP.
See Supplementary Sec.~\ref{sec:training-details} for more details.

\begin{algorithm}[!t]
\caption{Protein score model data-based training (single step)}\label{alg:protein_databased_training}
\begin{algorithmic}[1]
\Require Score model $\bfss^{\theta}_{\clD,t}(\bfR)$ to be trained, randomly sampled system $\clD$, a randomly sampled coarse-grained protein structure $\bfR_{\clD,0}^{(n)} = (\bfC_{\clD,0}^{(n)},\bfqq_{\clD,0}^{(n)})$ from the MD simulation data for system $\clD$, randomly sampled time step $t \in [0,\tau]$.
\State Sample $(\bfC_{\clD,t}^{(n)}, \bfqq_{\clD,t}^{(n)})$ from $(\bfC_{\clD,0}^{(n)}, \bfqq_{\clD,0}^{(n)})$ using \eqnref{R3-brownian-marginal-transition} and \eqnref{so3-brownian-marginal-transition};
\State Evaluate the data-based loss $\clL_{\data}$ as \eqnref{databasedloss-protein-so3} + \eqnref{databasedloss-protein-acarbon}; 
\State Update the model parameter $\theta$ by performing an optimization step on $\clL_{\text{data}}$ with respect to $\theta$.
\end{algorithmic}
\end{algorithm}

Next, we pick about 1000 protein complexes in PDBbind database~\cite{wang2005pdbbind}, and simulate them with GROMACS, together with about 200 proteins in GPCRmd~\cite{rodriguez2020gpcrmd} dataset to perform DiG training with simulation data (See more details in Supplementary Sec.~\ref{appx:md-prot-complex}). We further train the score model pre-trained by PIDP by minimizing a score matching loss from \eqnref{scoreloss} that directly supervises the score model with the empirical data distribution that approximates the equilibrium distribution, using protein structures $\bfR_{\clD,0}^{(n)}$ for each system $\clD$. See Supplementary Sec.~\ref{sec:model-details} and~\ref{sec:training-details} for more details on the model and training.

\begin{algorithm}[!t]
\caption{Protein structure sampling}\label{alg:protein_sampling}
\begin{algorithmic}[1]
\Require A trained score model $\bfss^{\theta}_{\clD,t}(\bfR)$, target protein system $\clD$.
\State Initialize random structure $\bfR_\tau:= (\bfC_\tau, \bfqq_\tau)$, where $\bfC_\tau \sim \clN(\bfzro, \sigma_\tau^2 \bfI)$, and $\bfqq_\tau \sim p_\Unif$ on $\so(3)$ defined in \eqnref{so3-unif}. $t_N := \tau$.
\For {$i$ in $N, \cdots, 1$}
\State $t_{i-1} := \frac{i-1}N \tau$;
\State $\bfC_{t_{i-1}} := \bfC_{t_i} + \frac12 (\sigma_{t_i}^2 - \sigma_{t_{i-1}}^2) \bfss^{(\bfC),\theta}_{\clD,t_i}(\bfC_{t_i},\bfqq_{t_i})$ (c.f. \eqnref{rev-ode-disc-acarbon});
\State Construct $\bfQ_{t_i}$ from $\bfqq_{t_i}$, and $\bfss^{(\bfQ),\theta}_{\clD,t_i}(\bfC_{t_i},\bfqq_{t_i})$ from $\bfss^{(\bfqq),\theta}_{\clD,t_i}(\bfC_{t_i},\bfqq_{t_i})$, using \eqnref{Q-and-q};
\State $\bfQ_{t_{i-1}} := \Exp \Big( \frac12 (\sigma_{t_i}^2 - \sigma_{t_{i-1}}^2) \bfss^{(\bfQ),\theta}_{\clD,t_i}(\bfC_{t_i},\bfqq_{t_i}) \Big) \bfQ_{t_i}$ (c.f. \eqnref{rev-ode-disc-SO3});
\EndFor
\State Return the sampled structure $\bfR_0 := (\bfC_0, \bfqq_0)$.
\end{algorithmic}
\end{algorithm}

After training, amino acid sequences serve as input descriptors, denoted as $\clD$, for sampling protein conformations via DiG. During the sampling procedure, an initial random structure is generated, which subsequently is transformed into a physically plausible conformation, as shown in Alg.~\ref{alg:protein_sampling}. 

\subsection{Ligand Structure Sampling around Binding Sites}
\label{sec:protein-ligand-sampling-details}

\newcommand{\Com}{\textrm{Com}}
\newcommand{\Rec}{\textrm{Rec}}
\newcommand{\Lig}{\textrm{Lig}}
\newcommand{\sol}{\textrm{sol}}

In contrast to the coarse-grained representation employed in protein conformation sampling, 
we train DiG of ligand structure sampling with all-atom (except hydrogens) representations, which offer a more precise description of atomic interactions between the binding site (pocket) and ligand during the sampling of ligand-binding structures with proteins. However, the time complexity and memory usage associated with attention layers in Transformer-based architectures exhibit a quadratic increase with respect to the number of input nodes. This becomes impractical when the atom count surpasses one thousand. Consequently, we restrict our model to incorporate only the atoms of the ligand and the protein atoms in close proximity to the pocket, using a distance threshold. This threshold is set to $10\,\AA$ for the side length.

DiG defines a distribution over the vector space $\bfRb := (\bfRb_\Rec,\bfRb_\Lig)$, where $\bfRb_\Rec$ and $\bfRb_\Lig$ are the absolute coordinates of the near-site receptor and ligand atoms, respectively. Since the receptor atom coordinates may have different distribution centers for different proteins while the diffusion process always starts from a zero-centered Gaussian, we use the near-site receptor atom coordinates $\bfRb_{\Rec}^\star$ from the crystal structure of the protein in the PDBBind database~\citep{wang2005pdbbind} as a reference structure, and let the model predict the residue. This effectively shifts the diffusion process to $\bfR := (\bfRb_\Rec - \bfRb_{\Rec}^\star, \bfRb_\Lig)$. DiG then generates the binding structures for the ligand and the protein pocket by reversing the diffusion process, as shown in \eqnref{rev-disc}.

To train DiG, we reuse the simulation data of protein complexes in PDBbind, as detailed in Supplementary Sec.~\ref{sec:training-details}. The data preprocessing and featurization follows~\cite{min2022predicting,he2022masked}. The atom representations are further embedded into real-valued embedding vectors for a Graphormer~\cite{ying2021transformers}.
In this context, performing PIDP is not feasible due to limitations imposed by the energy function. Specifically, DiG only considers atoms surrounding the binding site, while conventional force fields necessitate the inclusion of all atoms. As such, we sought to enhance ligand sampling performance within the pocket by conducting a pre-training task focused on binding structure prediction, utilizing the CrossDocked dataset~\cite{francoeur2020three}. Further information regarding model architecture and training can be found in Supplementary Sec.~\ref{sec:model-details} and~\ref{sec:training-details}.

\subsection{Catalyst-Adsorbate Sampling}
\label{sec:catalyst-adsorbate-sampling-details}

\renewcommand{\Cat}{\mathrm{Cat}}
\renewcommand{\Ad}{\mathrm{Ad}}
\newcommand{\atom}{\mathrm{atom}}

For catalyst-adsorbate sampling, DiG adopts the same input representation strategy employed in the OC20 dataset~\cite{chanussot2021open}, employing the descriptor $\clD$ to characterize the system. Specifically, besides the atom types $\clZ$, also provided from the OC20 dataset are the absolute coordinates $\bfRb^\star_\base$ for non-surface catalyst atoms, $\bfRb^\star_\Cat$ for surface catalyst atoms, and the initial absolute coordinates $\bfRb^\star_\Ad$ for adsorbate atoms prior to relaxation. Consequently, the system descriptor for this task was defined as $\clD := (\clZ, \bfRb^\star_\base, \bfRb^\star_\Cat, \bfRb^\star_\Ad)$.
The microscopic state of the system $\bfRb := (\bfRb_\Cat, \bfRb_\Ad)$ encompasses the absolute coordinates $\bfRb_\Cat$ of surface catalyst atoms and $\bfRb_\Ad$ of the adsorbate atoms.
Similar to the ligand-receptor sampling case, to ease the prediction of different distribution centers for different systems, we leverage the absolute coordinates in the descriptor to define the diffusion-process variable as relative coordinates, i.e., $\bfR := (\bfR_\Cat, \bfR_\Ad)$ where $\bfR_\Cat := \bfRb_\Cat - \bfRb^\star_\Cat$ and $\bfR_\Ad := \bfRb_\Ad - \bfRb^\star_\Ad$, whose distribution center is largely aligned across different systems.

During the reverse diffusion process, including the initial structure $\bfRb^\star_\Ad$ into the model input is found crucial for stable training. 
In the Graphormer model, in addition to the diffusion-variable $\bfR$ in the input, the initial structure $\bfRb_\Ad^\star$ is also encoded as an additional structural attention bias term, as in~\cite{shi2022benchmarking}. More specifically, the pairwise distance between atoms in $\bfRb^\star_\Ad$ was calculated, which is then encoded into a $K$-dimensional feature using $K$ radial basis function (RBF) kernels with learnable means and variances.
Likewise, the initial positions of the atoms are encoded as extra node features and incorporated into the node representation. By summing the pairwise features and aggregating them with the node features, per-atom features are updated and projected into the atom embedding dimension within the model. Further details regarding the structural attention bias can be found in Supplementary Sec.~\ref{sec:struct_attn_bias}.

Catalyst systems in OC20 are inorganic and lack bond information between atoms. However, adsorbates are organic, and the bonds between atoms in adsorbates can benefit the model in generating more physically accurate adsorbate structures. Consequently, we explicitly incorporated the 2D topology of adsorbates, featuring bonds connecting the atoms, within the model. Bonds are generated using the initial structure of the adsorbate and encoded in the same manner as Graphormer~\cite{ying2021transformers}. The spatial encoding, centrality encoding, and edge encoding of Graphormer are utilized alongside the encodings derived from 3D information. It is important to note that, in some instances, bonds in adsorbates may break upon adsorption to the catalyst surface. Explicitly encoding bond information does not imply the enforcement of bonded atoms to remain close in the sampled structures. Instead, the model is allowed to learn when to separate two bonded atoms.

The training and sampling of DiG adhere to the general description in the main text (Sec.~\ref{sec:framework}). 
The training loss is based on \eqnref{noiseloss}, and Alg.~\ref{alg:catalyst_adsorbate_training} shows the detailed training process.
For sampling new structures using DiG, in accordance with the training process, it is also conducted on the relative coordinates w.r.t the initial structure. Alg.~\ref{alg:catalyst_adsorbate_sampling} outlines the sampling process.

\begin{algorithm}
\caption{Catalyst-adsorbate score model training (single step)}\label{alg:catalyst_adsorbate_training}
\begin{algorithmic}[1]
\Require Noise-predicting model $\bfeps^\theta_{\clD,t}(\bfR)$ to be trained, randomly sampled time step $t \in [0,\tau]$, randomly sampled system with descriptor $\clD = (\clZ, \bfRb^\star_\base, \bfRb^\star_\Cat, \bfRb^\star_\Ad)$, let $\bfRb^\star := \left( \bfRb^\star_\Cat, \bfRb^\star_\Ad \right)$, randomly sampled catalyst-adsorbent structure $\bfRb_{\clD,0}^{(n)} = (\bfRb_{\clD,\Cat,0}^{(n)}, \bfRb_{\clD,\Ad,0}^{(n)})$ from the MD simulation data for this system $\clD$.
\State Let $\bfR_{\clD,0}^{(n)} := \bfRb_{\clD,0}^{(n)} - \bfRb^\star$;
\State Sample noise variable $\bfeps_t \sim \clN(\bfzro, \bfI)$ in the same dimension as $\bfR_{\clD,0}^{(n)}$;
\State $\bfR^{(n)}_{\clD,t} := \alpha_t \bfR_{\clD,0}^{(n)} + \sqrt{1 - \alpha_t^2} \bfeps_t$ (c.f. \eqnref{q_il0});
\State Evaluate the loss $\lrVert{\bfeps^\theta_{\clD,t}(\bfR^{(n)}_{\clD,t} + \bfRb^\star) - \bfeps_t}^2$ (c.f. \eqnref{noiseloss});
\State Update the model parameter $\theta$ by performing an optimization step on the loss with respect to $\theta$.
\end{algorithmic}
\end{algorithm}

\begin{algorithm}
\caption{Catalyst-adsorbate structure sampling}\label{alg:catalyst_adsorbate_sampling}
\begin{algorithmic}[1]
\Require A trained noise-predicting model $\bfeps^\theta_{\clD,t}(\bfR)$, the descriptor $\clD = (\clZ, \bfRb^\star_\base, \bfRb^\star_\Cat, \bfRb^\star_\Ad)$ of the target system, let $\bfRb^\star := \left( \bfRb^\star_\Cat, \bfRb^\star_\Ad \right)$.
\State Sample a noisy structure $\bfR_\tau \sim p_\simple = \clN(\bfzro, \bfI)$;
\For {$i$ in $N, \cdots, 1$}
\State $t_{i-1} := \frac{i-1}N \tau$;
\State Let $\beta_i := \frac{\tau}{N} \beta_{t_i}$, $\alpha_i := \prod_{j=1}^i \sqrt{1 - \beta_j}$;
\State Sample a noise variable $\bfeps_{t_i} \sim \clN(\bfzro, \beta_i \bfI)$ in the same dimension as $\bfR_{t_i}$;
\State $\bfR_{t_{i-1}} \!:= \! \frac{1}{\sqrt{1-\beta_i}} \Big(\! \bfR_{t_i} - \frac{\beta_i}{\sqrt{1-\alpha_i^2}} \bfeps^\theta_{\clD,t_i}(\bfR_{t_i} + \bfRb^\star) \!\Big) + \bfeps_{t_i}$ (c.f. Eqs.~(\ref{eqn:rev-disc}, \ref{eqn:noise-and-score}));
\EndFor
\State \textbf{Return} $\bfR_0 + \bfRb^\star$.
\end{algorithmic}
\end{algorithm}

As catalysts are periodic systems along the $x$ and $y$ directions, DiG expands the unit cell in these dimensions before feeding a system into the model. Providing an exact descriptor of the infinitely repeated system in the model is non-trivial; instead, DiG adopts a simple yet effective approach by establishing a local cutoff for the infinitely repeating system. Specifically, an atom outside the unit cell is included in the model only if its distance to any atom inside the unit cell is within a threshold. A distance threshold of $6\,\AA$ is used in the experiments. In each layer of the transformer model, the representation of a repeated atom outside the unit cell is enforced to be identical to the representation of the corresponding atom in the unit cell. More details about handling periodic boundary conditions can be found in Supplementary Sec.~\ref{sec:pbc}.

Lastly, to ensure a good initialization for the diffusion model, a model pre-trained on the IS2RS task of OC20 is used to initialize the weights, except for the time step embedding, which is dedicated to the diffusion task.

\subsection{Property-Guided Structure Generation}
\label{sec:inverse-design-diffusion-details}
\newcommand{\bfLb}{\bar{\bfL}}
\newcommand{\bfXb}{\bar{\bfX}}
\newcommand{\bfccb}{\bar{\bfcc}}
\newcommand{\bfllb}{\bar{\bfll}}
\newcommand{\bfrrb}{\bar{\bfrr}}
\newcommand{\bfzzb}{\bar{\bfzz}}

For modeling carbon polymorphs, the structural variable needs to represent the unit cell which defines a spatial period in the crystal, and the absolute coordinates $\bfXb$ of the carbon atoms in the unit cell.
The unit cell is a parallelepiped to guarantee periodicity, so it can be determined by the coordinates of 4 non-coplanar vertices, say $\bfccb_0$, $\bfccb_0 + \bfllb_x$, $\bfccb_0 + \bfllb_y$, and $\bfccb_0 + \bfllb_z$, where $\bfccb_0$ is the origin of the unit cell, and $\bfLb := \{\bfllb_x, \bfllb_y, \bfll_z\}$ are known as the lattice vectors. (The locations of the other 4 vertices can be determined as $\bfccb_0 + \bfllb_x + \bfllb_y$, $\bfccb_0 + \bfllb_x + \bfllb_z$, $\bfccb_0 + \bfllb_y + \bfllb_z$, and $\bfccb_0 + \bfllb_x + \bfllb_y + \bfllb_z$.) In the structure representation, the origin $\bfccb_0$ is fixed as $\bfzro$, so the unit cell is fully determined by the lattice vectors $\bfLb$.

Similar to the cases of protein-ligand sampling in Supplementary Sec.~\ref{sec:protein-ligand-sampling-details} and catalyst-adsorbate sampling in Supplementary Sec.~\ref{sec:catalyst-adsorbate-sampling-details}, we take the diffusion-process variable $\bfR$ as relative coordinates to release the burden of the diffusion model to also predict the distribution center for different systems.
For the lattice vectors, we introduce a reference lattice vector set $\bfLb^\star$ taken as the mean lattice vector set on the dataset, and diffuse the relative vectors $\bfL := \bfLb - \bfLb^\star$.
For the carbon-atom coordinates, we use their relative coordinates w.r.t the unit cell center $\bfzzb := \frac12 (\bfllb_x + \bfllb_y + \bfllb_z)$, which means $\bfX := \bfXb - \bfzzb$. The diffusion-process variable is then defined as $\bfR := (\bfX, \bfL)$.
The diffusion-variable part of the input to the diffusion model still requires the corresponding absolute coordinates $(\bfX, \bfL)$ of $\bfR$. For the descriptor part of the input, in addition to the number of carbon atoms $N_\atom$, we also include the reference lattice vectors $\bfLb^\star$, which is an informative feature similar to the discussion in Supplementary Sec.~\ref{sec:catalyst-adsorbate-sampling-details}.

The DiG model first learns the (unconditional) distribution of the structures of carbon polymorphs from a dataset created by random structure search (RSS). We take the noise-prediction form to learn the model (c.f. \eqnref{noise-and-score}). The training process is detailed in Alg.~\ref{alg:inverse_design_training}.
The DiG model is then asked to generate structure samples conditioned on a given property value, specifically a desired band gap value $c$ in our case. According to \eqnref{conditional-score}, this requires a property predictor/classifier. For this, we use a GNN model M3GNet~\cite{chen2022universal}, which provides the prediction for the band gap of a given structure in absolute coordinates. To evaluate a probability, we convert the regression task into a classification task by discretizing an inclusive range of the band gap value into $K$ intervals of length $1.0$, represented by $\clI_0 = [a_0, b_0], \cdots, \clI_{K-1} = [a_{K - 1}, b_{K - 1}]$. The property $c$ is then taken as the bin index. Using the predicted band gap value from M3GNet, we define the probability of a given $c$ as:
\begin{align}
    q_\clD(c \mid \bfXb, \bfLb) := \frac{ \exp\lrparen{ -\lrvert{ \mathrm{M3GNet}(\bfXb, \bfLb) - \frac{a_c + b_c}{2} }} }{ \sum_{k=0}^{K-1} \exp\lrparen{ -\lrvert{ \mathrm{M3GNet}(\bfXb, \bfLb) - \frac{a_k + b_k}{2} }} }.
    \label{eqn:band-gap-predictor}
\end{align}
This equation is used to construct the required conditional score $\bfss^\theta_{\clD,t} (\bfXb_t, \bfLb_t \mid c)$ following \eqnref{conditional-score}. The value $\frac{a_c + b_c}{2}$ is the target band gap for the interval $\mathcal{I}_c$.

\begin{algorithm}
\caption{Carbon structure score model training (single step)}\label{alg:inverse_design_training}
\begin{algorithmic}[1]
\Require Noise-predicting model $\bfeps^\theta_{\clD,t}(\bfXb,\bfLb)$ to be trained, descriptor $\clD = (N_\atom, \bfLb^\star)$, where $N_\atom$ is the number of carbon atoms, and $\bfLb^\star = \{\bfllb^\star_x, \bfllb^\star_y, \bfllb^\star_z\}$ is the reference lattice vectors;
randomly sampled time step $t \in [0,\tau]$, randomly sampled structure $(\bfXb_{\clD,0}^{(n)}, \bfLb_{\clD,0}^{(n)})$ from the dataset, where $\bfXb_{\clD,0}^{(n)}$ comprises coordinates of carbon atoms, and $\bfLb_{\clD,0}^{(n)} = \{\bfllb_{\clD,0,x}^{(n)}, \bfllb_{\clD,0,y}^{(n)}, \bfllb_{\clD,0,z}^{(n)}\}$ is the collection of lattice vectors.
\State Calculate the unit cell center $\bfzzb_{\clD,0}^{(n)} := \frac12 (\bfllb_{\clD,0,x}^{(n)} + \bfllb_{\clD,0,y}^{(n)} + \bfllb_{\clD,0,z}^{(n)})$;
\State Let $\bfX_{\clD,0}^{(n)} := \bfXb_{\clD,0}^{(n)} - \bfzzb_{\clD,0}^{(n)}$, $\bfL_{\clD,0}^{(n)} := \bfLb_{\clD,0}^{(n)} - \bfLb^\star$;
\State Sample the noise variable $\bfeps^{(\bfX)}_t \sim \clN(\bfzro, \bfI_{3 N_\atom})$ and $\bfeps^{(\bfL)}_t \sim \clN(\bfzro, \bfI_9)$;
\State Let $\bfX_{\clD,t}^{(n)} := \alpha_t \bfX_{\clD,0}^{(n)} + \sqrt{1-\alpha_t^2} \bfeps^{(\bfX)}_t$,
and $\bfL_{\clD,t}^{(n)} := \alpha_t \bfL_{\clD,0}^{(n)} + \sqrt{1-\alpha_t^2} \bfeps^{(\bfL)}_t$ (c.f. \eqnref{q_il0});
\State Let $\bfLb_{\clD,t}^{(n)} := \bfL_{\clD,t}^{(n)} + \bfLb^\star$ which is structured as $\{\bfllb_{\clD,t,x}^{(n)}, \bfllb_{\clD,t,y}^{(n)}, \bfllb_{\clD,t,z}^{(n)}\}$;
\State Let $\bfzzb_{\clD,t}^{(n)} := \frac12 (\bfllb_{\clD,t,x}^{(n)} + \bfllb_{\clD,t,y}^{(n)} + \bfllb_{\clD,t,z}^{(n)})$ and $\bfXb_{\clD,t}^{(n)} := \bfX_{\clD,t}^{(n)} + \bfzzb_{\clD,t}^{(n)}$;
\State Evaluate the loss $\lrVert{\bfeps^\theta_{\clD,t}(\bfXb_{\clD,t}^{(n)}, \bfLb_{\clD,t}^{(n)}) - (\bfeps^{(\bfX)}_t, \bfeps^{(\bfL)}_t)}^2$ (c.f. \eqnref{noiseloss});
\State Update the model parameter $\theta$ by performing an optimization step on the loss with respect to $\theta$.
\end{algorithmic}
\end{algorithm}

\begin{algorithm}
\caption{Sampling for carbon structure inverse design}
\label{alg:inverse_design_sampling}
\begin{algorithmic}[1]
\Require A trained noise-predicting model $\bfeps^\theta_{\clD,t}(\bfXb,\bfLb)$ which decomposes as $(\bfeps^{(\bfX),\theta}_{\clD,t}, \bfeps^{(\bfL),\theta}_{\clD,t})$ according to its output channels, descriptor $\clD = (N_\atom, \bfLb^\star)$ of the target system, where $N_\atom$ is the number of carbon atoms, and $\bfLb^\star = \{\bfllb^\star_x, \bfllb^\star_y, \bfllb^\star_z\}$ is the reference lattice vectors;
a trained property classifier $q_\clD(c \mid \bfXb, \bfLb)$, the desired property value $c$, guidance strength $\lambda_\mathrm{guide}$.
\State Sample a noisy initial structure $\bfX_\tau \sim \clN(\bfzro, \bfI_{3N_\atom})$, $\bfL_\tau \sim \clN(\bfzro, \bfI_9)$;
\For {$i$ in $N,\cdots,1$}
\State $t_{i-1} := \frac{i-1}N \tau$;
\State Let $\bfLb_{t_i} := \bfL_{t_i} + \bfLb^\star$ which is structured as $\{\bfllb_{t_i,x}, \bfllb_{t_i,y}, \bfllb_{t_i,z}\}$;
\State Calculate the unit cell center $\bfzzb_{t_i} := \frac12 (\bfllb_{t_i,x} + \bfllb_{t_i,y} + \bfllb_{t_i,z})$;
\State Let $\bfXb_{t_i} := \bfX_{t_i} + \bfzzb_{t_i}$ which is structured as $\{\bfXb_{t_i,a}\}_{a=1}^{N_\atom}$;
\State Let $\beta_i := \frac{\tau}{N} \beta_{t_i}$, $\alpha_i := \prod_{j=1}^i \sqrt{1 - \beta_j}$;
\State $\bfX_{t_{i-1}} := (2 - \sqrt{1-\beta_i}) \bfX_{t_i} - \frac{\beta_i}{2 \sqrt{1 - \alpha_i^2}} \bfeps^{(\bfX),\theta}_{\clD,t_i}(\bfXb_{t_i}, \bfLb_{t_i}) + \lambda_\mathrm{guide} \frac{\beta_i}{2} \nabla_{\bfXb_{t_i}} \log q_\clD(c \mid \bfXb_{t_i}, \bfLb_{t_i})$ (c.f. Eqs.~(\ref{eqn:inverse-design-rev-ode}, \ref{eqn:noise-and-score}, \ref{eqn:conditional-score}));
\State $\bfL_{t_{i-1}} = (2 - \sqrt{1-\beta_i}) \bfL_{t_i} - \frac{\beta_i}{2 \sqrt{1 - \alpha_i^2}} \bfeps^{(\bfL),\theta}_{\clD,t_i}(\bfXb_{t_i}, \bfLb_{t_i}) + \lambda_\mathrm{guide} \frac{\beta_i}{2} \big( \nabla_{\bfLb_{t_i}} \log q_\clD(c \mid \bfXb_{t_i}, \bfLb_{t_i}) + \frac12 \sum_{a=1}^{N_\atom} \nabla_{\bfXb_{t_i,a}} \log q_\clD(c \mid \bfXb_{t_i}, \bfLb_{t_i}) \big)$ (c.f. Eqs.~(\ref{eqn:inverse-design-rev-ode}, \ref{eqn:noise-and-score}, \ref{eqn:conditional-score}));
\EndFor
\State Let $\bfLb_0 := \bfL_0 + \bfLb^\star$ which is structured as $\{\bfllb_{0,x}, \bfllb_{0,y}, \bfllb_{0,z}\}$;
\State Calculate the unit cell center $\bfzzb_0 := \frac12 (\bfllb_{0,x} + \bfllb_{0,y} + \bfllb_{0,z})$;
\State Let $\bfXb_0 := \bfX_0 + \bfzzb_0$;
\State \textbf{Return} $(\bfXb_0, \bfLb_0)$.
\end{algorithmic}
\end{algorithm}

The sampling process starts from a standard-Gaussian sample as $\bfR$, and in each step $\bfR$ is converted to absolute coordinates using $\bfLb^\star$ and the corresponding unit cell center for the input to the model, and finally outputs the structure sample in absolute coordinates. 
Similar to the case of protein structure sampling as explained at the end of Supplementary Sec.~\ref{sec:acarbon-so3-diffusion-sampling-details}, for simulating the sampling process, instead of simulating the SDE in the fashion of \eqnref{rev-disc}, it achieves better results to simulate the equivalent ODE in \eqnref{rev-ode}. This is explained in Supplementary Sec.~\ref{sec:details-density}.
Under discretization, \eqnref{rev-ode} is written as:
\begin{align}
    \bfR_{t_{i-1}}
    = \bfR_{t_i} + \frac{\beta_i}{2} \lrparen*[\Big]{ \bfR_{t_i} + \bfss^\theta_{\clD,t_i}(\bfRb_{t_i} \mid c) },
\end{align}
where $h$ is the discretization step size, $\bfRb_t := (\bfXb_t, \bfLb_t)$ is the absolute coordinates corresponding to $\bfR_t$, and the plus sign is due to the simulation is reversed in time. Also recall $\beta_i := h \beta_{t_i}$.
Since $h$ hence $\beta_i$ is an infinitesimal, the weight for the $\bfR_{t_i}$ term can be formulated as:
$1 + \frac{\beta_i}2 = 2 - (1 - \frac{\beta_i}2) = 2 - \sqrt{1 - \beta_i} + o(h)$, which gives an alternative up to $o(h)$ which is acceptable since the disretization itself has $o(h)$ error. This then recovers the ODE simulation in~\citep{song2021score}. By leveraging \eqnref{conditional-score}, the simulation step becomes:
\begin{align}
    \bfR_{t_{i-1}}
    = (2 - \sqrt{1 - \beta_i}) \bfR_{t_i} + \frac{\beta_i}{2} \lrparen*[\Big]{ \bfss^\theta_{\clD,t_i}(\bfRb_{t_i}) + \nabla_{\bfR_{t_i}} \log q_\clD(c \mid \bfRb_{t_i}) },
\end{align}
where $q_\clD(c \mid \bfRb_{t_i})$ is given by \eqnref{band-gap-predictor}. Finally, by involing \eqnref{noise-and-score}, we can express the generation process using the noise-prediction model $\bfeps^\theta_{\clD,t}$:
\begin{align}
    \bfR_{t_{i-1}} = (2 - \sqrt{1-\beta_i}) \bfR_{t_i} - \frac{\beta_i}{2 \sqrt{1 - \alpha_i^2}} \bfeps^\theta_{\clD,t_i}(\bfRb_{t_i})
    + \frac{\beta_i}{2} \nabla_{\bfR_{t_i}} \log q_\clD(c \mid \bfRb_{t_i}).
   \label{eqn:inverse-design-rev-ode}
\end{align}
Note that the input to the $q_\clD(c \mid \bfRb_{t_i})$ model is absolute coordinates while the gradient is taken w.r.t the relative coordinates $\bfR_{t_i}$. To conduct this conversion, note that the conversion is
$\bfLb = \bfL + \bfLb^\star$, and $\bfXb = \bfX + \bfzzb$ where $\bfzzb := \frac12 (\bfllb_x + \bfllb_y + \bfllb_z)$ with $\bfLb = \{\bfllb_x, \bfllb_y, \bfllb_z\}$. So $\nabla_\bfX \log q_\clD(c \mid \bfXb, \bfLb)
= \nabla_{\bfXb} \log q_\clD(c \mid \bfXb, \bfLb)$,
and $\nabla_\bfL \log q_\clD(c \mid \bfXb, \bfLb)
= \nabla_{\bfLb} \log q_\clD(c \mid \bfXb, \bfLb) + \frac12 \sum_{a=1}^{N_\atom} \nabla_{\bfXb_a} \log q_\clD(c \mid \bfXb, \bfLb)$.
In practice, the strength of the property guidance can be tuned for better performance, as is a common practice in machine learning~\cite{ho2022classifier}. We hence introduce a $\lambda_\mathrm{guide}$ parameter. Alg.~\ref{alg:inverse_design_sampling} details the sampling process.

It is important to note that, in the conditional generation, the classifier can be fully decoupled from the training of the diffusion model. As a result, our model can serve to generate any demanded properties when provided with a corresponding property prediction model. This flexibility allows the approach to be adapted for various applications and desired properties, given an appropriate predictive model.

\subsection{Accelerating Inference}
\label{appx:accel-infer}

The inference procedure of DiG can be viewed as gradually removing the noise from Gaussian random variables to sample clear conformations in an approximated equilibrium distribution. Although it exhibits several orders of magnitude speedup for the cases in Sec.~\ref{sec:result} compared to traditional simulation methods, the inference is still potentially expensive since it generally needs to go over all time steps. Fortunately, the inference time can be further saved with recently developed methods~\cite{songdenoising, baoanalytic, ludpm, lu2022dpm, karraselucidating}. For example, Ref.~\citep{ludpm} shows that the analytic solution of the diffusion ordinary differential equations (sampling of DiG can be alternatively viewed as solving the corresponding diffusion ordinary differential equations) can significantly accelerate the inference, where high-quality samples can be drawn in around 10 steps, resulting in a further 50 to 100 folds speedup.
These recent advances demonstrate the potential for even more efficient inference procedures in the context of DiG and similar models. By combining these methods with the existing framework, it becomes increasingly feasible to generate desired structures with specific properties in a faster and more efficient manner.

\subsection{Evaluation methods} \label{appx:eval_method}

\paragraph{Protein Conformation Sampling}

As detailed in Sec.~\ref{sec:result-prot-sampling}, this study aims to demonstrate the applicability of using the DiG method to sample protein distributions. To this end, simulations of two proteins from the SARS-CoV-2 virus are used to approximate their actual distributions in equilibrium states. The molecular dynamics (MD) simulation trajectories of the receptor-binding domain (RBD) of the spike protein and the main protease are extracted from a public dataset\footnote{\url{https://covid.molssi.org/simulations/}}. For RBD, there are 2995 independent MD simulations with 1.8 ms of aggregate simulation time. For main protease, the aggregated simulation time is 2.6 ms from 5688 independent trajectories. Both sets of simulations are performed in the constant pressure and temperature (NPT) conditions at 310 K and 1 atm pressure.

To facilitate the comparison between distributions obtained from MD simulations and DiG generations, time-lagged independent component analysis (TICA) is utilized to project the simulated structures onto a low-dimensional manifold~\citep{perez2013identification, schwantes2013improvements}. This projection enables the visualization of the conformational distribution in the low-dimension space, such as the one spanned by the two TICA coordinates (Fig.~\ref{fig:2}a). The TICA projection analysis is carried out as the following: first, the backbone conformation is featurized with the $\textit{cosine}$ and $\textit{sine}$ values of backbone torsion angles; then standard TICA projection is executed using PyEmma~\citep{scherer_pyemma_2015}, with lag times of 10 ns and 2 ns for RBD and main protease respectively. We first parametrized the TICA transformation matrix to MD simulation structures to obtain the probability distribution as references, then the same transformation is applied to structures generated by DiG. From MD simulation trajectories, about 1.8 million structures of each system are used for TICA analysis. Furthermore, to reduce the influence of disordered regions at protein termini, the terminal residues are excluded during this TICA analysis and projection to the low-dimensional conformational space. For the distribution comparison in the reduced 2D space, we focus on the populated regions, which correspond to metastable states. The regions with very low probability in the distribution map are not included for detailed comparison, in order to focus on the functional relevant conformations.

For a further comparative analysis between DiG and atomistic MD simulation, representative structures for both proteins are obtained from MD simulations by clustering analysis.
Initially, cluster centroid coordinates of all dominant meta-stable clusters in the 2D TICA space (Fig.~\ref{fig:2}a) are estimated for each protein. Following this, 1000 MD simulated structures near cluster centroids are sampled for each structure cluster. The first 16 TICA components of these structures are used as features to cluster the simulation structures into 4 sub-clusters using the K-Means algorithm. Finally, the centroid structure of the largest sub-cluster was extracted by MDTraj~\citep{McGibbon2015MDTraj} and taken as the representative structure of the respective meta-stable state.

In order to assess the degree of conformity of the distributions, we compute quantitative metrics on the 2D TICA space. Particularly, we devise a $G \times G$ grid to uniformly cover the populated regions of the 2D TICA space, wherein the grid is labelled as positive if there exists at least one simulation structure within the corresponding TICA range. This grid is referred to as the ``groundtruth'' grid. With structures sampled by DiG or other sampling methods, we can construct a similar ``sampled'' grid following the same procedure.

In the coverage analysis in Supplementary Sec.~\ref{appx:additional-results}, we employ four distinct types of metrics, namely:

\begin{align}
    \mathrm{Accuracy} &= \frac{\mathrm{TP} + \mathrm{TN}}{\mathrm{TP} + \mathrm{FP} + \mathrm{TN} + \mathrm{FN}}, \\
    \mathrm{Precision} &= \frac{\mathrm{TP}}{\mathrm{TP} + \mathrm{FP}}, \\
    \mathrm{Recall} &= \frac{\mathrm{TP}}{\mathrm{TP} + \mathrm{FN}}, \\
    \mathrm{F1} &= \frac{2 \times \mathrm{Precision} \times \mathrm{Recall}}{\mathrm{Precision} + \mathrm{Recall}}.
\end{align}

In the above equations, the grid labelled as the ``groundtruth'' is considered to be true, whereas the one labelled as ``sampled'' is considered to be predicted. The values TP, FP, TN, and FN represent the values of true positives, false positives, true negatives, and false negatives, respectively.

To quantitatively assess the similarity of structures to their reference conformations, we employ two metrics: the template modeling score (TM-score)~\citep{zhang2004scoring} and the root mean square deviation (RMSD). TM-score is a normalized measure of structural similarity between two  conformations, with a score of 1 indicating a perfect match; and RMSD calculates the average distance between the paired atoms of two optimally superimposed structures. In our evaluations, we restrict our RMSD calculations to the alpha carbon atoms in protein structure comparison, and for all non-hydrogen atoms for ligand structure comparison. These metrics provide a quantitative means of gauging the accuracy of our protein conformation samples relative to their experimental counterparts. 

The quality of DiG-generated structures was assessed using the TM-scores, by comparing each generated structure against crystal structures (6M0J for RBD and 6LU7 for main protease). For RBD, the mean value of TM-score is 0.84 and all structures have TM-score larger than 0.8, indicating highly similar structures; while in the case of main protease, the TM-score is more spread, with about $94\%$ structures with TM-score $>$ 0.5. We adopt a criterion suggested by~\citep{xu2010significant}, using TM-score = 0.5 as a cutoff to remove the structures that are dissimilar to the experimental model. For the downstream analysis, such as structure distributions, only structures with TM-score $>$ 0.5 are used to reduce noises from incorrect predictions.

\paragraph{Catalyst-Adsorbate Sampling}

After training the model to find different adsorption configurations, we traverse the initial positions of the adsorbate by shifting the original initial structure of adsorbate in the dataset along the $x$ and $y$ vectors of the unit cell. Specifically, we equally divide the unit cell in $x$ and $y$ dimensions into a grid of $15 \times 15$ points. Without changing the conformation and height in the $z$ dimension of the initial structure of the adsorbate, we shift its coordinates in the $x$ and $y$ dimensions to match these $15 \times 15$ points, thus creating $15 \times 15$ different initial structures of the system.
For each initial position, we use DiG to sample $10$ structures. The sampled structures are then verified by DFT relaxation with VASP~\cite{hafner2008ab}. In VASP, we allow both the catalyst surface and the adsorbate to move, which is consistent with our model and the dataset. The structures generated by our model are close to the relaxed structures, as described in Sec.~\ref{sec:result}. For some initial positions, we are able to find multiple adsorption sites within the $10$ sampled structures. Fig.~\ref{fig:catalyst_adsorbate} shows such a case, where two of the adsorption configurations are sampled from the same initial structure.
Note that the training dataset contains only very short MD trajectories. The capability of our model is also limited by the dataset. With longer MD trajectories that traverse more structures, our model should be able to find more adsorption sites from an arbitrary initial structure. More discussions can be found in Supplementary Sec.~\ref{appx:limitation}.

The probability density map of DiG is generated with single-atom adsorbates. To plot the map of probability density, we equally divide the unit cell along $x$ and $y$ dimensions as mentioned previously, resulting in a $20 \times 20$ grid. We place the adsorbate above each grid point, creating $20\times 20$ structures for each fixed height of the adsorbate atom. We traverse 10 different heights for the adsorbate atom, ranging from $0\,\AA$ to $1\,\AA$ from the surface of the catalyst. Atoms in the catalyst surface are kept as the initial positions when density evaluation. In total, we obtain $4000$ structures.
We calculate the log-likelihood of our model on these structures, given an initial structure where the adsorbate atom is $2\,\AA$ from the catalyst surface above the center of the catalyst surface in the unit cell. The equation below summarizes the calculation of the probability density map:
\begin{align}
    p(x_i, y_j \mid x_0, y_0, z_0) = \max_{k}{p_{\mathrm{model}} (x_i, y_j, z_k \mid x_0, y_0, z_0)},
\end{align}
where $i,j,k\in \{1,\cdots,20\}\times\{1,\cdots,20\}\times\{1,\cdots,10\}$, $z_0=2\,\AA$ is distance from the initial position of the atom to the highest point of the catalyst surface, and $(x_0, y_0)$ is the center of the catalyst surface in the unit cell. Finally, in the probability density map, we plot the log values of the probabilities above.
In the energy map from VASP, for each $i,\, j$ we plot the negative of the energy of the relaxed structure, with the $x_i, y_j$, and the catalyst surface fixed. Starting from an initial height $z_k=2\,\AA$ from the catalyst surface, we used DFT to relax along the $z$ dimension to obtain the minimum energy upon the grid point $(x_i,y_i)$. 

\paragraph{Property-Guided Structure Generation}
To measure the ability of modeling the carbon polymorph structures, we use the \texttt{StructureMatcher} in the Pymatgen~\cite{ong2013python} package to calculate the ratio that a sampled structure matches a structure in the training dataset. We use hyperparameters \texttt{stol=0.5}, \texttt{angle\_tol=10} and \texttt{ltol=0.3} for the \texttt{StructureMatcher}, where \texttt{stol} is the tolerance for the displacement of atom positions, \texttt{angle\_tol} controls the difference in lattice vector angles between the matched structures, and \texttt{ltol} is the tolerance for the difference in matched lengths of lattice vectors. The \texttt{get\_rms\_dist} method is used to calculate the RMSD, which is normalized by the average free length per atom.

\section{Model Details} \label{sec:model-details}
\subsection{Descriptors of Molecular Systems}

We consider four types of molecular systems in previous sections: proteins, protein-ligand, catalyst-adsorbate, and carbon polymorphs. A descriptor $\clD$ is used for each system that captures the relevant features of the molecular structure and can be processed by DiG. The descriptor $\clD$ in the four systems is first processed into node representations $\clV$ describing the feature of each system-specific individual element, and a pair representation $\clP$ describing inter-node features.
Note that in some systems, the descriptor $\clD$ also contains structural features, which are treated as part of the node representation $\clV$.
The $\{\clV, \clP\}$ representation is the direct input from the descriptor part to the Graphormer model, as illustrated in Fig.~\ref{fig:1}.

For protein systems, we follow AlphaFold~\citep{jumper2021highly} and adopt a coarse-grained representation that uses the position of the alpha-carbon atom and the orientation of each residue (see Supplementary Sec.~\ref{sec:acarbon-so3-diffusion-sampling-details}). The node representation $\clV$ is a sequence of feature vectors that are generated by the Evoformer module in~\citep{jumper2021highly}, which takes as input the amino acid sequence and the multiple-sequence alignment (MSA) of the protein. The pair representation $\clP$ is composed of two matrices: the first represents all pairwise interactions between residues, which is also produced by Evoformer, and the second represents the lengths of all residue pairs on the amino acid sequence. The node and pair representations are learnable embeddings in~\citep{jumper2021highly}, but we hold them fixed in this work to avoid additional computational cost from fine-tuning the Evoformer. See more discussions about the limitation of fixing parameters of Evoformer can be found in Supplementary Sec.~\ref{appx:limitation}.

For protein-ligand binding systems, we use an all-atom representation that includes atoms from both the protein and the ligand. The features are obtained following the method in~\citep{min2022predicting}. To be specific, the node representation $\clV$ for the protein part consists of the types of atoms around the binding pocket and also the positions of these atoms $\bfRb_\Rec^\star$ in the crystal structure of the protein, and the ligand part consists of a graph of the skeletal formula of the ligand. The pair representation $\clP$ is also composed of two parts: one for the intra-molecular bonds and one for the inter-molecular interactions. The intra-molecular bonds are represented by feature embeddings of the chemical bonds between the atoms of the protein or the ligand, and the inter-molecular interactions are represented by feature embeddings of interactions like the hydrogen bonding between the protein and the ligand, as defined in~\citep{min2022predicting}.

For catalyst systems, we use an all-atomic representation that includes both the catalyst surface and the adsorbates. The node representation $\clV$ consists of the types of atoms of the catalyst and adsorbed molecules as well as their positions $\bfRb_\Cat^\star, \bfRb_\Ad^\star$ in the initial structure from the OC20 dataset~\citep{chanussot2021open}. The features of the nodes are only embeddings of the atomic type and position, following the method in~\citep{shi2022benchmarking}. The pair representation $\clP$ is only defined for the adsorbates, which consists of feature embeddings of the chemical bonds between the atoms of the adsorbates.

For carbon polymorphs, we use an all-atomic representation that only includes carbon atoms. The node representation $\clV$ consists of the initial embedding of the carbon element. The only difference among the systems is the number of carbon atoms.
There is no pair representation $\clP$ for the carbon polymorphs, as the chemical bonds are not pre-defined and may change during the diffusion process.

\subsection{Backbone Architecture}

The deep learning models used in DiG are extended by our previously proposed Graphormer~\cite{ying2021transformers,shi2022benchmarking}, which is a Transformer-based graph neural network~\cite{vaswani2017attention}, and could efficiently capture the topological information while keep the powerful expressiveness from the Transformer architecture.
The model is composed of a few concatenated so-called attention layers and feed-forward layers.
Each attention layer takes the hidden node representation $\clH$ as the input tokens of the Transformer and uses the pair representation $\clP$ as a learnable attention bias for the attention mechanism.
Formally, given a hidden node representation $\clH=\{\bfhh_1,\cdots,\bfhh_I\}$, where $I$ is the number of nodes, and a pair representation $\clP=\{\bfP_{\imath\jmath}\}_{\imath,\jmath=1}^I$, where $\bfP_{\imath\jmath}$ is the learnable embedding of the edge features from node $\imath$ to node $\jmath$, Graphormer computes the attention score $\bfA_{\imath\jmath}$ from node $\imath$ to node $\jmath$ as:

\begin{align}\label{eqn:graphormer-bias}
    \bfA_{\imath\jmath}=\frac{(\bfhh_\imath \bfW^{(Q)})(\bfhh_\jmath \bfW^{(K)})\trs}{\sqrt{d}} + \bfP_{\imath\jmath}\trs \bfww^{(\clP)},
\end{align}
where $\bfW^{(Q)}, \bfW^{(K)}$ are the ``query'' and ``key'' linear projections for the node representation, $\bfww^{(\clP)}$ is a learnable weight vector for the pair representation, and $d$ is the dimension of the query and key vectors. The hidden node representation $\clH$ in the first layer is $\clV$. The attention bias term $\bfP_{\imath\jmath}\trs \bfww^{(\clP)}$ enables the model to learn the importance of the pair representation for the attention mechanism. The attention score is then normalized by a softmax function over all nodes and used to compute the attention output following the invariant point attention mechanism in AlphaFold~\citep{jumper2021highly} for protein systems, or standard Transformer architecture~\cite{vaswani2017attention} for other molecular systems.

\subsection{Structural Attention Biases}
\label{sec:struct_attn_bias}

Besides the descriptor $\clD$ input of the molecular systems, the Graphormer model also needs to process the geometric structure input $\bfR$ and produce a physically finer structure. To more informatively encode the geometric information during the diffusion process, we also introduce a structural representation for the input $\bfR$, which is used to help Graphormer to capture the spatial and rotational relationships among the nodes and refine the noisy structures to more physically realistic ones.

For all molecular systems in full-atom representation, we follow~\citep{shi2022benchmarking} to encode the Euclidean distance $d_{\imath\jmath}$ between the positions of node $\imath$ and node $\jmath$ as a bias term $b_\phi(d_{\imath\jmath})$, where $\phi$ is a learnable parameter. The distance encoding bias is added to the attention score in \eqnref{graphormer-bias} to modulate the attention based on the distance between nodes.
For protein systems that use the coarse-grained representation $\bfR = (\bfC, \bfQ)$ (see Supplementary Sec.~\ref{sec:acarbon-so3-diffusion-sampling-details}), we adopt the invariant point attention mechanism in~\citep{jumper2021highly} to construct the corresponding attention score, which has been shown to be indispensable for capturing the local rotational invariance feature of the protein structures.

\subsection{Equivariant Graphormer}

\begin{algorithm}[t]
\caption{Equivariant Vector Prediction}\label{alg:equi-graphormer}
\begin{algorithmic}[1]
\Require Node 3D positions $\{\bfR_1, \cdots, \bfR_I\}$,
node representation $\bfH = [\bfhh_1\trs, \cdots, \bfhh_I\trs]\trs \in \bbR^{I \times C}$, attention bias $\bfE^\mathrm{attn} \in \bbR^{I \times I}$;
linear projector matrices $\bfW^{(Q)}, \bfW^{(K)} \in \bbR^{C \times d}$, $\bfW^{(V)} \in \bbR^{C \times C}$, $\bfww^{(F)} \in \bbR^C$;
\State Calculate the relative positions $\bfE^\mathrm{rel} \in \bbR^{I \times I \times 3}: \bfE^\mathrm{rel}_{\imath\jmath} := \bfR_\imath - \bfR_\jmath$;
\State $\bfQ = \bfH \bfW^{(Q)}, \bfK = \bfH \bfW^{(K)}, \bfV = \bfH \bfW^{(V)}$;
\State $\bfA = \frac{\bfQ \bfK\trs}{\sqrt{d}} + \bfE^\mathrm{attn}$;
\State Calculate $\bfF \in \bbR^{I \times 3}:
\bfF_\imath := \sum_{\jmath=1}^I \mathrm{softmax}(\bfA_{\imath:})_\jmath ({\bfww^{(F)}}\trs \bfV_\jmath) \bfE^\mathrm{rel}_{\imath\jmath}$;
\State \textbf{Return} $\bfF$;
\end{algorithmic}
\end{algorithm}

Due to the score model interpretation (gradient of log-density function), the output of the Graphormer model is required to be equivariant w.r.t the $\bfR$ input, which requires a proper design for processing the $\bfR$ input.
To ensure the rotational equivariance of the Graphormer model, we add one equivariant attention layer~\cite{shi2022benchmarking} as the 3D vector output head, which produces geometric vectors that are equivariant to any rotation transformations in 3D Euclidean space on the input, as detailed in Alg.~\ref{alg:equi-graphormer}. Specifically, we first compute the attention matrix $\bfA$ and the transformed invariant features $\bfH = [\bfhh_1\trs, \cdots, \bfhh_I\trs]\trs$ as in the previous layers, and then the vector output is obtained by attentively aggregating the relative position information and the invariant features. Since we only apply scalar multiplication and linear combination operations on the vector features, the resulting vector $\bfF$ is naturally equivariant to the $\SO(3)$ group of rotations. Moreover, $\bfF$ is translation-invariant because it only depends on the relative positions between two nodes. This design strategy for processing vector features is similar to those used in previous works on protein modeling~\cite{jing2020learning} and quantum chemistry~\cite{schutt2021equivariant}.

\subsection{Periodic Boundary Condition}
\label{sec:pbc}
In catalyst-adsorbate systems and carbon polymorphs, atoms in a 3D unit cell are periodically repeated. Therefore, radius graphs with periodic boundary conditions are constructed to represent the systems, where each atom in one single cell (the centric cell) will connect with its neighboring atoms within a pre-defined cutoff distance. Since atoms are periodically repeated, the same atom in different cells may appear repeatedly in one graph as different nodes. To avoid that the same atom has different node representations in the network, typically a multi-graph will be constructed for message-passing neural networks (MPNNs), where one node represents one atom, and multiple edges between nodes represent the interactions with the same atom in different cells. In this way, information on neighboring atoms will be aggregated by MPNNs with multiple times through each edge.

Differently, message aggregation is done by attentively weighted sum on full graph in Graphormer, and interactions between atoms are encoded into spatial distance embeddings acting as attention bias. Multi-graph will lead to a summation of multiple biases in the distance embedding space, which might be projected to a new distance, and would not reflect multiple interactions with the same atom in different cells. Therefore, to reflect the multiple interactions correctly while enforce one representation for the same atom in different cells, we use a cross-attention sub-layer to aggregate information from all atoms in the radius graph into the atoms in the centric cell as shown in Alg.~\ref{alg:cap}.

\begin{algorithm}
\caption{Handling Periodic Boundary Condition}\label{alg:cap}
\begin{algorithmic}[1]
\Require Atom positions in the centric unit cell $\widetilde{\clX} := \left\{ \widetilde{\bfxx}_\imath \right\}_{\imath=1}^I$.
\Require Lattice vectors $\bfll_x, \bfll_y, \bfll_z$. Cutoff distance $D_\text{cut}$.
\Require Atom representation $\left\{ \bfhh(\widetilde{\bfxx_\imath}) \mid \widetilde{\bfxx}_\imath \in \widetilde{\clX} \right\}$ in current layer.
\Ensure Atom representation $\left\{ \bfhh'(\widetilde{\bfxx_\imath}) \mid \widetilde{\bfxx}_\imath \in \widetilde{\clX} \right\}$ after attention.
\State ${\clX} \leftarrow \displaystyle \left\{ \widetilde{\bfxx} + l \bfll_x + m \bfll_y + n \bfll_z \mid l,m,n \in \bbZ, \widetilde{\bfxx} \in \widetilde{\clX} \right\}$;
\State $\clX_D \leftarrow \displaystyle \left\{ \bfxx \in \clX \mid \exists \widetilde{\bfxx}'\in \widetilde{\clX},\ \Vert \bfxx - \widetilde{\bfxx}' \Vert \le D_\text{cut} \right\}$;
\State $\forall \bfxx \in \clX_D, \widetilde{\bfxx}:= \bfxx + l \bfll_x + m \bfll_y + n \bfll_z$, s.t. $\widetilde{\bfxx} \in  \widetilde{\clX},\ l,m,n \in \bbZ$;
\State $\bfA_{\imath\jmath} \leftarrow \frac{\left(\bfhh(\widetilde{\bfxx}_\imath) \bfW^{(Q)} \right)\left(\bfhh_\jmath(\widetilde{\bfxx}_\jmath) \bfW^{(K)} \right)\trs}{\sqrt{d}} + b_\phi\left( \left\Vert \bfxx_\imath - \bfxx_\jmath \right\Vert  \right),\ \forall \bfxx_\imath, \bfxx_\jmath \in \clX_D$;
\State $\bfhh'(\widetilde{\bfxx}_\imath) \leftarrow \sum_\jmath \mathrm{softmax}(\bfA_{\imath:})_\jmath \left(\bfhh(\widetilde{\bfxx_\jmath}) \bfW^{(V)} \right)$.
\end{algorithmic}
\end{algorithm}

The basic idea is that learnable node embeddings are only assigned to atoms in the centric cell, and the embeddings for the same atom appearing in neighboring cells are its replicas. The attention bias encoded from pair-wise atom distances is used to tell replicas of atoms in different cells.  Therefore, the representation of each node in the centric cell will be updated by the correlation and interaction with all atoms in the radius graph.

In modeling the carbon polymorphs, with the lattice vectors, we can expand the unit cell and handle periodic boundary conditions as described in Alg.~\ref{alg:cap}. However, we find that explicitly encoding the vertices in the unit cell as tokens in the Graphormer encoder is very helpful for sampling physical structures. Thus, each vertex of the unit cell is treated as an atom with a special type in the model. In other words, for a structure with $n$ atoms in the unit cell, the model will take $n + 8$ tokens, and expand the tokens according to the PBC handling method. Alg.~\ref{alg:inverse_design_training} and~\ref{alg:inverse_design_sampling} describe the details of the training and sampling processes, respectively.

\section{Training Details} \label{sec:training-details}

\subsection{Protein Conformation Sampling} \label{sec:training-details-protein}

\subsubsection*{\textit{Training Pipeline and Dataset}}

Our training process for the protein system consists of three stages: initialization, physics-informed diffusion pre-training (PIDP), and data-based training using simulation data.
In the first stage which aims to provide a good initialization to stabilize PIDP training, we collect all experimental structures from the Protein Data Bank (PDB)~\citep{wwpdb2019protein} before December 25, 2020 as training data and employ Alg.~\ref{alg:protein_databased_training} for model training.
Such experimental structures are widely used in structure prediction methods, in which case a dataset is organized following the pattern $(\clD, \bfR_{\clD})$, where each amino-acid sequence $\clD$ is paired with one experimental structure $\bfR_{\clD}$.
In contrast, to provide distributional information during the training of DiG, we organize the structures to construct a physical distribution dataset, in which each data point follows the pattern $(\clD, \{\bfR_{\clD}^{(n)}\}_{n=1}^{N_\data})$ where each amino acid sequence $\clD$ is paired with a \emph{set} of structures $\{\bfR_{\clD}^{(n)}\}_{n=1}^{N_\data}$. To prepare this dataset, we adopt all the sequence identity clusters from PDB obtained by MMSeqs2~\cite{steinegger2018clustering}, and include all the available experimental structures for each cluster. Following AlphaFold~\cite{jumper2021highly}, we filter out structures from PDB that have a resolution worse than $9\,\AA$. This eliminates about 0.2\% of structures. For proteins longer than 256 amino acids, we divide them into segments of length no longer than 256 amino acids. In each training step, we randomly draw clusters and structures within each drawn cluster with equal probability. We would like to remark that although this physical structure distribution is hard to verify to obey the equilibrium distribution, it can still provide rich information about the different modes of the equilibrium distribution.

In the second stage, to prepare the relevant structures for evaluating the PIDP loss, we run short MD simulations for about 1000 proteins, for which the details can be found in Supplementary Sec.~\ref{appx:simulation-details}. We randomly pick 100 simulated structures for each protein as the relevant structures. The training process follows Alg.~\ref{alg:protein_pidp_training}.
In the final stage, we use a simulation dataset consisting of the above simulation dataset, and 238 simulation trajectories randomly picked from the GPCRmd dataset~\citep{rodriguez2020gpcrmd}. The GPCRmd dataset contains short simulations of various classes of G protein-coupled receptors (GPCRs). The training process follows Alg.~\ref{alg:protein_databased_training}.

\subsubsection*{\textit{PIDP Training}}

As mentioned in the main text (Sec.~\ref{sec:framework}), the structures $\{\bfR_0^{(m)}\}_{m=1}^M$ for evaluating the PIDP loss are ideally grid points (as in finite-element methods) spanning the structure space, but this is unaffordable since the number of grid points increases exponentially with the dimension of the space, which is typically exceedingly high for molecular systems. But we only need to supervise the model on a low-dimensional manifold of physically relevant structures (with low energy).
Due to the grid point nature, these structures do not have to follow the equilibrium distribution, but only need to demonstrate the manifold. This then enables a wide range of affordable methods to prepare such structures, such as perturbation around experimentally observed structures~\citep{zhang2021prody}, and short MD simulation structures.
In practice, we found protein structures perturbed by~\citep{zhang2021prody} lead to overly large energy gradient which hinders effective optimization and cannot be easily mitigated by e.g. gradient clipping (see Supplementary Sec.~\ref{sec:limitation-energy-function} for more details about the limitation of energy function). For this reason, we adopt structure samples from short MD simulation trajectories for PIDP training. This is much cheaper than generating structures following equilibrium distribution by long enough MD simulations.
The short MD simulations provide structures that can be seen in a physical process thus demonstrating the relevant manifold, and the information of equilibrium distribution is provided by the energy function.

For evaluating the energy function (or its gradient, i.e., force field), we use OpenMM to compute the full-atom force using the Amber force field as $-\nabla E(\bfRb)$, which also serves for calculating the coarse-grained forces $-\nabla_\bfC E$ and $-\nabla_\bfqq E$ through Alg.~\ref{alg:protein_pidp_full_to_cg_ff}. We use PDBFixer~\cite{eastman2013openmm} to fix the input protein structure files before processed by OpenMM to avoid potential failures.

We find that the range of magnitude of the calculated forces varies drastically, which poses a significant challenge in optimizing the PIDP loss in \eqnref{pinnloss}.
We hence retain only the structures that have a force magnitude within the smallest three orders of magnitude.
To further address the optimization challenge, we also modify the loss terms for matching the score model at $t=0$ to the force field in \eqnsref{pinnloss-protein-so3,pinnloss-protein-acarbon} by only matching their directions.

We also find that for loss terms for $t>0$ (or $i>0$) in the PIDP losses \eqnsref{pinnloss-protein-so3,pinnloss-protein-acarbon}, different sampling time steps $t$ (or $i$) result in significant differences in the scale of the loss, ranging from 0 to $1\e{6}$. For stable and effective training, we rescale and clip these losses.
Specifically, with $\ell^{(\bfC)}_t$ and $\ell^{(\bfQ)}_t$ denoting the PIDP loss terms at time step $t$ of alpha-carbon coordinates and residue orientations, we find $\ell^{(\bfC)}_t$ and $\ell^{(\bfQ)}_t$ increase exponentially with $t$, with exponents $\rho^{(\bfC)}$ and $\rho^{(\bfQ)}$. To avoid an excessively large loss, we scale these losses by:
$\tilde{\ell}^{(\bfC)}_t := \ell^{(\bfC)}_t / (\rho^{(\bfC)})^{1-\mathrm{clip}(t)}$, and
$\tilde{\ell}^{(\bfQ)}_t := \ell^{(\bfQ)}_t / (\rho^{(\bfQ)})^{1-\mathrm{clip}(t)}$,
where $\mathrm{clip}(t) := \min(0.05\tau, t)$.

Furthermore, we find that a balanced combination of the losses in \eqnref{dsmloss} and \eqnref{pinnloss} is essential for generating more physical structures after PIDP training.

These training methods, implemented to ensure more stable optimization, may compromise the accuracy of the distribution learned by the model. These limitations will be discussed further in Supplementary Sec.~\ref{appx:limitation}.

\subsubsection*{\textit{Hyperparameter Choices}}
In protein training, we discretize the diffusion-process time variable $t \in [0, \tau]$ into $i \in \{0,1,\cdots,N\}$ discrete time steps.
The noise scales $\sigma_i$ in \eqnref{fwd-cont-so3} and \eqnref{fwd-cont-acarbon} (or in Algs.~\ref{alg:protein_pidp_training} and~\ref{alg:protein_databased_training} for training, and Alg.~\ref{alg:protein_sampling} for sampling) at the corresponding discretized time step $t_i$ are taken in the form $\sigma_i = (\sigma_{\min})^{(1 - t_i/\tau)}(\sigma_{\max})^{t_i/\tau}$. The parameters $\sigma^{(\bfC)}_{\min}$ and $\sigma^{(\bfC)}_{\max}$ for the diffusion process on alpha-carbon coordinates and $\sigma^{(\bfQ)}_{\min}$ and $\sigma^{(\bfQ)}_{\max}$ for the residue orientation are detailed in Supplementary Tab.~\ref{tab:protein_hps}, together with all other hyperparameters.
To strike a balance between computational efficiency and performance, we train for at least one epoch at each stage and halt training when the rate of loss reduction noticeably decelerates.

\begin{table}[]
    \centering
    \caption{Hyperparameters of protein model. Different hyperparameters are used in different stages.}
    \begin{tabular}{c|c|c|c}
        \toprule
         Hyperparameter & Initialization & PIDP & Data Training \\
         \midrule
         Model depth & \multicolumn{3}{c}{12} \\
         Hidden dim (Single) & \multicolumn{3}{c}{768}\\
         Hidden dim (Pair) & \multicolumn{3}{c}{256} \\
         Hidden dim (Feed Forward) & \multicolumn{3}{c}{1024} \\
         Number of Heads & \multicolumn{3}{c}{32} \\
         Optimizer & \multicolumn{3}{c}{Adam} \\
         Learning rate schedule & \multicolumn{3}{c}{Inverse Square Root}\\
         Peak Learning rate & 1E-03 & 1E-05 & 1E-05 \\
         Dropout $p$ & 0.1 & 0.0 & 0.1 \\
         Adam $(\beta_1, \beta_1)$ & (0.9, 0.999) & (0.9, 0.999) & (0.9, 0.999) \\
         Adam $\epsilon$ & 1E-06& 1E-06& 1E-06 \\
         Weight decay & 1E-02& 1E-02& 1E-02 \\
         Warmup ratio & 0.06& 0.06& 0.06 \\
         Batch size & 128& 32& 128 \\
         Diffusion steps $N$ & 500& 500& 500 \\
         $(\sigma^{(\bfC)}_{\min}, \sigma^{(\bfC)}_{\max})$ & (0.1, 35) & (0.1, 35) & (0.1, 35) \\
         $[\sigma^{(\bfQ)}_{\min}, \sigma^{(\bfQ)}_{\max}]$ & (0.02, 1.65) & (0.02, 1.65) & (0.02, 1.65) \\
         Boundary weight $\lambda_1$ & - & 5 & - \\
         Hutchinson $N_\mathrm{est}$ & - & 20 & - \\
         \bottomrule
    \end{tabular}
    \label{tab:protein_hps}
\end{table}

\subsection{Ligand Structure Sampling around Binding Sites}

Two-stage training was performed in ligand sampling. The first stage employs CrossDocked~\citep{francoeur2020three} as a binding structure prediction task. The CrossDocked dataset contains docked conformations of varying quality, so we filter out all complexes whose RMSD between the docked and the experimental crystal structures is larger than $2.5\,\AA$. The second stage employs the simulation data (Sec. ~\ref{appx:md-prot-complex}), where we set the threshold as $6\,\AA$, and 1 ns as the sampling stride. For data-based training, we collect data from CrossDocked~\citep{francoeur2020three} and MD simulations to explore the most concerned part in the conformational space. The CrossDocked dataset contains docked conformations of varying quality, so we filter out all complexes whose RMSD between the docked and the experimental crystal structures is larger than $2.5\,\AA$. For MD simulation data, we set the threshold as $6\,\AA$, and 1 ns as the sampling stride. We conduct a quality screening on the simulation data, by filtering out trajectories that there are ligand's atoms around the protein laying within $5\,\AA$. After filtering, the MD simulation dataset contains 1157 protein-ligand complex trajectories, and we split 80\% for training, 10\% for evaluation, and 10\% for testing. All hyperparameters are listed in Supplementary Tab.~\ref{tab:ligand_hps}.

\begin{table}[]
    \centering
    \caption{Hyperparameters of protein-ligand binding model used in CrossDocked datasset and MD dataset training.}
    \begin{tabular}{c|c|c}
        \toprule
         Hyperparameter & \multicolumn{2}{c}{CrossDocked \& MD Training} \\
         \midrule
         Model depth & \multicolumn{2}{c}{12} \\
         Hidden dim (Model) & \multicolumn{2}{c}{768}\\
         Hidden dim (Feed Forward) & \multicolumn{2}{c}{768} \\
         Number of Heads & \multicolumn{2}{c}{32} \\
         Optimizer & \multicolumn{2}{c}{Adam} \\
         Peak learning rate & \multicolumn{2}{c}{0.0002}\\
         Warmup ratio & \multicolumn{2}{c}{0.06}\\
         Learning rate schedule & \multicolumn{2}{c}{Linear Decay}\\
         Dropout $p$ & \multicolumn{2}{c}{0.1} \\
         Adam $(\beta_1, \beta_2)$ & \multicolumn{2}{c}{(0.9, 0.98)} \\
         Adam $\epsilon$ & \multicolumn{2}{c}{1E-08} \\
         Weight decay & \multicolumn{2}{c}{0.0} \\
         Batch size & \multicolumn{2}{c}{64} \\
         Diffusion steps $N$ & \multicolumn{2}{c}{500} \\
         Diffusion $\beta$ schedule & \multicolumn{2}{c}{Sigmoid}\\
         Diffusion $\beta$ start & \multicolumn{2}{c}{1E-07} \\
         Diffusion $\beta$ end & \multicolumn{2}{c}{0.02} \\
         EMA decay & \multicolumn{2}{c}{0.9999} \\
         EMA fp32 & \multicolumn{2}{c}{true} \\
         Clip norm & \multicolumn{2}{c}{10.0} \\
         \bottomrule
    \end{tabular}
    \label{tab:ligand_hps}
\end{table}

\subsection{Catalyst-Adsorbate Sampling}
Our model captures the distribution of the structure of an adsorbate on a catalyst surface. DiG predicts the distribution conditioned on an initial structure, which is the first frame in a relaxation trajectory provided in the OC20 dataset~\cite{chanussot2021open}. To train DiG, we use the MD part of the OC20 dataset following Alg.~\ref{alg:catalyst_adsorbate_training}. 20,000 systems of the MD dataset are separated for validation. Before this data-based training, the Graphormer model is first pretrained on the IS2RS task of OC20. Detailed hyperparameters for the IS2RS pretraining are listed in Supplementary Tab.~\ref{tab:is2rs_md_hps}. In the data-based training, we use a peak learning rate of $2\e{-4}$, maximum number of epochs 300, warm-up ratio $6\%$, a batch size of 64, and the number of diffusion steps $N = 5000$. The beta schedule follows a sigmoid form:
\begin{equation}
    \beta_i = \frac{1}{1+\exp{(12(0.5 - i / N))}} (\beta_{\mathrm{end}} - \beta_{\mathrm{start}}) + \beta_{\mathrm{start}},
\end{equation}
with $\beta_{\mathrm{start}} = 1\e{-7}$ and $\beta_{\mathrm{end}} = 2\e{-3}$ where $i \in \{0,\cdots,N\}$ is the diffusion time step. The training is stopped after 86 epochs. We use a cutoff value of $6\,\AA$ for PBC handling. Supplementary Tab.~\ref{tab:is2rs_md_hps} summarizes detailed hyperparameters for training of DiG for catalyst-adsorbate sampling. 

\begin{table}[thb]
    \centering
    \caption{Hyperparameters of backbone model for pretraining on Open Catalyst IS2RS and training on Open Catalyst MD dataset.}
    \begin{tabular}{c|c|c}
        \toprule
         Hyper Parameter & IS2RS pretraining & MD training \\
         \midrule
         Model depth & \multicolumn{2}{c}{12} \\
         Hidden dim (Model) & \multicolumn{2}{c}{768}\\
         Hidden dim (Feed Forward) & \multicolumn{2}{c}{768} \\
         Number of Heads & \multicolumn{2}{c}{32} \\
         Optimizer & \multicolumn{2}{c}{Adam} \\
         Learning rate & \multicolumn{2}{c}{0.0002}\\
         Warmup ratio & \multicolumn{2}{c}{0.06}\\
         Learning rate schedule & \multicolumn{2}{c}{Linear Decay}\\
         Dropout $p$ & \multicolumn{2}{c}{0.1} \\
         Adam $(\beta_1, \beta_2)$ & \multicolumn{2}{c}{(0.9, 0.98)} \\
         Adam $\epsilon$ & \multicolumn{2}{c}{1E-08} \\
         Weight decay & \multicolumn{2}{c}{0.0} \\
         PBC Cutoff & \multicolumn{2}{c}{6.0} \\
         Batch size & 1024 & 64 \\
         Diffusion steps $N$ & - & {5000} \\
         Diffusion $\beta$ schedule &  - & {Sigmoid}\\
         Diffusion $\beta$ start &  - & {1E-07} \\
         Diffusion $\beta$ end &  - & {0.002} \\
         \bottomrule
    \end{tabular}
    \label{tab:is2rs_md_hps}
\end{table}

\subsection{Property-Guided Structure Generation}
For training the DiG on an unconditional distribution, we use $15,697$ structures of carbon polymorphs generated from ab initio random structural search (RSS) at the DFT level (PBE/plane-wave basis, with an energy cutoff of $520$ eV) with a range of number of atoms from 2 to 24, following the method in~\cite{lu2021computational}. Only the relaxed structures, i.e., the final frames in relaxation processes, are taken for training.
We remark that for the inverse design task, the distribution of the structures at local energy minima does not follow an equilibrium distribution. However, the primary goal here is to generate structure candidates with reasonable stability and targeted properties. In this context, the relaxed structures from random structure search can well represent the low energy structure manifold.
The number of training epochs is $50,000$, with a peak learning rate $2\e{-4}$, batch size $4096$, and a warm-up ratio $6\%$.
For the reference lattice vector set $\bfLb^\star$, we use the mean lattice vector set over the dataset, which is close to three orthogonal vectors with a length of $4\,\AA$. 
As the initial structure in the catalyst-adsorbate model, this reference lattice vector structure is also encoded into the model with an additional attention bias term. Tab.~\ref{tab:inv_hps} summarizes the hyperparameters for training the diffusion model for property-guided structure generation. We use a cutoff value of $20\,\AA$ for PBC handling.

\begin{table}[thb]
    \centering
    \caption{Hyperparameters of diffusion model training for property-guided structure sampling.}
    \begin{tabular}{c|c}
        \toprule
         Hyperparameter & Data Training \\
         \midrule
         Model depth & 12 \\
         Hidden dim (Model) & 768\\
         Hidden dim (Feed Forward) & 768 \\
         Number of Heads & 32 \\
         Optimizer & Adam \\
         Learning rate & 0.0002\\
         Warmup ratio & 0.06\\
         Learning rate schedule & Linear Decay\\
         Dropout $p$ & 0.1 \\
         Adam $(\beta_1, \beta_2)$ & (0.9, 0.98) \\
         Adam $\epsilon$ & 1E-08 \\
         Weight decay & 0.0 \\
         PBC Cutoff & 20.0 \\
         Batch size & 1024 \\
         Diffusion steps $N$ & {500} \\
         Diffusion $\beta$ schedule &  {Sigmoid}\\
         Diffusion $\beta$ start &  {1E-07} \\
         Diffusion $\beta$ end &  {0.02} \\
         \bottomrule
    \end{tabular}
    \label{tab:inv_hps}
\end{table}

\section{Molecular Simulation and Energy Evaluations} \label{appx:simulation-details}

\subsection{Molecular Dynamics Simulation for Protein-Ligand Complexes}
\label{appx:md-prot-complex}

We generate MD simulation data for complex systems selected from the PDBbind v2020~\cite{su2018comparative} using an automatic pipeline called protocolGromacs\footnote{\url{https://github.com/tubiana/protocolGromacs}}. It utilizes GROMACS~\cite{van2005gromacs} as the backend engine with a common simulation setting for all complexes, providing a capability of high-throughput MD simulations. Specifically, this pipeline comprises four stages: preparation, minimization, equilibration, and production simulations. In the system preparation stage, a protein topology is generated with pdb2gmx with the amber99sb-ildn~\cite{lindorff2010improved} force field with the tip3p explicit water model; the ligand parameter and topology are generated with acpype~\cite{sousa2012acpype}. For cases with missing atoms/residues in the PDB files, PDBFixer~\cite{eastman2013openmm} is applied to complete the molecules. Then, a cubic simulation box is used with a minimum distance of 1.2 nm between the protein-ligand complex and the box boundaries. Finally, a pre-equilibrated system of 216 water molecules is repeated over the simulation box to provide the solvated environment. To neutralize charged systems, appropriate ions ($\mathrm{Na}^+$ or $\mathrm{Cl}^-$, depending on the net charge of the solute molecules) are applied by replacing randomly selected water molecules. Once the simulation box is prepared, an energy minimization process is carried out to remove the atomic clashes and optimize the geometry of all molecules. In the equilibration simulation stage, a thermostat is applied to heat the system from 0 to 300K within 100 ps. The heated system is then further equilibrated to 1 bar in an NPT ensemble for another 100 ps. During the equilibration stage, the bonds for molecules are constrained. For the final production, the leap-frog algorithm~\cite{van1988leap} is used for integrating Newton’s equations of motion and a Particle Mesh Ewald (PME)~\cite{essmann1995smooth} method is used for calculating long-range electrostatic interactions. The LINCS~\cite{hess1997lincs} algorithm is adopted for resetting all bonds to their correct lengths after an unconstrained update. Finally, the production is performed for 100 ns with an integration time step of 2 fs.

Following the above protocol, we generate MD simulation trajectories for 1500 protein-ligand complexes. To facilitate model training, each 100-ns simulation trajectory from the production run is divided into segments of length 1 ns, resulting in 100 trajectory segments for each complex system. The protein-ligand complex is mapped to the center of the primary simulation box by applying periodic boundary conditions to ensure the integrity and connectivity of the molecules. 

\subsection{Energy Evaluations in Physics-Informed Diffusion Pre-training} \label{sec:energy-eval-in-pidp}

For physics-informed diffusion pre-training (PIDP), the energy and gradients are evaluated using OpenMM~\cite{eastman2017openmm} following the settings used in Folding@home~\citep{zimmerman2021sars}. The amber14sb force is used for the ablation study in Supplementary Sec.~\ref{appx:ablation-study}. The Generalized Born solvent model~\cite{mongan2004constant} is used for solvation energy calculation. 

\subsection{Density Functional Theory Computation}

We use DFT for the grid search of adsorbate configurations on catalyst surfaces, for verifying the adsorbate configuration distributions predicted by DiG as shown in \figref{catalyst_adsorbate}. They are carried out using VASP6.3 with setups compatible with OC20~\citep{chanussot2021open}. Specifically, periodic boundary conditions and projector-augmented wave pseudopotentials are adopted with plane-wave electron kinetic energy cut-off of 350 eV. Generalized gradient approximation and the revised Perdew-Burke-Ernzerhof (RPBE) functional are employed~\citep{hammer1999improved, kresse1996efficient}. The Monkhorst-Pack grid is used to sample the reciprocal space. For the electronic degree of freedom, the convergence criteria for self-consistent computations is set to $1\e{-3}$ eV/atom. For ionic degree of freedom, the relaxations are carried out only on the atomic coordinates with the lattice being fixed. Convergence is considered to be reached when the Hellmann-Feynman forces are smaller than 0.02 eV/\r{A}.

\section{Additional results} \label{appx:additional-results}

\subsection{Protein Conformation Sampling}

\begin{table}[thb]
    \centering
    \caption{Protein systems utilized in this paper. Reference structure 1 is denoted in cyan, while reference structure 2 is denoted in brown in Fig.~\ref{fig:2}b.}
    \begin{tabular}{c|c|c|c}
        \toprule
         Protein & Ref. 1 & Ref. 2 & TMscore \\
         \midrule
         Adenylate Kinase & 4ake (chain A) & 1ake (chain A) & 0.6899 \\
         Lmrb & DEER-AF & 6t1z (chain A) & 0.7600 \\
         human B-Raf kinase & 6uan (chain A) & 3skc (chain A) & 0.9235 \\
         D-ribose & 3dri (chain A) & 1urp (chain A) & 0.7187 \\
         \bottomrule
    \end{tabular}
    \label{tab:protein-systems}
\end{table}

We present a list of proteins employed to demonstrate the efficacy of DiG in generating multiple conformations in \figref{2}b in Table~\ref{tab:protein-systems}. The table provides a detailed overview of the protein systems utilized in this study, including the reference structures denoted in {\color{cyan}cyan} and {\color{brown}brown}, respectively, and the conformational differences between the two reference structures, measured in TMscore.

For both RBD and main protease proteins, there are some structures in the PDB dataset used for DiG model training. We project those structures onto the reduced 2D space spanned by the first two TICA coordinates, and show the results in Supplementary Fig.~\ref{fig:figs1}. Clearly, multiple states are observed in the maps indicated by the diamonds, but these structures together only represent a small fraction of the structure space. In contrast, the DiG-generated structures show much broader overlaps with MD simulations in the structure space. In the case of RBD, DiG even predicted a new region (lower right, cluster-IV in the main text), which has no experimental determined structure. This particular region was significantly sampled by MD simulation. We can observe clear correspondence for the regions sampled by DiG and MD simulations. The diverse structures reveal more information about the functional states of proteins.

Besides the qualitative comparison of the distributions generated by DiG and those sampled by MD simulations. We provide a detailed quantitative measurement of the overlaps of explored regions by these two methods. Taking the regions with MD simulation structures as references, the coverage analysis is carried out following a standard binary classification approach (see Supplementary Fig.~\ref{fig:figs1}). The distributions are first converted to masked binary maps (divided to $50 \times 50$ regions), then the two sets of binary maps (DiG results vs. MD simulation results) are compared. The accuracy, precision, recall, and F1-score are computed for each case, at different sampling data sizes. There are 50,000 structures generated by DiG for each protein. We see that the coverage (recall) increases as more structures are included for the analysis, while the precision level is kept at high levels. In order to compare with the MD simulation data, we took two approaches to draw samplings: (1) take structures consecutively from simulation trajectories; (2) take structures randomly from all simulation trajectories (i.i.d sampling). We observed that the sampling coverage and efficiency of DiG are better than MD simulations. 

\begin{figure}[!p]
    \centering
    \includegraphics[width=\textwidth]{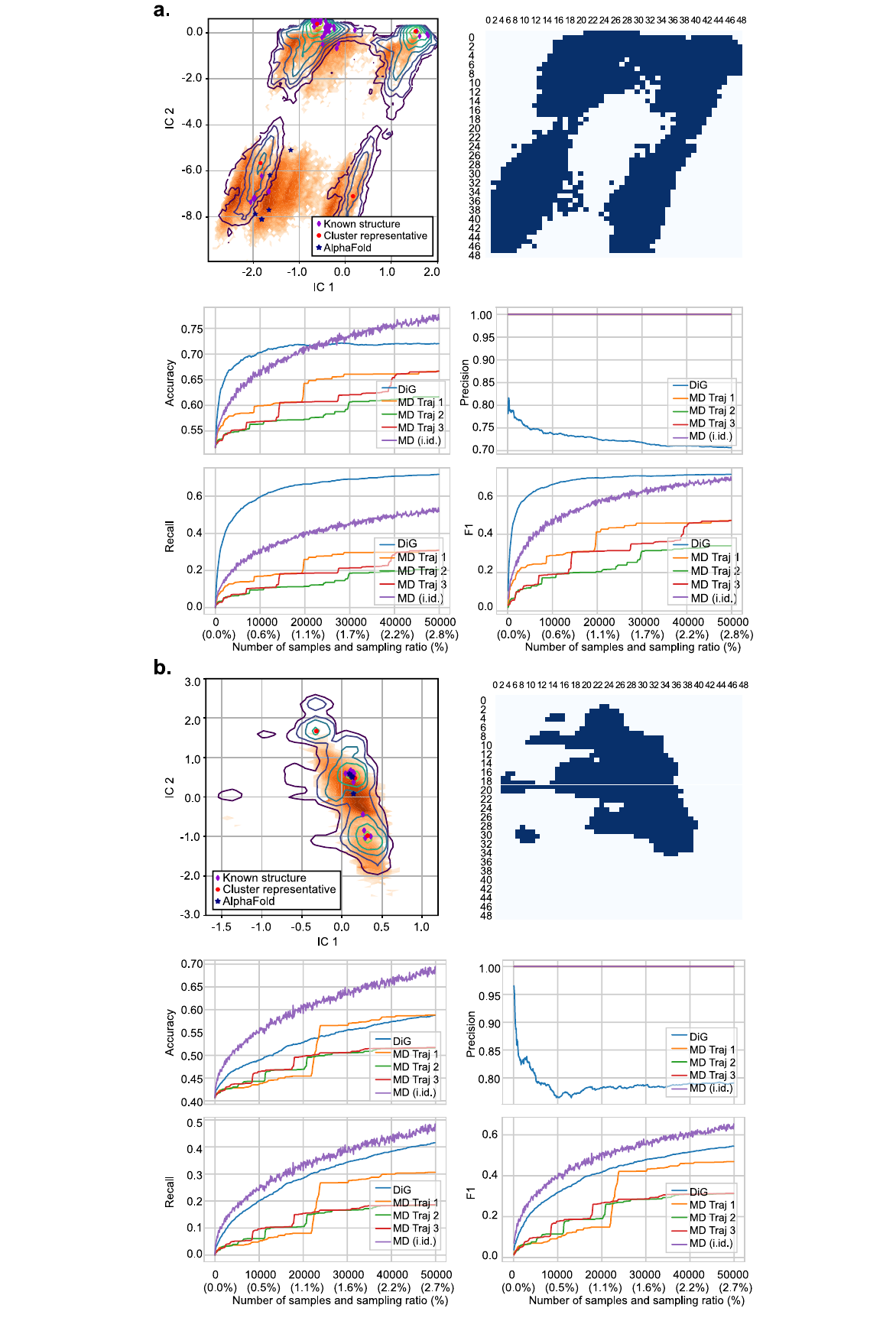}
    \caption{\textbf{Protein structure distributions and sampling coverages.}} 
\end{figure}
\begin{figure} [t!]
    \captionsetup{labelformat=adja-page}
    \ContinuedFloat
    \caption{\textbf{a.} Results for the RBD of SARS-CoV-2 spike protein. Experimentally determined structures are mapped to the reduced 2D space, indicated using the purple diamond symbols. On the top right panel, the same space is divided into 50x50 grids, which are classified into explored (blue) and unexplored (white) sub-regions, depending on the presence of MD simulation structures. The accuracy, precision, recall, and F1-score are shown as a function of sampling size (or the ratio to the whole dataset). The cluster in the lower-right region has no experimental structures, indicating a new state revealed DiG, which is consistent with MD simulations. \textbf{b.} Results for the main protease following the same analysis protocol and representations. In both plots \textbf{a} and \textbf{b}, blue star symbols indicate the AlphaFold predicted structures in the 2D space, and red circles show the cluster centers of MD simulation structures.}
    \label{fig:figs1}
\end{figure}

For two representative ligand-protein systems shown in Fig.~\ref{fig:ligand}b, structures from 100 ns MD simulations are superposed to show the variation in ligand structures (see Supplementary Fig.~\ref{fig:figs2}).

\begin{figure}[!ht]
    \centering
    \includegraphics[width=\textwidth]{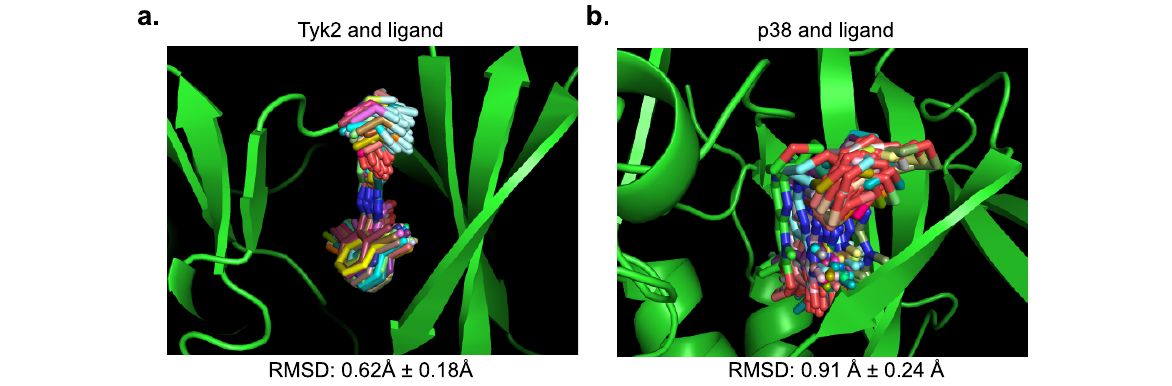}
    \caption{\textbf{Ligand structures observed in MD simulations.} \textbf{a.} For the case of Tyk2 target, the binding pocket is deep and well confined, MD simulations show highly similar ligand structures and binding poses. The RMSD values compared to the crystal structure is small. \textbf{b.} the MD simulation results are shown for the case of P38 protein. Similar to the DiG results, the ligand structure exhibits larger variations compared to the cases of tyk2. For both proteins, the same ligands shown in Figure 2 in the main text are used in the simulations, and the simulation duration is 100 ns for both systems.}
    \label{fig:figs2}
\end{figure}

The ligand structure generations are carried out for a set of 16 proteins, each paired with various numbers of ligands. In total, there are 409 ligand-protein systems in this testing dataset. In Supplementary Fig.~\ref{fig:figs3}, five proteins, each with four different ligands, are shown to illustrate the best structures generated by DiG. The ligand binding poses and atomic structures generated by DiG exhibit diversity and are correlated with the characteristics of protein pockets.

\begin{figure}[!ht]
    \centering
    \includegraphics[width=\textwidth]{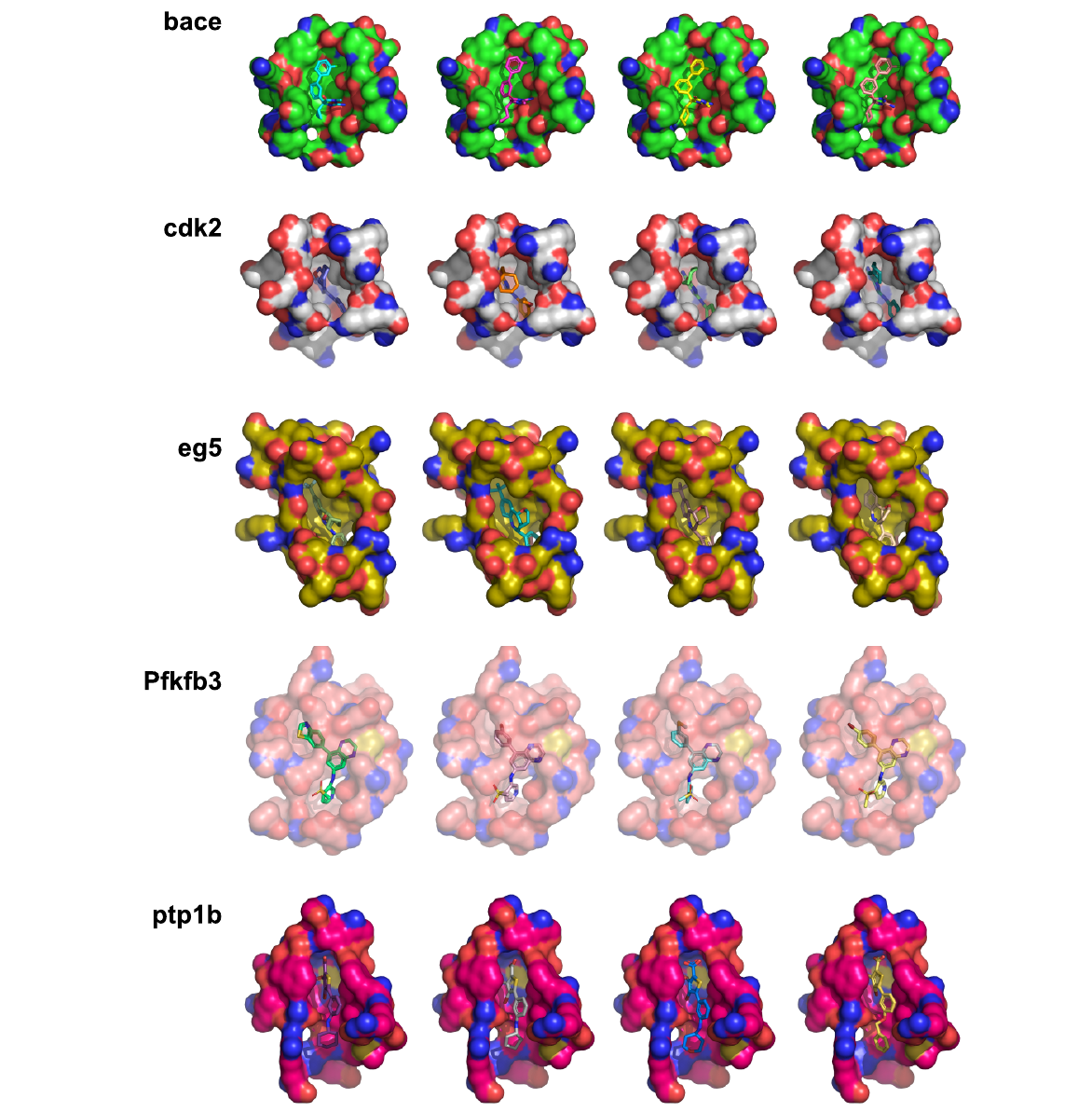}
    \caption{\textbf{Ligand structure generation in the binding pocket of target proteins.} The target names are indicated in the figure, with each row showing different ligands with its best binding poses to the same target protein. Here, best binding poses is defined as the most similar structure to the experimental observations.}
    \label{fig:figs3}
\end{figure}

\subsection{Catalyst-Adsorbate Sampling}

\begin{figure}[!ht]
    \centering
    \includegraphics[width=\textwidth]{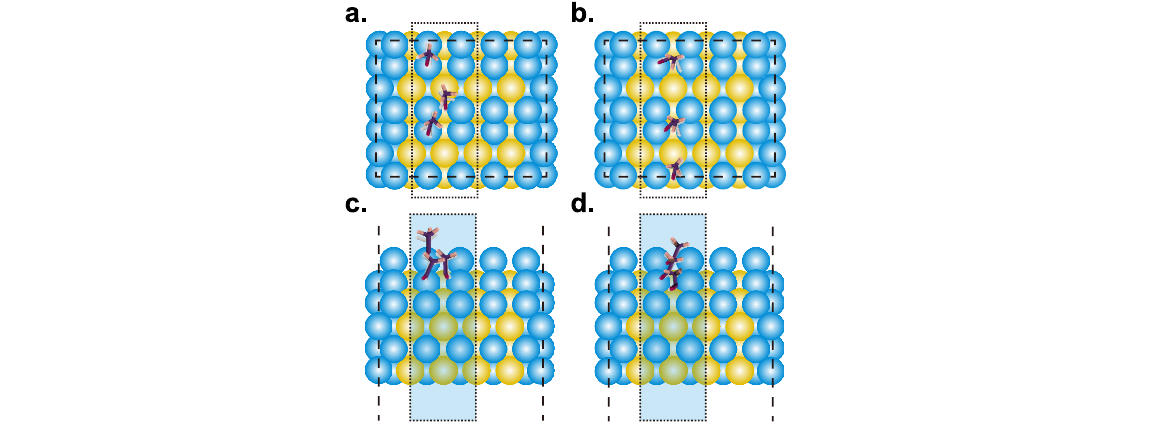}
    \caption{Additional Results for catalyst surface adsorption. All adsorption configurations found by DiG, with the configurations from model in color and the configurations from grid search with DFT in white. The number of adsorption sites is 6 in total. We divide all the sites into 2 groups in (a)(c) and (b)(d), and show both the top view in (a)(b) and the front view in (c)(d). }
    \label{fig:figs4}
\end{figure}

Supplementary Fig.~\ref{fig:figs4} shows top and front views of all adsorption configurations generated by DiG, which covers all the adsorption configurations found by DFT relaxation for this system by traversing initial positions of the adsorbate. Note that for most systems, with a very short MD trajectory as provided in the dataset, it can only cover one or two structures close to these relaxed configurations. Thus, the result shows the ability of DiG to generate to unseen systems from very short MD trajectories in the training set.

\subsection{Ablation Study} \label{appx:ablation-study}

We conduct an ablation study to investigate the effect of various components in the training pipeline of DiG for protein systems. The training pipeline consists of three stages: (i) initialization from experimental data, (ii) physics-informed diffusion pre-training (PIDP), and (iii) training with simulation data.

\begin{figure}[!ht]
    \centering
    \includegraphics[width=\textwidth]{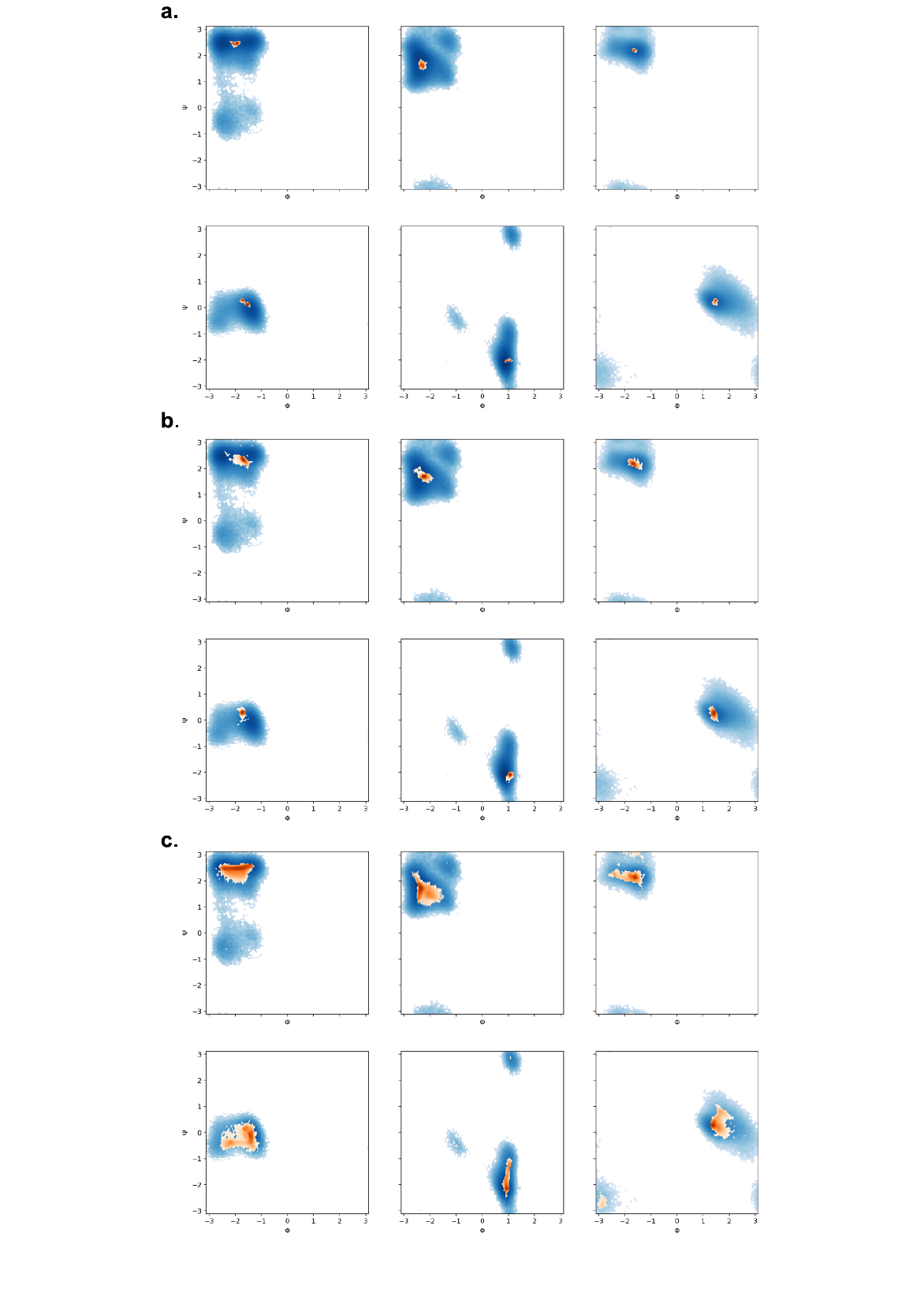}
    \caption{Ramachandran plots of the sampled structures at different stages of the optimization process (initialization, PIDP, and simulation data training) compared with the reference MD structures.}
    \label{fig:appx-pidp-ablation}
\end{figure}

The main protease of SARS-CoV-2 is selected as a case study since it has long simulation trajectories (2.6 ms)~\citep{zimmerman2021sars} and has been studied in Section~\ref{sec:result-prot-sampling}. We investigate the importance of each stage of the training pipeline by evaluating the quality of structures sampled at different stages. The quality is measured by comparing the torsion angle distribution of sampled structures with that of the simulation trajectories. We use Ramachandran plots to visualize the distribution of torsion angles of the 6 amino acids corresponding to the first 10 TICA components of the simulation trajectories~\citep{RAMACHANDRAN196395}. 

Supplementary Fig.~\ref{fig:appx-pidp-ablation} summarizes the results of the ablation study. We obtain approximately 100,000 filtered results from each of the three stages. The blue area represents the ground-truth simulation distribution, which encompasses a relatively large and diverse range of torsion angles, reflecting the dynamic and flexible nature of the protein. Supplementary Fig.~\ref{fig:appx-pidp-ablation}a displays the distribution of structures sampled by DiG initialized from experimental data, i.e., the PDB protein structure dataset~\citep{wwpdb2019protein} (see Supplementary Sec. \ref{sec:training-details-protein}). The distribution is highly concentrated on a single point, indicating that the sampled structures are only fit to experimental structures and do not capture the protein's dynamics. Supplementary Fig.~\ref{fig:appx-pidp-ablation}b illustrates the distribution of structures sampled by DiG after PIDP, which improves the generated distribution towards the equilibrium distribution using the energy function. We find that PIDP training may generate some failure cases with very high RMSDs or low TMscores compared to the crystal structure. Therefore, we filter out results with RMSD higher than $10\,\AA$ and TMscore lower than 0.6. 

However, PIDP alone is still insufficient to fully capture the equilibrium distribution.  
Supplementary Fig.~\ref{fig:appx-pidp-ablation}c displays the distribution of structures sampled by DiG after further training with simulation data, where simulation structures serve as direct signals to supervise DiG's learning. The distribution becomes even more similar to the equilibrium distribution, demonstrating DiG's ability to learn from simulation data and generate realistic and diverse structures.

All sampled structures at various stages of the training pipeline are physically reasonable and are similar to low energy values, as verified by the structural quality metrics. The results show that DiG can effectively combine information from both energy function and simulation data to produce high-quality structures reflecting the protein systems' equilibrium distribution.

\subsection{Reproducibility}

\begin{figure}[!ht]
    \centering
    \includegraphics[width=\textwidth]{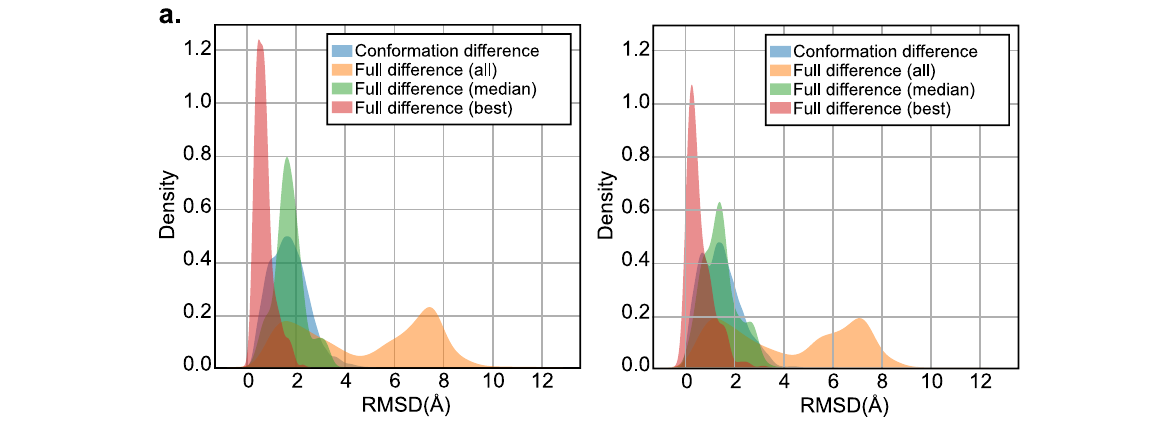}
    \caption{Reproducibility experiments. Taking the ligand structure generation as an example, different DiG models were trained by varying training parameters. The results of the two trained DiG models show high similarity in terms of ligand structure differences compared to crystal structures.}
    \label{fig:s6}
\end{figure}

In this section, we investigate the reproducibility of DiG training, under different initialization, and different random seeds which affect the order of data batches fed into the model during the training. We conduct this experiment on the protein-ligand systems and compare the distribution of the generated ligand structures with respect to the crystal structures. Fig.~\ref{fig:s6} shows the histograms of RMSD statistics for ligand structures generated by DiG. The left panel is identical to Fig.~\ref{fig:ligand}a, where the results are obtained from the model checkpoint with the training pipeline described in Supplementary Sec.~\ref{sec:protein-ligand-sampling-details}, which is used to generate all results for protein-ligand systems in this paper. The initialization of this model is from pre-training on the CrossDock dataset. The right panel shows the results from another model checkpoint without any pre-training but a random initialization. Other hyperparameters except the random seeds are kept the same in the two training processes. We observe that the distributions from the two model checkpoints are very similar, suggesting that DiG training is robustly reproducible.

\section{Limitations} \label{appx:limitation}

\subsection{Limitations on Data Quantity and Quality}
\label{appx:limit-data}
One of the major challenges and limitations of DiG is the scarcity of data for training and evaluating the deep learning models for equilibrium distribution prediction. The ground truth data of equilibrium distribution of different molecular systems are not easily available, as they require massive computational resources and time to generate by molecular dynamics simulation or other methods. Therefore, we only have access to very little data for some molecular systems, and the data quality and quantity may not be sufficient to support the learning and generalization of DiG.

For example, for the catalyst systems, we use the Open Catalyst dataset~\citep{chanussot2021open}, which contains DFT-based molecular dynamics simulations of catalyst-adsorbate systems for only 80 or 320 femtoseconds. However, this simulation time is too short to capture the dynamics and transitions of the systems, and the structures may not move significantly from their initial positions. Thus, we need to traverse initial positions of the adsorbate to find out all the adsorption configurations within a unit cell.  With longer MD trajectories, the model should be able to sample more adsorption configurations from a single initial position of adsorbate. Moreover, the adsorption configurations in the MD trajectories, which start from the relaxed configurations, tend to have low energies, and high-energy configurations are rare in the dataset. Thus, the density for high-energy configurations estimated by the learned DiG may be inaccurate.

For property-guided structure generation, we use a dataset of 15,697 carbon crystal structures~\citep{lucarbondataset} to train the model, which is also very limited compared to the huge space of possible carbon polymorphs. The generative model trained on such data may not be able to recover all the structures in the dataset, let alone generalize to unseen carbon polymorphs. For example, our conditionally generated structures (including 2, 4, 6, 8 carbon atoms per unit cell) only match $88.33\%$ of the structures in the dataset with the same numbers of atoms, using the \texttt{StructureMatcher} from Pymatgen.

For the protein conformation sampling, we collect MD simulations of about only 1000 proteins, each with 100 nanoseconds of simulation time. However, this amount of data may not be enough to cover the diversity and complexity of different protein structures and functions, especially for large and complex proteins that may have longer time scales and more energy barriers for conformational changes. Furthermore, the desired equilibrium distribution for protein model training is not well represented by the available data. Although some simulated data have sufficient length to approximate the equilibrium distribution, they are too scarce to support the neural network models with robust generalization and practical accuracy. Therefore, we resort to using a larger number of experimental structures, such as the PDB dataset, as the initial training data for the models. However, these data influence the final distribution learned by the models. We observe that these data lead to more accurate structures and better generalization, but also to a more concentrated learned distribution. Besides the issue of simulation length, the accuracy of simulation may also be problematic. We find that some structures simulated by molecular dynamics in some systems deviate too much from the experimental structures in terms of structural accuracy. Moreover, in some systems molecular dynamics is highly sensitive to the initial state, and different initial states can result in different distributions in practice. These factors compromise the use of simulated data as approximations of the equilibrium distribution and cause the model to learn inaccurate distributions.

Similar issues also exist in protein-ligand training, where the simulation time does not warrant equilibrium distributions. The simulation trajectories used for ligand-structure model training are limited to 100 ns. Yet, we observed ligand dissociation from the pocket in some of the systems, which were excluded from model training. 
Moreover, our training set only covers a small fraction of systems. In the CrossDocked and our self-generated MD simulation datasets, there are about only 1000 unique proteins, which may affect the generalization of our model.

The data scarcity also affects the evaluation of DiG, as we do not have enough ground truth data of equilibrium distribution to compare with the predictions of DiG. Therefore, we have to rely on indirect metrics, such as the energy function, the structural quality metrics, or properties, to measure the quality and diversity of the generated structures. However, these metrics may not fully capture the accuracy and reliability of the equilibrium distribution prediction, and may have some limitations or biases themselves. For example, the structural quality metrics may not account for the dynamic and stochastic nature of the molecular structures, and may have some dependencies on the reference structures or the alignment methods. The property prediction may not be sensitive to the subtle changes or variations of the molecular structures, and may have some noise or uncertainty in the measurements or the models.

Therefore, we acknowledge that data scarcity is a serious limitation for DiG, and we hope that more and better data of equilibrium distribution of molecular systems can be generated and shared in the future, to enable more robust and reliable learning and evaluation of DiG and other equilibrium distribution prediction methods.

Data limitations also affect the training of the property predicting model $q_\clD(c \mid \bfR)$, which guides the structure generation based on properties. 
In the sampling process for a desired property value $c$ as described by \eqnsref{rev-disc,conditional-score}, the structures in early stages (i.e., large $i$ or $t$) are nearly random noise, for which the predictor model $q_\clD(c \mid \bfR)$ are not typically trained on, making its contribution $\nabla_\bfR \log q_\clD(c \mid \bfR)$ less controlled. This may not be as harmful as it appears, since even with an oracle property predictor, what the early stage of sampling does is still refining the random structure to physically reasonable ones in which the predictor model contribution $\nabla_\bfR \log q_\clD(c \mid \bfR)$ does not dominate. The effect of this process also largely aligns with the desired property since for which a random structure is unlikely to achieve. 
However, we do observe that in some cases small perturbance to the structure causes significant changes in the band gap prediction using the M3GNet predictor, which makes sampling structures with demanded band gap more challenging. This also adds to the evidence of overfitting of the predictor to the limited stable structures. We also find that the model produces more unphysical structures that violate the geometric or energetic constraints when conditional generation. For example, a conditionally generated structure close to a graphite may not have perfect bond angles of exactly $120^\circ$. The band gap predictor model is trained on stable carbon polymorphs. But structures in the denoising process can be quite unstable. Thus the gradients from the predictor used to guide the denoising process may be inaccurate and does not always lead to physical structures. Adding guidance from an energy prediction model in the conditional generation process may guide to more physical structures. Training the property predictor with more abundant carbon polymorphs can also improve the quality of conditionally generated structures.
Thus, it would still be helpful to generate labeled data on more noisy structures and train the property predictor model on them, so that the model could take effect earlier in the sampling process to better guide the structure to the desired property.

\subsection{Limitations on Energy Function} \label{sec:limitation-energy-function}

\newcommand{\CG}{\mathrm{CG}}
\newcommand{\FG}{\mathrm{FG}}

In applications involving proteins, we adopt the common choice of a coarse-grained representation for proteins to reduce the dimensionality of the problem while maintaining most of the structural features. This nevertheless incurs challenges from the energy function (equivalently, force field) side: we have to convert a full-atom force field to the coarse-grained level. This is required in the PIDP training as shown in \eqnsref{pinnloss-protein-so3,pinnloss-protein-acarbon}, while employing an established coarse-grained force field is neither suitable since it is unnecessarily the coarse-grained version of the full-atom force field used in the simulation to generate the dataset. Such a conversion is conducted in Alg.~\ref{alg:protein_pidp_full_to_cg_ff}, but this is not precise for coarse-graining for statistical use.
Specifically, if denoting the invertible transformed full-atom coordinate $\bfRb$ and energy function as $(\bfR_\CG, \bfR_\FG)$ and $E(\bfR_\CG, \bfR_\FG)$ where $\bfR_\CG$ denotes the coarse-grained coordinates ($(\bfC, \bfqq)$ in Alg.~\ref{alg:protein_pidp_full_to_cg_ff}) and $\bfR_\FG$ the fine-grained details ($\bfX$ excluding $\bfC$ in Alg.~\ref{alg:protein_pidp_full_to_cg_ff}), the required coarse-grained energy (equilibrium free energy) under temperature $T$ would be:
\begin{align}
    E_\CG(\bfR_\CG) = -k_\rmB T \log \int \exp\lrbrace{-\frac{E(\bfR_\CG, \bfR_\FG)}{k_\rmB T}} \dd \bfR_\FG.
    \label{eqn:cg-eng-and-full-atom-eng}
\end{align}
In practice, this integral is hard to evaluate, and is a long-standing problem in statistical mechanics and Bayesian statistics.
Even in the case of Alg.~\ref{alg:protein_pidp_full_to_cg_ff} where we have access to the full-atom coordinates of a query structure, the estimation is still an approximation. To see this, the gradient (negative force) of \eqnref{cg-eng-and-full-atom-eng} can be written as:
\begin{align}
    & \nabla_{\bfR_\CG} E_\CG(\bfR_\CG)
    = \bbE_{p_T(\bfR_\FG \mid \bfR_\CG)} [\nabla_{\bfR_\CG} E(\bfR_\CG, \bfR_\FG)], \\
    & p_T(\bfR_\FG \mid \bfR_\CG) := \frac{ \exp\lrbrace{-\frac{E(\bfR_\CG, \bfR_\FG)}{k_\rmB T}} }
    { \int \exp\lrbrace{-\frac{E(\bfR_\CG, \bfR_\FG)}{k_\rmB T}} \dd \bfR_\FG },
    \label{eqn:cg-grad}
\end{align}
so in principle, the coarse-grained gradient is an average of the full-atom gradient over samples from $p_T(\bfR_\FG \mid \bfR_\CG)$.
Under this perspective, the rigid body assumption that Alg.~\ref{alg:protein_pidp_full_to_cg_ff} is based on can be understood as assuming $p_T(\bfR_\FG \mid \bfR_\CG)$ only concentrates on one value of $\bfR_\FG$ (i.e., a Dirac delta distribution), meaning $\bfR_\FG$ can be uniquely determined from the given $\bfR_\CG$. In the algorithm, this $\bfR_\FG$ is provided from the corresponding full-atom coordinates.
This is a good approximation if the true $p_T(\bfR_\FG \mid \bfR_\CG)$ distribution indeed concentrates at the determined $\bfR_\FG$ value; otherwise (e.g., there are very flexible residue or in relatively high temperature), a more precise coarse-graining method for the energy function is required (e.g.~\citep{koehler2023flow}).

Moreover, in PIDP training \eqnref{pinnloss}, although the samples $\{\bfR_{\clD,0}^{(m)}\}_{m=1}^M$ for evaluating the loss can be taken as any that are relevant to the problem in principle, overly loosely chosen structures may cause numerical difficulties as the corresponding gradient energy gradient would be too large. This is the limiting issue from using normal mode perturbed structures hence we have to resort to MD structures.
A possible approach to mitigate this limitation is using a ``milder'' energy function, which does not increase its value as steeply on off-equilibrium structures. For PIDP training, the energy function only needs to indicate a very small probability, and it does not matter much how small it is, as all small values almost equally indicate a vacuum.

\subsection{Limitations on Model Architecture and Scale}

Another crucial limitation of DiG is the model restriction, which is resulted from the compromise between the model capacity and the required computational resource. The model capacity determines the expressiveness and generalization ability of the deep learning models for equilibrium distribution prediction, while the computational resource determines the availability and speed of the training and inference processes. In this work, we have to face the constraint of the computational resource and make some choices that may affect the performance of DiG.

For example, for the protein systems, we use a 12-layer Graphormer with about 80M learnable parameters, which is relatively small considering the complexity and diversity of protein structures and distributions. The model capacity of DiG may not be enough to capture the intricate and high-dimensional energy landscapes and distributions of protein systems, and this can be evidenced by the structural quality of the generated protein structures. We observe that smaller models are easily outperformed by larger models. For example, a 4-layer Graphormer with about 10M learnable parameters only produces a median TM-socre~\cite{zhang2004scoring} of 0.46 on the PDB validation dataset, while a 12-layer Graphormer can easily reach more than 0.8.

Besides, we fix the parameters of the pre-trained Evoformer module in AlphaFold, which is used to extract the features from the protein sequence and the MSA. Evoformer is a powerful and sophisticated module that can encode rich and informative features for protein structure prediction, but it is also computationally expensive and complex, and hasn't been fine-tuned during the training of DiG for predicting equilibrium distribution. This may lead to a significant performance drop since the frozen parameters of Evoformer restrict the expressiveness of DiG very much. In Supplementary Fig.~\ref{fig:figs1}a and~\ref{fig:figs1}b, the high-density regions of MD simulation are perfectly aligned with both known structures and predicted structures by AlphaFold, but there is an observable shift of the high-density regions generated by DiG. We suspect that the shift is due to the limitation of model capacity. Specifically, a fixed Evoformer implemented in DiG does not perform as well as a learnable Evoformer that is used in AlphaFold. A similar observation is that, although for RBD protein, both AlphaFold and DiG can generate high-quality structures with TMscores $> 0.8$ in all cases, their performances are different for the case of main protease, where AlphaFold could generate high-quality structures, but structures with TMscores $>0.8$ generated by DiG only accounted for about $6.8\%$ of all generated structures (The average TMscore of all generated structures of main protease by DiG is about 0.64). This performance gap between AlphaFold and DiG can be possibly caused by the fixed Evoformer. If so, the performance of DiG could be significantly improved if we can fine-tune the Evoformer module with the data and objective of DiG.

Moreover, in this work, we mainly focus on algorithm development, but not on model architecture development. We mainly use the existing deep learning architectures, such as Graphormer and Evoformer. More advanced and specialized architectures that can better exploit the 3D conformational information and the physics principles of molecular systems may improve the performance and efficiency of DiG.

Therefore, we acknowledge that the model architecture restriction is a serious limitation for DiG in the current implementation, and this will be resolved in the future with enhanced capacity of advanced models.

\end{appendices}



\end{document}